%% file: mainDyn_rvsd.tex
\documentclass[a4paper,11pt]{article}
\pdfoutput=1 
\usepackage{jheppub} 
\usepackage[T1]{fontenc} 
\usepackage{epsfig,amsmath}
\usepackage{hepnames,hepunits}
\usepackage{bm}

\usepackage{orcidlink}

\title{Collinear and TMD distributions with dynamical soft-gluon resolution scale}

\author[a,b]{F.~Hautmann\,\orcidlink{0000-0001-7563-687X}}
\author[a]{L.~Keersmaekers\,\orcidlink{0000-0003-3752-9568}}
\author[a]{A.~Lelek\,\orcidlink{0000-0001-5862-2775}}  
\author[a,c]{S.~Sadeghi Barzani\,\orcidlink{0009-0005-2342-269X}}
\author[d]{S.~Taheri Monfared\,\orcidlink{0000-0003-2988-7859}}

\affiliation[a]{University of Antwerp, Belgium}
\affiliation[b]{University of Oxford, United Kingdom}
\affiliation[c]{Shahid Beheshti University, Tehran, Iran}
\affiliation[d]{Deutsches Elektronen-Synchrotron DESY, Germany}

\abstract{Soft-gluon resolution scales 
characterize parton branching Monte Carlo implementations of the evolution equations for  
parton distribution functions in Quantum Chromodynamics (QCD). We examine scenarios 
with dynamical, i.e., branching-scale dependent, resolution scale, and discuss physical implications 
for both collinear and transverse-momentum dependent (TMD) distributions. We perform the first 
determination of  parton distributions with dynamical resolution scale, at next-to-leading order (NLO) in 
perturbation theory,  
from fits to precision deep-inelastic scattering  measurements from HERA. We present an application  
of TMD distributions with dynamical resolution scale  
to Drell-Yan lepton-pair transverse momentum spectra at the LHC 
and lower-energy experiments, and comment on the extraction of 
non-perturbative intrinsic-$k_T$ parameters from Drell-Yan data at small transverse momenta.} 

\keywords{Parton Distributions, Parton Showers, QCD}

\begin{document} 
\begin{flushright}
DESY-25-031 
\end{flushright}

\maketitle

\flushbottom
\unboldmath

\input{sec-introDyn_rvsd.tex}

\input{sec-pbDyn_rvsd}

\input{sec-zmaxfit1_rvsd.tex}

\input{sec-zmaxfit2_rvsd.tex}

\input{sec-concDyn_rvsd.tex}

\vskip 0.3 cm
\input{acknDyn.tex}

\bibliographystyle{JHEP}
\bibliography{bibliogDyn_rvsd}

\end{document}

%% file: sec-introDyn_rvsd.tex
\section{Introduction}
\label{Intro}

The ever increasing precision of 
experimental measurements at present 
and future particle 
colliders~\cite{Azzi:2019yne,LHeC:2020van,FCC:2018byv,Proceedings:2020eah,CEPCPhysicsStudyGroup:2022uwl} poses high demands   
on the performance of 
parton shower Monte Carlo event generators~\cite{Buckley:2019kjt}, 
which are essential 
to provide realistic event simulations and guide collider-physics analyses. 

A long-standing issue in the use of QCD branching algorithms for 
parton showering in hadronic collisions concerns the 
treatment of parton distribution functions (PDFs), 
and more precisely 
the 
possible systematic bias induced by the potential mismatch 
between the (backward) shower evolution of the 
initial state~\cite{Bengtsson:1987kr,Marchesini:1987cf} and the  
PDF evolution 
equations~\cite{Gribov:1972ri,Altarelli:1977zs,Dokshitzer:1977sg}. 

In fact, on one hand shower algorithms make use of PDFs  obtained 
from global fits to experimental data based on the evolution 
equations~\cite{Gribov:1972ri,Altarelli:1977zs,Dokshitzer:1977sg}; 
on the other hand, they incorporate features 
which are not present in these evolution equations and can give rise to 
potential differences. These may come from 
the choice of the 
ordering variable in the branching algorithm; the scale in the strong coupling 
$\alpha_s$, and in particular 
 the dependence of  the coupling on the kinematics 
at each emission vertex in the parton decay chain;  
 the showering scale $q_0$, 
 representing  the minimum transverse momentum with 
which any parton emitted in the partonic cascade can be resolved.   
As a result, even if at a given mass 
scale  the PDF from global fit and 
the PDF used in the shower coincide, 
a possible mismatch may arise at a different mass scale due to evolution.  
See e.g.~\cite{Jadach:2003bu,Tanaka:2003gk,Kurihara:2002ne,Tanaka:2005rm,Golec-Biernat:2006qye,Jadach:2010ew,Kusina:2010gp,Jadach:2012vs,Nagy:2014oqa,Nagy:2020gjv,Hautmann:2017xtx,Hoche:2017hno,prestel:2020,vanBeekveld:2023chs,vanBeekveld:2022ukn,vanBeekveld:2022zhl,Mendizabal:2023mel,Jung:2024eam,Frixione:2023ssx,Jung:2025mtd} for a sample of works 
which have investigated  PDF issues in shower event generators.  

In particular,  the showering scale $q_0$ 
 implies, by energy-momentum conservation, 
an upper bound on the fraction  $z$  of 
longitudinal momentum transferred at each emission.  
This defines, for any shower algorithm, 
a resolution scale in $z$   for the soft gluons emitted in the branching. 
No such upper bound in $z$, in contrast, enters the 
 PDFs which are extracted from global fits based on 
``inclusive'' evolution 
equations~\cite{Gribov:1972ri,Altarelli:1977zs,Dokshitzer:1977sg}, and 
used in the shower.  
See~\cite{Hautmann:2017xtx,Mendizabal:2023mel,Frixione:2023ssx} for recent 
studies focusing on  soft-gluon resolution effects. 

An additional systematic effect arises from 
the transverse momentum recoils along the parton decay chain.  
Kinematic reshuffling transformations are employed to deal with the 
recoils in   shower Monte Carlo 
generators~\cite{Bellm:2015jjp,Bahr:2008pv,Bierlich:2022pfr,Sjostrand:2014zea,Sherpa:2019gpd,Gleisberg:2008ta}.   This gives rise  to shifts in the 
 longitudinal momentum fractions 
 $z$~\cite{Dooling:2012uw,Hautmann:2012dw},  which can 
  bias the use of PDFs in 
 the  shower generators, and 
 ultimately even affect the soft-gluon region.  
 An alternative method  
for incorporating  transverse momentum recoils 
in branching algorithms is to 
bring in 
  transverse momentum dependent 
(TMD) PDFs~\cite{Angeles-Martinez:2015sea,Rogers:2015sqa,Boussarie:2023izj}.  
Through TMD PDFs, one can take into account  consistently    both the 
transverse momenta produced in the shower evolution   and   
the nonperturbative ``intrinsic'' transverse momentum distributions 
characterizing hadronic states at low mass scales. 
From this perspective, the  longitudinal-shift effect  
observed in~\cite{Dooling:2012uw,Hautmann:2012dw} 
due to the transverse recoils 
underlines  the potential interplay of TMD dynamics with 
 the soft-gluon resolution scale in $z$, and in particular with 
the systematics possibly induced 
by the use of collinear  PDFs in parton showers. 

 The 
parton branching (PB) approach proposed 
in Refs.~\cite{Hautmann:2017xtx,Hautmann:2017fcj,BermudezMartinez:2018fsv}, 
is designed 
to consistently treat both collinear and TMD distributions. 
As noted earlier, any  
branching algorithm  will be characterized by the choices of 
the ordering variable, the scale in the strong coupling $\alpha_s$, the soft-gluon 
resolution scale. 
As regards ordering variables, 
by examining transverse momentum, virtuality and angle   
 Ref.~\cite{Hautmann:2017fcj} finds that, while all three can be made to work  
 for the case of collinear PDFs,  the extension to TMDs can only be 
  carried through in a well-prescribed manner by using angular ordering. 
  As regards the strong coupling, by examining 
 the branching scale and the emitted transverse momentum as 
 scales in  $\alpha_s$ it is found that,  while both choices 
 are able to describe inclusive deep inelastic scattering (DIS) structure 
 functions~\cite{BermudezMartinez:2018fsv}, it is only the coupling evaluated at 
 the transverse-momentum scale which is able  to describe 
 the Drell-Yan (DY) di-lepton transverse momentum 
 distributions~\cite{BermudezMartinez:2019anj,BermudezMartinez:2020tys} 
 at small transverse momenta,   and the di-jet azimuthal 
 correlations~\cite{Abdulhamid:2021xtt,BermudezMartinez:2022tql} at 
 large azimuthal angles.  

 As regards the soft-gluon resolution scale, 
the PB 
 approach~\cite{Hautmann:2017xtx,Hautmann:2017fcj} 
 can be applied either with a fixed (i.e., constant) 
  resolution scale or with a running (i.e., branching-scale dependent, or 
 ``dynamical''~\cite{Hautmann:2019biw})  resolution scale. 
 These two scenarios have relevant implications for both 
 theoretical and phenomenological aspects. On the theory side, 
   the former scenario, once it is 
   combined  with the strong coupling evaluated at the 
   branching scale, gives, upon 
    integration over all transverse momenta,    
   the DGLAP~\cite{Gribov:1972ri,Altarelli:1977zs,Dokshitzer:1977sg} evolution 
   equations~\cite{Hautmann:2017fcj}. 
   The latter scenario, once it is 
   combined  with the strong coupling evaluated at the 
   emitted transverse momentum scale, gives, upon  
    integration over all transverse momenta,   
   the CMW~\cite{Marchesini:1987cf,Catani:1990rr,Webber:1986mc} coherent-branching 
   equations~\cite{Hautmann:2019biw}. 

On the phenomenology side, the PB TMD approach with a 
fixed soft-gluon resolution scale and  
strong coupling at the transverse momentum scale    
has been  successfully applied 
to    simultaneously describe  DIS inclusive structure 
functions~\cite{BermudezMartinez:2018fsv} and 
DY transverse momentum 
 distributions~\cite{BermudezMartinez:2019anj,BermudezMartinez:2020tys,Bubanja:2023nrd}. 
In particular, it has been used to make a determination of the nonperturbative 
  ``intrinsic'' 
  transverse-momentum parameter~\cite{Bubanja:2023nrd} from 
  fits to DY experimental 
  measurements 
  across a wide range in DY mass and center-of-mass energy.  
 This determination 
   features  a distinctive energy behavior 
   of the intrinsic 
  transverse-momentum   
   at variance with 
   determinations 
  from tuning~\cite{CMS:2024eprint,CMS:2024ydx,CMS:2019csb,Skands:2014pea,CMS:2020dqt,Buckley:2009bj} of 
  parton shower Monte Carlo 
  generators~\cite{Bellm:2015jjp,Bahr:2008pv,Bierlich:2022pfr,Sjostrand:2014zea}.  
We discuss  these intrinsic-$k_T$ studies 
 (and their possible  interpretation in terms of Sudakov 
  effects~\cite{Martinez:2024mou,Martinez:2024twn,Lelek:2024kax}) 
   later in Sec.~\ref{intrins}.

  Phenomenological implications of PB TMD scenarios with 
  dynamical soft-gluon resolution scale, on the other hand,  have not yet been 
  explored. It is the purpose of the present paper to start such 
  investigations. To this end, 
  we  analyze the precision DIS structure function 
  data~\cite{H1:2015ubc} and perform fits to these data 
  using the implementation of the PB TMD approach in the 
 open-source   fitting platform 
  \verb+xFitter+~\cite{xFitter:2022zjb,Alekhin:2014irh}.  
 We investigate whether good $\chi^2$ 
values can be achieved,  at next-to-leading order (NLO) in perturbation theory, 
using dynamical resolution scales. We find 
that this is the case; thus, from the best fit  we extract 
for the first time 
collinear and TMD PDF sets with dynamical resolution scales. 
Similarly to the case of fixed resolution 
in~\cite{BermudezMartinez:2018fsv},  
we provide the  dynamical-resolution set 
including experimental and model uncertainties. 

Next, we turn to DY transverse momentum spectra. Using the 
NLO  PB sets extracted from DIS fits, we compute theoretical 
predictions for DY transverse momentum 
by applying the method~\cite{BermudezMartinez:2019anj} 
to match NLO DY matrix elements, obtained   from the 
MADGRAPH5$\_$AMC@NLO  program~\cite{Alwall:2014hca} (denoted hereto as MCatNLO),  with 
 PB TMD parton distributions and parton showers implemented in the 
 Monte Carlo event generator   {\sc Cascade}3~\cite{CASCADE:2021bxe}.   
By comparing these predictions with the measurements~\cite{CMS:2022ubq} 
of DY transverse momentum at the Large Hadron Collider (LHC), 
we illustrate that a good description of DY $p_T$ can be obtained 
with dynamical resolution scale, and perform a fit  of  the intrinsic transverse momentum parameter.   
This is the  first  extraction of intrinsic $k_T$ using the PB TMD approach 
in the presence of dynamical resolution scales. We further perform fits to 
DY transverse momentum measurements from lower-energy experiments at Tevatron, RHIC 
and fixed target, obtaining  a good description of the data and extracting 
the intrinsic transverse momentum parameter at different energies.

The collinear and TMD distributions determined from 
the DIS and DY analyses presented in this paper have wide applicability.  
First, 
the ability to properly describe inclusive DIS structure functions 
in a  scenario with dynamical resolution scale (that is, ``shower-like'') 
is an important new result. 
The collinear PDFs with dynamical resolution scale 
extracted from DIS fits in this paper  
are suitable for use in shower Monte Carlo event generators. 

Second, this paves the way for investigations, 
  in the same framework,  of  non-inclusive  
observables describing the detailed structure of DIS final states 
both at HERA~\cite{H1:2024aze,H1:2021wkz}  and at future 
lepton-hadron colliders~\cite{LHeC:2020van,Proceedings:2020eah}.  
 Monte Carlo event generators for DIS are    currently    
being developed~\cite{vanBeekveld:2023chs,Banfi:2023mhz}. 
A distinctive feature characterizing the 
 framework of the present paper is the ability  to  
simultaneously  describe the inclusive DIS structure functions  
and non-inclusive DIS final states.  
The  {\sc Cascade} parton-shower 
Monte Carlo~\cite{CASCADE:2021bxe,CASCADE:2010clj} 
may be used for  these applications  
(see e.g.~\cite{Hautmann:2008vd,Hautmann:2007yok} for 
earlier results on DIS jets).  

Third, the extraction of 
the intrinsic-$k_T$ distribution in TMD PDFs performed in this paper  
from DY transverse momentum measurements 
can  help elucidate the role of the 
 soft-gluon resolution scale and nonperturbative Sudakov form factor~\cite{Martinez:2024mou,Martinez:2024twn}  
 in determining the energy dependence  of  
the transverse-momentum Gaussian width~\cite{Bubanja:2023nrd,CMS:2024eprint}. 

The paper is organized as follows. In Sec.~\ref{sec-pb}
we briefly review the PB TMD formalism, discussing the  
 evolution equations, the nonperturbative contributions and current 
 results on the extraction of intrinsic-$k_T$ parameters from experimental 
 measurements. In Sec.~\ref{sec-zmaxfit1} we perform our fits to 
 precision DIS data using dynamical resolution scales. We present results 
 for integrated and TMD distributions, and  discuss their  
 dependence  on the scale of the minimum transverse momentum 
 for resolved parton emission.  In   Sec.~\ref{sec-zmaxfit2} we describe the 
 application of the TMD distributions with dynamical resolution scale 
 to DY transverse momentum spectra. We give conclusions in 
 Sec.~\ref{sec:concl}.

%% file: sec-pbDyn_rvsd.tex
\section{Branching evolution}
\label{sec-pb}

The analysis performed in this paper is based on 
the formulation of the PB TMD method given in 
Refs.~\cite{Hautmann:2017fcj,Hautmann:2019biw}. 
We refer to these papers for 
the description and details of the method. 
This section provides a short    
recap of its main elements, concentrating 
on the features which will be central to   
 the work presented in the following 
sections. 

Subsec.~\ref{subsec2-1} is devoted to discussing the 
TMD branching evolution and its phase space in terms 
of two branching variables: the branching scale 
$\mu$ (which is related to the kinematic variables 
according to the ordering condition, e.g., angular 
ordering) and the 
longitudinal momentum fraction $z$ (which is 
related to the rapidity of the radiated partons);  
or, alternatively, the longitudinal fraction $z$ 
and the radiated transverse 
momentum $q_\perp$. 

In this subsection we introduce 
the running, or dynamical, resolution scale 
$ z_{\rm{dyn}} (\mu)$. This characterizes the rapidity 
evolution in the branching. 
We illustrate three relevant transverse-momentum 
scales in the branching: 
the intrinsic transverse momentum 
parameter $q_s$, describing the Gaussian width 
of the TMD distribution at low evolution scales; 
the minimum transverse momentum $q_0$ with which 
parton emissions can be resolved, representing the 
showering scale  in the  parton cascade associated
with the evolution; the transverse-momentum scale $q_c$ 
entering the strong coupling $\alpha_s$ according to the 
angular ordering~\cite{Webber:1986mc,Dokshitzer:1987nm,Bassetto:1983mvz} 
and ``pre-confinement'' picture~\cite{Amati:1980ch,Bassetto:1983mvz}. 
The scale $q_c$ is  
 of the same order as $q_0$, but it is conceptually distinct 
from it, and it does not necessarily have the  same numerical value.  

Subsec.~\ref{intrins}    summarizes recent determinations of 
the intrinsic transverse momentum 
parameter $q_s$ from experimental DY measurements at small transverse 
momenta. Here we 
compare results obtained using 
the PB TMD method~\cite{Bubanja:2023nrd} and results obtained from 
the tuning of collinear shower Monte Carlo 
generators~\cite{CMS:2024eprint}.   
We comment on the role of the showering scale and 
soft-gluon resolution  in this comparison.  
This discussion provides a further motivation  for  the 
investigations of the running, dynamical   resolution scale   
which will follow in the next sections. 

\subsection{PB TMD evolution equations and nonperturbative contributions}
\label{subsec2-1}

The PB approach~\cite{Hautmann:2017xtx,Hautmann:2017fcj}  gives 
TMD evolution equations of the schematic form
\begin{eqnarray}
\label{eq:tmdevol} 
 {A}_a ( x, {\bf k}, \mu^2 ) &=& 
 \Delta_a  (\mu^2, \mu_0^{2} )
  {A}_a ( x, {\bf k}, \mu_0^2 )
  \\ 
 &+& 
 \sum_b\int \frac{\textrm{d}^2{\boldsymbol \mu}^{\prime}}{\pi {\mu}^{\prime 2}} 
   \int  \textrm{d}z 
\   {\cal E}_{a b}  [ \Delta ;  P^{(R)};  \Theta   
] 
{A}_b ( x / z ,  {\bf k} + a(z){\boldsymbol \mu}^\prime, \mu^{\prime 2}) \; ,  
\nonumber   
\end{eqnarray} 
where $
A_a  ( x, {\bf k}, \mu^2 )$ is the 
TMD distribution of flavor $a$, carrying the longitudinal momentum  fraction $x$ of the hadron's momentum and  transverse momentum ${\bf k}$ 
at the evolution scale $\mu$; $ \Delta_a (\mu^2, \mu_0^{2} ) $  
 is the Sudakov form factor for the no-emission probability of flavor $a$ 
 from scale $\mu_0$ to scale $\mu$;  ${\cal E}_{a b} $ are the evolution kernels, which 
 are given as  functionals of the Sudakov form factors $\Delta_a$ and of the 
 real-emission splitting functions   $ P_{ab}^{R} $~\cite{Hautmann:2017fcj}, and depend 
 on phase-space constraints collectively denoted by $\Theta$ in Eq.~(\ref{eq:tmdevol}). 
 The functions that appear in the evolution kernels can be computed 
  as perturbation expansions in powers of the strong coupling $\alpha_s$. The 
 explicit expressions of these expansions for all flavor channels 
 are given to two-loop order in Ref.~\cite{Hautmann:2017fcj}.   
 
\begin{figure}[htbp]
\centering 
\includegraphics[width=0.5\textwidth]{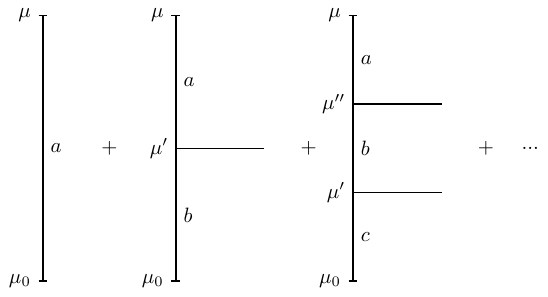}
\caption{Solution of the branching equation by iteration.}  
 \label{fig:iteration}
\end{figure}

 The evolution 
in Eq.~(\ref{eq:tmdevol})
is expressed in terms of two branching variables,  
$z$ and ${\boldsymbol \mu}^\prime$: $z$ is the longitudinal momentum transfer  
at the  branching, while $ \mu^\prime = \sqrt{ {\boldsymbol \mu}^{\prime 2}}$ is the mass 
scale at which the branching occurs. The parton decay chain 
arising from the evolution is pictured in   
Fig.~\ref{fig:iteration} in terms of multiple branchings, $i = 1 , 2 , \dots $. 
 The longitudinal momentum transfers $z_i$ control the 
 rapidities $y_i$ of 
the partons emitted along the parton decay chain,   
$y_i  \sim \ln 1/z_i$.  The mass scales of the branching 
$ \mu_i^\prime$, on the other hand, 
are related to the branching's kinematic variables according to the ordering 
condition.\footnote{Rapidity and mass evolution for TMD distributions can also be  
formulated in a CSS~\cite{Collins:1984kg,Collins:2011zzd}, rather 
than PB, approach. See 
e.g.~\cite{Scimemi:2018xaf,Hautmann:2020cyp}.} 
This is 
specified by the function $a(z)$ in the last factor on the right hand side of 
Eq.~(\ref{eq:tmdevol}). Transverse momentum ordering, virtuality ordering and 
angular ordering are examined in Refs.~\cite{Hautmann:2017xtx,Hautmann:2017fcj}. 
The angular-ordered branching is motivated by the treatment of the endpoint 
region in the TMD case~\cite{Hautmann:2017fcj,Hautmann:2007uw}. 
For angular ordering, one has  $a(z) = 1-z$, and the branching scale  
$ \mu_i^\prime$ is related to the transverse 
momentum $q_{\perp , i} $ of the parton emitted at the $i$-th branching by 
\begin{equation} 
\label{angord} 
\mu_i^\prime  = q_{\perp , i }  / (1-z_i) . 
\end{equation}

    The initial  
evolution scale in Eq.~(\ref{eq:tmdevol}) 
is denoted by $\mu_0$, which is taken to be  
$ \mu_0 >   \Lambda_{\rm{QCD}}  $. It is usually 
of order $\mu_0 \sim {\cal O}$(1 GeV).  
  The   distribution  
  $ {A}_a  ( x, {\bf k}, \mu_0^2 )$ 
  at scale $\mu_0$ in the first term on the right hand side of 
Eq.~(\ref{eq:tmdevol})   represents the intrinsic   $k_T$  distribution. 
   This is  a nonperturbative 
boundary condition to the evolution equation, which  can 
 be arrived at  from comparisons of 
theory predictions to experimental data.  Following 
previous 
applications of the PB TMD method~\cite{BermudezMartinez:2018fsv,Bubanja:2023nrd},  
for simplicity in the calculations that follow 
we will parameterize  $  { {\cal A}}_a(x,{\bf k},\mu^2_0)   $ 
in the form 
\begin{equation}
\label{TMD_A0}
  { {\cal A}}_a(x,{\bf k},\mu^2_0)  =  f_{a} (x,\mu_0^2)  
\cdot \exp\left(-| k_T^2 | / 2 \sigma^2\right) / ( 2 \pi \sigma^2) \; , 
\end{equation}
with $ | k_T^2 |  = {\bf k}^2 $ and  the width of the Gaussian transverse-momentum 
distribution given by  
$ \sigma  =  q_s / \sqrt{2} $,   
independent of parton flavor  and $x$, where 
$q_s$ is the intrinsic-$k_T$ parameter.  

As noted earlier, 
the evolution kernels $ {\cal E}_{a b} $ 
 and the  Sudakov form factors $\Delta_a$  in Eq.~(\ref{eq:tmdevol})  
 depend on kinematic 
 constraints   
  providing the phase space of the branchings 
 along the parton cascade. These can be represented, using the 
 ``unitarity'' picture of QCD evolution~\cite{Webber:1986mc}, 
by separating resolvable and non-resolvable branchings 
 in terms of a  resolution scale to classify 
 soft-gluon emissions~\cite{Hautmann:2017xtx}.  Employing  the branching 
variables $( z , { \mu}^\prime ) $, applications 
of  the PB TMD method have  either taken 
the soft-gluon resolution  parameter 
  to be a finite constant  close to the 
 kinematic limit  $z =1$  
 or allowed for a running, ${ \mu}^\prime$-dependent 
 resolution. In the first case, one has the fixed soft-gluon resolution parameter 
 \begin{equation} 
\label{fixed-zM} 
 z_M = 1 - \varepsilon \;\; , 
\end{equation} 
where $\varepsilon$ is chosen to be a constant $ \ll 1 $. 
In the second case, let $q_0$ be 
the minimum transverse momentum with which any emitted parton can be resolved, 
so that $ q_{\perp , i } \geq q_0$ for any $i$. Using the 
angular-ordering relation in Eq.~(\ref{angord}),  the condition 
for resolving soft gluons is given by $z_i \leq z_M (\mu_i^\prime)$, where the 
running  soft-gluon resolution parameter is 
 \begin{equation} 
\label{running-zM} 
 z_M (\mu^\prime) = 1 - q_0 / \mu^\prime \equiv z_{\rm{dyn}}   \;\; .  
\end{equation} 
Eq.~(\ref{running-zM})  defines the dynamical resolution scale   
  $ z_{\rm{dyn}}   $  
  associated with the angular 
ordering~\cite{Webber:1986mc,Marchesini:1987cf,Catani:1990rr,Hautmann:2019biw}. The transverse momentum scale $q_0$ represents the showering scale, down to which 
resolved parton emissions take place.  Like the scale $\mu_0$, the showering 
scale $q_0$ lies in the region above 
$ \Lambda_{\rm{QCD}}  $, and is usually of order $q_0 \sim {\cal O}$(1 GeV).

\begin{figure}[h]
  \begin{centering}
    \includegraphics[width=0.49\textwidth]{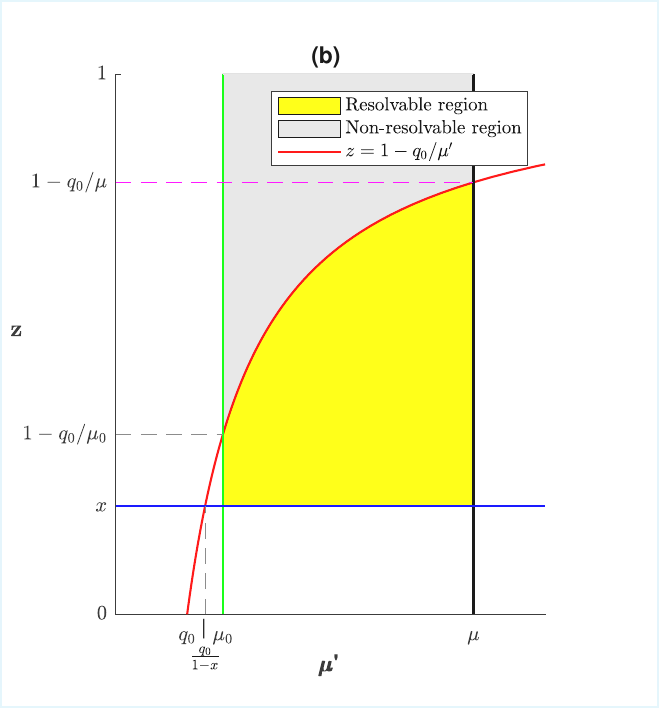}
    \includegraphics[width=0.49\textwidth]{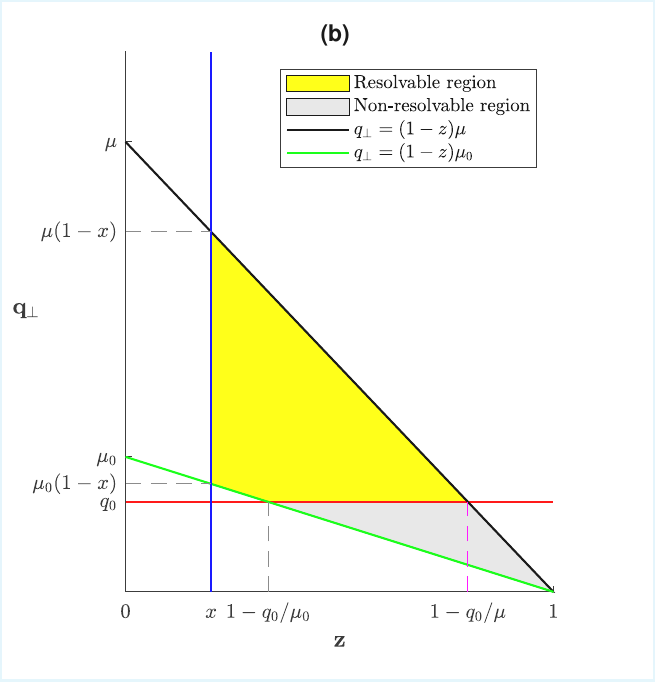}
    \caption{The regions of resolvable and non-resolvable branchings   
   (from   Ref.~\protect\cite{Hautmann:2019biw}),    for $ q_0 /  \mu_0 < 1 - x $,  mapped 
    on  the $(\mu^\prime, z)$ plane (left) and  the $(z , q_\perp)$ plane (right).}
    \label{fig:resolv-region}
  \end{centering}
\end{figure}

Fig.~\ref{fig:resolv-region} depicts the resolvable and 
nonresolvable regions in the partonic branching 
phase space~\cite{Hautmann:2019biw},  mapped  on the 
branching variables $( z ,  \mu^\prime ) $ 
in the left hand side panel and on $ ( q_\perp , z) $ in the right hand side panel. 
The pictures refer to the case in which the ratio of the 
showering scale $q_0$ and starting evolution scale $\mu_0$ is smaller than 
 $1 - x$,   $ q_0 /  \mu_0 < 1 - x $. 
 Analogous pictures are given in Ref.~\cite{Hautmann:2019biw} 
for the complementary case $ q_0 / \mu_0 \geq 1 - x $.

The green-colored line in either panel 
(on the left or on the right of Fig.~\ref{fig:resolv-region}) 
represents the starting evolution scale,  $\mu^\prime =  \mu_0$, at which 
 the nonperturbative intrinsic $k_T$ distribution in 
 Eq.~(\ref{TMD_A0}) is defined. 
The red-colored line in the left hand side panel 
represents the dynamical, running resolution scale $ z_{\rm{dyn}}$ in 
Eq.~(\ref{running-zM}); equivalently, the 
red-colored line in the right hand side panel corresponds to 
 $ q_{\perp  } = q_0$, where $q_0$ is the showering scale. 
 The yellow-colored domains 
in either panel  correspond to the resolvable radiation region, which is treated 
perturbatively in Eq.~(\ref{eq:tmdevol}). The grey-colored domains 
in either panel, on the other hand, involve infrared-sensitive 
radiative processes with $z >  z_{\rm{dyn}}$ and small transverse 
momenta below the scale $q_0$, requiring the modeling of 
nonperturbative contributions.

Both the green-colored line and 
 grey-colored region, in either panel of Fig.~\ref{fig:resolv-region}, 
 are thus controlled  by 
nonperturbative effects. 
The former corresponds to the intrinsic TMD distribution, and  the latter 
corresponds to the nonperturbative component of the Sudakov 
form factor. 
The former may be viewed as a boundary condition to 
the evolution, while the latter as an integral part of the evolution.  
\footnote{An analogous picture can be given   
in the CSS~\cite{Collins:1984kg,Collins:2011zzd} context:  see 
e.g.~\cite{Hautmann:2020cyp,Hautmann:2021ovt}.  
Nonperturbative Sudakov effects, corresponding to the grey-colored regions of  Fig.~\ref{fig:resolv-region},  are then 
embodied by the rapidity evolution Collins-Soper~\cite{Collins:1981uk,Collins:1981va}  kernel.}    

A possible approach to applying the PB TMD method 
is to use fixed resolution scale 
$z_M$ as in Eq.~(\ref{fixed-zM}), and extend 
the evolution kernels  $ {\cal E}_{a b} $ of Eq.~(\ref{eq:tmdevol}) 
into the grey-colored regions in  Fig.~\ref{fig:resolv-region} 
with a suitable treatment of the strong coupling 
$\alpha_s$~\cite{BermudezMartinez:2018fsv,BermudezMartinez:2019anj,BermudezMartinez:2020tys,Bubanja:2023nrd}. Two common scenarios for 
$\alpha_s$ are the following: i) the coupling is evaluated 
at the branching scale,  $\alpha_s =  \alpha_s ( {\mu}^{\prime 2} )$; 
ii) the coupling is evaluated at the transverse momentum of the emission, 
$\alpha_s = \alpha_s ( q_\perp^2 ) 
=  \alpha_s ( { \mu}^{\prime 2}  (1-z)^2 )$. 
Case~i)  corresponds to 
DGLAP
evolution~\cite{Gribov:1972ri,Altarelli:1977zs,Dokshitzer:1977sg} , while 
case~ii) corresponds to angular ordering, 
e.g.~CMW evolution~\cite{Marchesini:1987cf,Catani:1990rr}.\footnote{Angular 
ordering is also used in the KMR approach~\cite{Kimber:1999xc,Martin:2009ii}. 
See~\cite{Hautmann:2019biw,Guiot:2024oja,Aslan:2022zkz,Guiot:2022psv,Nefedov:2020ugj,Guiot:2019vsm,Golec-Biernat:2018hqo}   
for recent discussions of both infrared and ultraviolet issues in this case.}     
While  either one can be used for 
totally inclusive cross sections~\cite{BermudezMartinez:2018fsv}, 
case ii) is required for the description 
 of differential transverse-momentum distributions and 
 angular 
 correlations~\cite{BermudezMartinez:2019anj,BermudezMartinez:2020tys,Abdulhamid:2021xtt,BermudezMartinez:2022tql}. 

We observe  
in Fig.~\ref{fig:resolv-region} 
that  the grey-colored regions 
do not require any further 
treatment of the strong coupling in case i) (as $\mu^\prime \geq \mu_0$), while 
they do in case ii) (as $q_\perp \lesssim q_0$). 
To this end, the strong coupling is modeled according to 
a ``pre-confinement'' picture~\cite{Amati:1980ch,Bassetto:1983mvz} 
as 
 \begin{equation} 
\label{freeze} 
\alpha_s = \alpha_s(\max(q^2_{c},{ q}_{\perp}^2)), 
\end{equation}
where $q_c $ is a semi-hard 
scale on the order of the GeV.    

This approach,  
characterized by the fixed $z_M$ in 
Eq.~(\ref{fixed-zM}) and the strong coupling in Eq.~(\ref{freeze}), 
 has been shown  to give rise to a successful description of 
inclusive DIS structure functions, DY transverse-momentum 
distributions and di-jet angular correlations~\cite{BermudezMartinez:2018fsv,BermudezMartinez:2019anj,BermudezMartinez:2020tys,Abdulhamid:2021xtt}. 
In fact, it has been used to make a determination~\cite{Bubanja:2023nrd} 
 of the  intrinsic-$k_T$ parameter 
$q_s$ in Eq.~(\ref{TMD_A0}) from fits 
to DY experimental measurements across a wide range in 
masses and center-of-mass energies. 

In the present paper, we explore a different approach. Namely, we  
take the running resolution scale 
$z_M$ as in Eq.~(\ref{running-zM}), and  investigate its implications for 
the description of DIS and DY processes. 
On one hand, this is of interest in its own right, as it allows one 
to investigate the role of the boundary between resolvable and unresolvable 
branchings, and deepen the study of nonperturbative effects.  
For instance, the extraction of the intrinsic-$k_T$ distribution  is expected to be sensitive 
to the modeling of soft-gluon resolution~\cite{Bubanja:2023nrd}. On the other 
hand, from the standpoint of general-purpose shower 
Monte Carlo generators~\cite{Bellm:2015jjp,Bierlich:2022pfr} 
the set-up with the running resolution scale (\ref{running-zM})   
may be helpful in  investigating open issues 
about the use of PDFs in collinear 
 showers~\cite{Mendizabal:2023mel,Frixione:2023ssx} and 
 the showering scale. 

We will proceed in a similar fashion to what is done in the 
literature for the fixed-$z_M$ case. That is, 
we  first apply the dynamical-scale scenario in Eq.~(\ref{running-zM}) 
to DIS: in Sec.~\ref{sec-zmaxfit1} we perform fits to precision data 
for DIS structure functions, determining new collinear and TMD PDF sets.  
These results are timely, given the currently planned DIS 
experiments~\cite{LHeC:2020van,Proceedings:2020eah}  
(with  accompanying efforts to develop new Monte Carlo event generators 
for DIS~\cite{vanBeekveld:2023chs,Banfi:2023mhz}). 
 Next, we apply the newly determined TMD distributions 
   to DY transverse momentum spectra: in  
 Sec.~\ref{sec-zmaxfit2} we compare theoretical predictions 
 obtained using  dynamical  resolution scale
 with  experimental measurements~\cite{CMS:2022ubq} of 
 DY transverse momentum,
 and examine the intrinsic-$k_T$ parameter $q_s$. 
This can be regarded as a first step toward a full fit covering DY data from experiments at 
all available energies and DY masses. 

Before we move on to this 
in Secs.~\ref{sec-zmaxfit1} and \ref{sec-zmaxfit2}, we 
conclude the present section with an overview of the 
current status of extractions of intrinsic transverse momentum from 
DY data, based on 
PB TMD calculations~\cite{Bubanja:2023nrd,Zhan:2024lym} on 
one hand, and on tuning of 
general-purpose Monte Carlo 
generators~\cite{CMS:2024eprint,CMS:2024ydx} on the 
other hand. The comparison of the results from these analyses underlines the 
potential impact of the showering scale and soft-gluon resolution on intrinsic $k_T$ 
distributions. It thus reinforces the motivation for the studies of 
dynamical resolution scale in the sections that follow.

\subsection{Extractions of intrinsic transverse momentum from data}
\label{intrins}

Recently, two analyses  
 of intrinsic transverse momentum distributions  from 
 DY  measurements at small transverse momenta have been carried 
 out  which highlight features of 
 the approach described in the previous subsection and of collinear showering  
approaches. The two analyses are that of 
 Ref.~\cite{Bubanja:2023nrd}, based on the 
  implementation of the PB TMD method in the 
 Monte Carlo event generator  {\sc Cascade3}, and 
 that of 
 the CMS Collaboration~\cite{CMS:2024eprint,CMS:2024ydx}, based 
on underlying-event tunes  of Monte Carlo event generators     
    {\sc Pythia8}  and {\sc Herwig7}. 
In this subsection we summarize and compare the results.

\begin{figure}[h]
  \begin{centering}
    \includegraphics[width=0.49\textwidth]{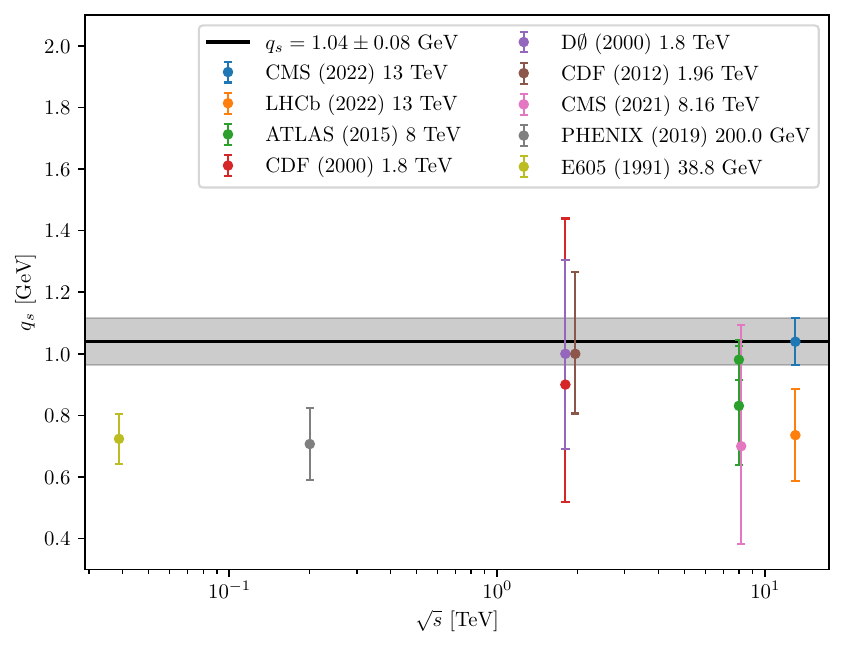}
    \includegraphics[width=0.49\textwidth]{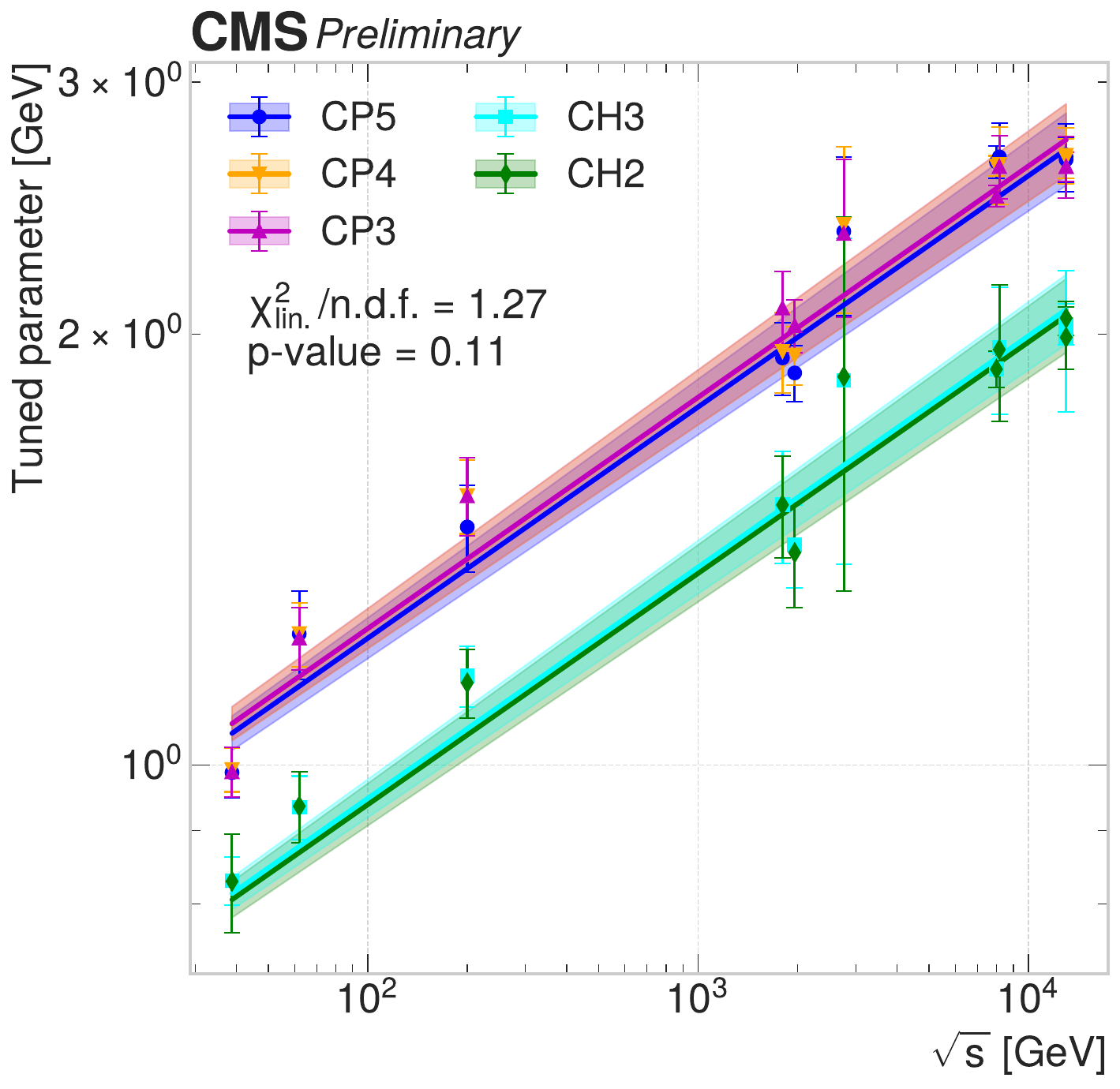}
    \caption{Extraction of intrinsic $k_T$ parameter from measurements of DY transverse momentum 
    distributions at varying    center-of-mass energies. 
    The plot on the left hand side is taken from 
    Ref.~\protect\cite{Bubanja:2023nrd}, and is based on predictions using the TMD parton branching  
    implemented in the Monte Carlo event generator  {\sc Cascade3}. 
   The plot on the right hand side is taken from Ref.~\protect\cite{CMS:2024ydx}, and is based 
   on underlying-event tunes of the Monte Carlo event generators     
    {\sc Pythia8}  (CP3, CP4, CP5) and {\sc Herwig7} (CH2, CH3).}
    \label{fig:extract-intrins}
  \end{centering}
\end{figure}

Ref.~\cite{Bubanja:2023nrd} uses the 
PB TMD approach~\cite{Hautmann:2017xtx,Hautmann:2017fcj} described 
above in this section, with 
the fixed $z_M$ in Eq.~(\ref{fixed-zM}), angular-ordered 
$\alpha_s$ in Eq.~(\ref{freeze}), and TMD PDF set  
PB-NLO-2018-Set2~\cite{BermudezMartinez:2018fsv}. 
It computes predictions for DY transverse momentum distributions 
by using the technique~\cite{BermudezMartinez:2019anj} for 
matching DY matrix elements at NLO obtained from the 
MCatNLO program~\cite{Alwall:2014hca} with TMD parton distributions and 
showers implemented in the Monte Carlo 
generator {\sc Cascade3}~\cite{CASCADE:2021bxe}. 
As discussed earlier, in this 
approach  TMD distributions are systematically introduced and evolved,  
and at the starting scale of evolution they are 
parameterized (see Eq.~(\ref{TMD_A0})) and fitted to experimental data. 
The analysis~\cite{Bubanja:2023nrd} extracts the intrinsic $k_T$ parameter 
$q_s$ by performing fits to DY transverse momentum 
data~\cite{CMS:2022ubq,ATLAS:2015iiu,LHCb:2021huf,D0:1999jba,CDF:1999bpw,CDF:2012brb,PHENIX:2018dwt,CMS:2021ynu,Moreno:1990sf}, 
 including a detailed treatment of statistical, correlated and uncorrelated 
uncertainties.\footnote{It is shown in  Refs.~\cite{Zhan:2024lym,Zhan:2025wup} that 
an  extraction of the $q_s$ parameter similar to that in Ref.~\cite{Bubanja:2023nrd}  
 can be  carried out by using, rather  than the differential cross section,  simpler observables defined from 
 ratios of $p_T$ bins in a coarse-grained binning scenario.}  
The CMS data~\cite{CMS:2022ubq} at center of mass energy 
$\sqrt{s} = $ 13 TeV cover DY invariant masses from 50 GeV to 1 TeV. The other 
data cover energies from 13 TeV down to 38 GeV.   
The results~\cite{Bubanja:2023nrd} for $q_s$ versus energy $\sqrt{s}  $ are 
reported in Fig.~\ref{fig:extract-intrins} in the left hand side panel. 

The CMS analysis~\cite{CMS:2024eprint,CMS:2024ydx} is 
based on  NLO predictions from MCatNLO~\cite{Alwall:2014hca}  
matched with {\sc Herwig7}~\cite{Bellm:2015jjp} and with   
  {\sc Pythia8}~\cite{Sjostrand:2014zea}   parton showers and underlying  
event models. It uses the 
{\sc Herwig7} tunes CH2 and CH3~\cite{CMS:2020dqt}
and  {\sc Pythia8} tunes CP3, CP4 and CP5~\cite{CMS:2019csb} 
for the underlying event. Although TMD distributions are not used, both 
the    {\sc Herwig7} and {\sc Pythia8}  calculations introduce 
intrinsic $k_T$ distributions  parameterized as Gaussian distributions with 
tunable parameters. The analysis~\cite{CMS:2024eprint,CMS:2024ydx} 
extracts the intrinsic $k_T$ parameter from DY transverse momentum 
measurements at the LHC, Tevatron, RHIC and fixed-target experiments. 
The results~\cite{CMS:2024ydx} for the intrinsic $k_T$ parameter 
versus energy $\sqrt{s}  $ are reported in Fig.~\ref{fig:extract-intrins} in 
the right hand side panel. 

 The left and right panels of Fig.~\ref{fig:extract-intrins}
indicate a striking difference in the energy dependence of the intrinsic $k_T$ 
parameter between the TMD parton branching and collinear shower calculations. 
While the right panel shows a strong rise with energy,\footnote{The  rise of intrinsic $k_T$ 
with energy observed in the right panel of Fig.~\ref{fig:extract-intrins} 
confirms earlier results obtained  with   
{\sc Herwig}~\cite{Gieseke:2007ad} and {\sc Pythia}~\cite{Sjostrand:2004pf}.} 
no strong rise is found in the left  panel. 

It is suggested in~\cite{Bubanja:2023nrd} that 
this difference in the energy behavior of the intrinsic $k_T$ 
can be attributed to nonperturbative Sudakov 
effects~\cite{Martinez:2024mou,Martinez:2024twn} arising from the region near the soft-gluon resolution 
boundary.  
With reference to Fig.~\ref{fig:resolv-region},  this means that 
different behaviors of 
intrinsic-$k_T$ distributions along the green-colored line 
in Fig.~\ref{fig:resolv-region}  can reflect 
  different treatments of soft radiative processes populating 
  the grey-colored regions~\cite{Hautmann:2019biw}. \footnote{A related observation has been made in the  context of 
  CSS  studies in Ref.~\cite{Hautmann:2020cyp}, by examining   correlations     
  between the TMD distribution (corresponding to the green-colored line in Fig.~\ref{fig:resolv-region}) and 
the Collins-Soper kernel (receiving nonperturbative contributions from the grey-colored regions in 
Fig.~\ref{fig:resolv-region}).}     
 
  In particular, the picture proposed in~\cite{Bubanja:2023nrd}    suggests that the 
  transition from the nearly-constant behavior of $q_s$ with 
energy in the left panel of Fig.~\ref{fig:extract-intrins} to the power-like rise 
in the right panel of Fig.~\ref{fig:extract-intrins} 
can be related to the change in resolution scales  from 
Eq.~(\ref{fixed-zM}) to Eq.~(\ref{running-zM}), and that this 
can be studied as a function of the showering scale 
$q_0$, entering the dynamical soft-gluon  scale $ z_{\rm{dyn}}$ in 
Eq.~(\ref{running-zM}).\footnote{The idea 
proposed  in~\cite{Bubanja:2023nrd}  is further explored  
in~\cite{Bubanja:2024puv,Bubanja:2024crf,Raicevic:2024yqq,Raicevic:2024obe,Monfared:2024vgc,TaheriMonfared:2024vbr}.}  

The potential 
interplay of $q_s$ and $q_0$ provides an important motivation for the 
systematic studies of dynamical $ z_{\rm{dyn}}$ which are 
at the center of this work.

%% file: sec-zmaxfit1_rvsd.tex
\section{Fits to DIS precision data with dynamical resolution scale}
\label{sec-zmaxfit1}

In this section we present the PB TMD fits to precision DIS 
structure function data~\cite{H1:2015ubc} from HERA, using 
dynamical resolution scales. 
First results from these fits have appeared in~\cite{Barzani:2022msy}. 
The collinear and TMD PDFs obtained from the fits will be made available 
in the TMDlib library~\cite{Abdulov:2021ivr,Hautmann:2014kza}. 

\subsection{Fit procedure and general settings}
\label{Sec:Fit}

The software to fit PB distributions to data from cross sections' measurements within 
the \texttt{xFitter} package~\cite{xFitter:2022zjb,Alekhin:2014irh} was developed 
in Refs.~\cite{Hautmann:2013tba,Jung:2012hy,Hautmann:2014uua,Jung:2024uwc} and 
further applied in Ref.~\cite{BermudezMartinez:2018fsv}. 
The fit procedure is based on the construction of kernels obtained 
from the Monte Carlo solution of the PB equation discussed in the 
previous section, and the use of TMD distributions 
as well as integrated TMD distributions (iTMD), resulting from the 
integration over transverse momenta of the TMDs.   
We refer the reader to the above-mentioned  
references~\cite{Hautmann:2013tba,Jung:2012hy,Hautmann:2014uua,Jung:2024uwc,BermudezMartinez:2018fsv}  for    details on the fit procedure. 

In what follows, we will apply this procedure with the only difference,  
compared to the above references, that the kernels will depend on the 
running soft-gluon resolution scale given by the dynamical $z_M$ 
in Eq.~(\ref{running-zM}).  We will perform fits for different values 
of the showering scale $q_0$ in Eq.~(\ref{running-zM}). 
We will use NLO  splitting functions and 
two-loop running coupling $\alpha_s$ in the kernels, unless 
stated otherwise.  We will set the scale 
$q_c$ in the running coupling in Eq.~(\ref{freeze}) equal to 
$q_0$, unless stated otherwise. 
The result of the fits will be the NLO determination 
of the initial distributions in Eq.~(\ref{TMD_A0}),  parameterized at 
the starting scale $\mu_0$ of the evolution. 

Following the strategy of Ref.~\cite{BermudezMartinez:2018fsv}, 
adapted from the HERAPDF2.0 fits~\cite{H1:2015ubc}, 
the initial parameterizations have the form 
\begin{eqnarray}
xg(x) &=& A_gx^{B_g}(1-x)^{C_g} - A^{\prime}_gx^{B^{\prime}_g}(1-x)^{C^{\prime}_g} \;, \nonumber \\
xu_{v}(x) &=& A_{u_{v}}x^{B_{u_v}}(1-x)^{C_{u_v}}(1+E_{u_{v}}x^2) \;, \nonumber \\
xd_{v}(x) &=&  A_{d_{v}}x^{B_{d_v}}(1-x)^{C_{d_v}}  \;, \nonumber \\
x\overline{U}(x) &=& A_{\overline{U}}x^{B_{\overline{U}}}(1-x)^{C_{\overline{U}}}(1+D_{\overline{U}}x)  \;, \nonumber \\
x\overline{D}(x) &=& A_{\overline{D}}x^{B_{\overline{D}}}(1-x)^{C_{\overline{D}}}\;, 
\label{fitparametrization}
\end{eqnarray}
where the notation for 
parton distributions and parameters follows that of Ref.~\cite{H1:2015ubc}. 
Additional constraints  are  applied at the initial scale $\mu_0$: 
$x\overline{U}= x\overline{u}$ and $x\overline{D}= x\overline{d} +x\overline{s}$.   For the initial strange-quark distribution,  we apply $x\overline{s} = f_sx\overline{D}$ with $f_s=0.4$.  
We set 
$B_{\overline{U}} = B_{\overline{D}}$ and $A_{\overline{U}} = A_{\overline{D}}(1-f_s)$. 
The normalization parameters $A_{u_{v}}$, $A_{d_{v}}$, $A_{g}$ and $A_{g^{\prime}}$ are constrained by the quark number and momentum sum rules. $C^{\prime}_g$ is fixed to $25$ as in the HERAPDF2.0 set~\cite{H1:2015ubc}. We use the same hard-scattering coefficient functions and we treat heavy flavors in the same way as was done in 
Refs.~\cite{H1:2015ubc,BermudezMartinez:2018fsv}.   The intrinsic transverse momentum is generated from the Gaussian distribution with the width $\sigma=q_s/\sqrt{2}$ with $q_s=0.5\;\rm{GeV}$ if not written explicitly otherwise.

The fit is performed 
with the inclusive DIS data from 
the HERA H1 and ZEUS combined 
measurement~\cite{H1:2015ubc} in the range 
of  $3.5 < Q^2 < 50000 \; \textrm{GeV}^2$,  $8\cdot10^{-5}< x < 0.65$. 
The number of data points included in the fit is $1131$. 

The $\chi^2$ is calculated within the \texttt{xFitter} package. 
As in Ref.~\cite{BermudezMartinez:2018fsv}, 
systematic shifts and the treatment of 
correlated and uncorrelated systematic uncertainties are 
included in the $\chi^2$ definition. 
A total of $162$ systematic uncertainties and uncertainties from 
the procedure of combining H1 and ZEUS data are treated 
as correlated.

\begin{figure}[!htb]
\begin{center}
\includegraphics[width=8cm]{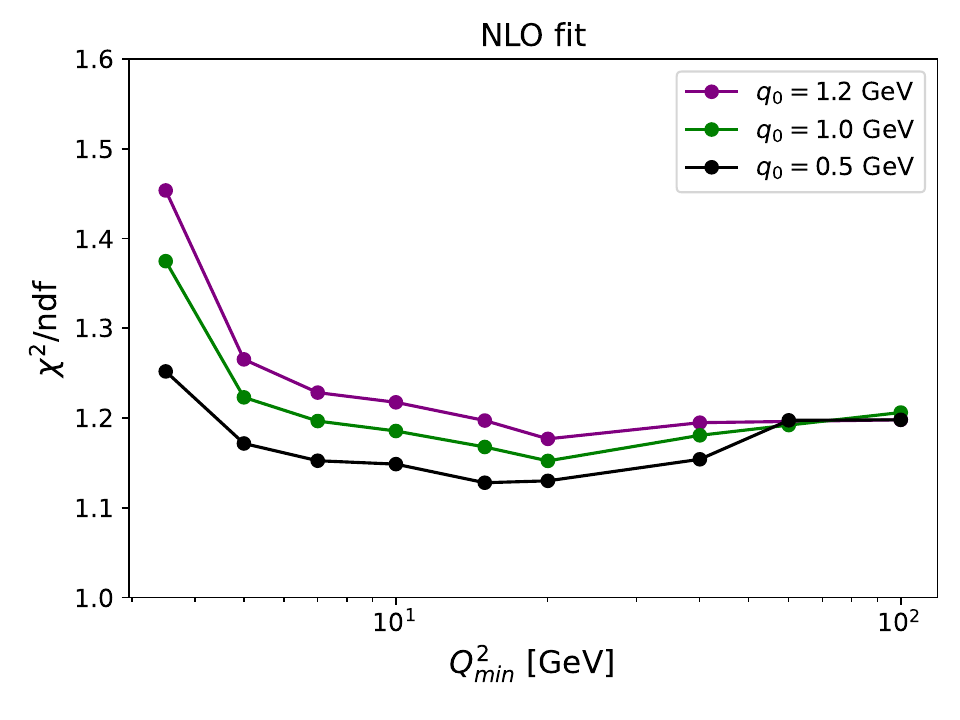}
\end{center}
\caption{The $\chi^2/\rm{n.d.f.}$ value of the NLO dynamical-$z_M$ fit 
  to DIS precision data~\protect\cite{H1:2015ubc}  
as a function of the minimum $Q^2$ of 
the data included in the fit, for different values of $q_0$.}
\label{fig:chi2vsQ2minNLO}
\end{figure}

The experimental uncertainties on the parton 
distributions are obtained  by
using the Hessian method~\cite{Pumplin:2001ct}, 
implemented in \texttt{xFitter},  with $\Delta\chi^2=1$.
 The model uncertainties are obtained by varying the charm ($m_c= 1.41, 1.47, 1.53 \;\textrm{GeV}$) and the bottom mass ($m_b=4.25, 4.5, 4.75 \;\textrm{GeV}$), and the starting evolution scale ($\mu_0^2= 1.6, 1.9, 2.2 \;\textrm{GeV}^2$)  where the 
 numbers in brackets are the lower, central and upper values respectively.

Further refinements of these fits can be envisioned but 
are not included in the present study. For instance, one could imagine 
fitting the intrinsic-$k_T$ scale $q_s$ as well as the scale $q_0$. 
Also, one could study uncertainties due to the 
parameterization choice, and the impact of 
H1 and Zeus combined heavy-quark  data~\cite{H1:2018flt,H1:2012xnw}.  
We leave these developments to future investigations.

\subsection{Description of the data and fitted distributions}
\label{subsec:NLO-DynZm-fits}

We now present  the results of the PB 
fits with dynamical resolution scale,  described in the previous subsection, at NLO. 
We present fits for three different values of the 
showering scale $q_0$: 1.2 GeV, 1 GeV and 0.5 GeV. 
We will refer to these fits as PB-NLO-2025-DynZm, with suffixes
q0-1\_2, q0-1\_0, q0-0\_5  and they will be available in TMDlib.

 The values of $\chi^2/\rm{n.d.f.}$ (number of degrees of freedom)
for these three fits are as follows: 
1.45 (for $q_0=1.2\;\rm{GeV}$); 1.37 (for $q_0=1 \;\rm{GeV}$); 
1.26 (for $q_0=0.5\;\rm{GeV}$).

To examine the behavior of the fits in different kinematic regions of $Q^2$, 
we also perform fits on data subsets characterized by different 
values of the lowest $Q^2$, $Q^2_{\rm{min}}$. The  results for the 
 $\chi^2/\rm{n.d.f.}$ values are reported in Fig.~\ref{fig:chi2vsQ2minNLO}  
 as a function of $Q^2_{\rm{min}}$,  for the fits corresponding 
to the three different values of $q_0$.

\begin{figure}[!htb]
\begin{minipage}{0.31\linewidth}
\includegraphics[width=5.0cm]{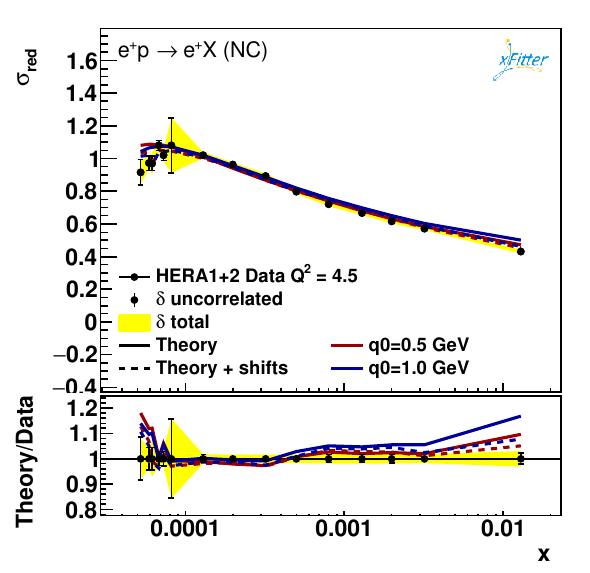}
\end{minipage}
\hfill
\begin{minipage}{0.31\linewidth}
\includegraphics[width=5.0cm]{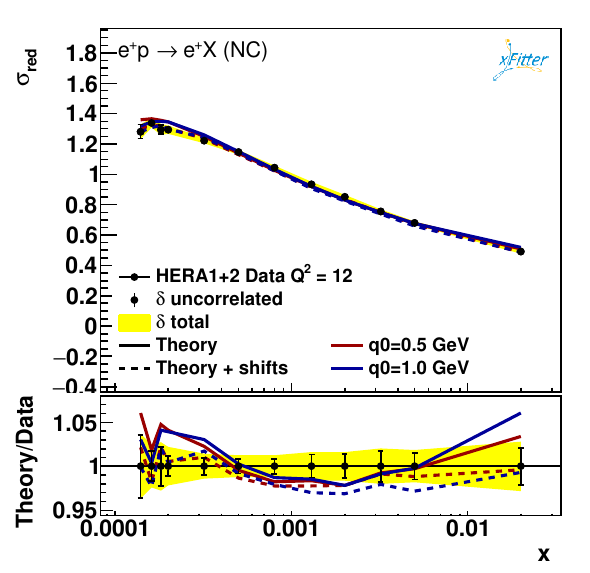}
\end{minipage}
\hfill
\begin{minipage}{0.31\linewidth}
\includegraphics[width=5.0cm]{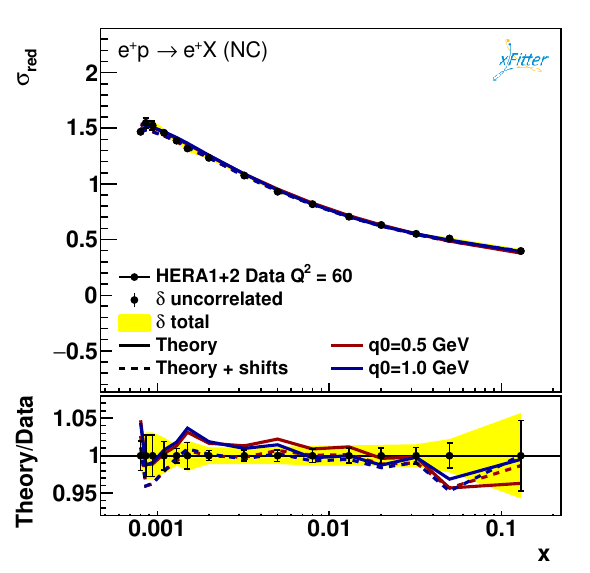}
\end{minipage}
\hfill
\caption[]{Predictions for the inclusive DIS reduced cross section 
obtained  using dynamical $z_M$ 
with $q_0=0.5\;\rm{GeV}$ and $q_0=1.0\;\rm{GeV}$,  
compared to HERA data~\protect\cite{H1:2015ubc}. 
 }
\label{fig:HERAF2}
\end{figure}

 Fig.~\ref{fig:chi2vsQ2minNLO} shows that good $\chi^2$ values 
 are obtained from the NLO fits 
 for all three values of $q_0$, throughout the kinematic range in $Q^2$.  
 The $\chi^2$ improves 
 as $q_0$ decreases in the range of the chosen values. 
 We will analyze this behavior further in the next two subsections, 
 both by examining the impact of the perturbative order on the fit 
 (through a comparison with leading-order (LO) fits 
 in Subsec.~\ref{app:remarks})   and by examining the role of the 
 coupling $\alpha_s$ in Eq.~(\ref{freeze})   
 in the low transverse momentum region 
 (through a comparison with a fit keeping $q_c \neq q_0$ in 
 Subsec.~\ref{subsec:q0notqc}).

\begin{table}[]
\begin{minipage}{0.5\textwidth}

\begin{tabular}{|l|l|l|l|l|l|l|l|l|}
\multicolumn{9}{c}{$q_0=0.5\;\rm{GeV}$}                                                                                                                                                                                                                                                   \\ 
\hline
                          & A      & B       & C      & D   & E       & A'      & B'      & C'   \\ \hline
$xg$                        & 0.49 & -0.10 & 0.86 &     &         & -0.08 & -0.22 & 25.0 \\ \hline
$xu_v$                     & 3.29 & 0.70  & 4.37 &     & 17.95 &         &         &      \\ \hline
$xd_v$                     & 3.29 & 0.87  & 3.45 &     &         &         &         &      \\ \hline
$x\overline{U}$ & 0.16 & -0.14 & 6.49 & 0.0 &         &         &         &      \\ \hline
$x\overline{D}$ & 0.27 & -0.14 & 9.04 &     &         &         &         &      \\ \hline
\end{tabular}

\end{minipage}\\

\begin{minipage}{0.5\textwidth}

\begin{tabular}{lllllllll}
\multicolumn{9}{c}{$q_0=1.0\;\rm{GeV}$}                                                                                                                                                                                                                                                   \\ \hline
\multicolumn{1}{|l|}{}                & \multicolumn{1}{l|}{A}      & \multicolumn{1}{l|}{B}       & \multicolumn{1}{l|}{C}      & \multicolumn{1}{l|}{D}      & \multicolumn{1}{l|}{E}       & \multicolumn{1}{l|}{A'}     & \multicolumn{1}{l|}{B'}      & \multicolumn{1}{l|}{C'}      \\ \hline
\multicolumn{1}{|l|}{$xg$}            & \multicolumn{1}{l|}{1.99} & \multicolumn{1}{l|}{-0.19} & \multicolumn{1}{l|}{7.37} & \multicolumn{1}{l|}{}       & \multicolumn{1}{l|}{}        & \multicolumn{1}{l|}{0.72} & \multicolumn{1}{l|}{-0.24} & \multicolumn{1}{l|}{25.0} \\ \hline
\multicolumn{1}{|l|}{$xu_v$}          & \multicolumn{1}{l|}{4.54} & \multicolumn{1}{l|}{0.74}  & \multicolumn{1}{l|}{5.43} & \multicolumn{1}{l|}{}       & \multicolumn{1}{l|}{17.12} & \multicolumn{1}{l|}{}       & \multicolumn{1}{l|}{}        & \multicolumn{1}{l|}{}        \\ \hline
\multicolumn{1}{|l|}{$xd_v$}          & \multicolumn{1}{l|}{2.35} & \multicolumn{1}{l|}{0.72}  & \multicolumn{1}{l|}{3.69} & \multicolumn{1}{l|}{}       & \multicolumn{1}{l|}{}        & \multicolumn{1}{l|}{}       & \multicolumn{1}{l|}{}        & \multicolumn{1}{l|}{}        \\ \hline
\multicolumn{1}{|l|}{$x\overline{U}$} & \multicolumn{1}{l|}{0.21} & \multicolumn{1}{l|}{-0.09} & \multicolumn{1}{l|}{7.83} & \multicolumn{1}{l|}{4.56} & \multicolumn{1}{l|}{}        & \multicolumn{1}{l|}{}       & \multicolumn{1}{l|}{}        & \multicolumn{1}{l|}{}        \\ \hline
\multicolumn{1}{|l|}{$x\overline{D}$} & \multicolumn{1}{l|}{0.34} & \multicolumn{1}{l|}{-0.09} & \multicolumn{1}{l|}{6.17} & \multicolumn{1}{l|}{}       & \multicolumn{1}{l|}{}        & \multicolumn{1}{l|}{}       & \multicolumn{1}{l|}{}        & \multicolumn{1}{l|}{}        \\ \hline
\end{tabular}

\end{minipage}

\begin{minipage}{0.5\textwidth}
\begin{tabular}{|l|l|l|l|l|l|l|l|l|}
\multicolumn{9}{c}{$q_0=1.2\;\rm{GeV}$}                                                                                                                                                                                                                                                   \\ \hline
                & A      & B       & C       & D      & E       & A'     & B'      & C'     \\ \hline
$xg$            & 4.89 & -0.12 & 12.25 &        &         & 3.29 & -0.09 & 25.000 \\ \hline
$xu_v$          & 4.30 & 0.71  & 5.52  &        & 16.90 &        &         &        \\ \hline
$xd_v$          & 2.43 & 0.72  & 3.92  &        &         &        &         &        \\ \hline
$x\overline{U}$ & 0.24 & -0.07 & 78.37  & 5.36 &         &        &         &        \\ \hline
$x\overline{D}$ & 0.40 & -0.07 & 7.59  &        &         &        &         &        \\ \hline
\end{tabular}
\end{minipage}

\caption{Parameter values of the fitted initial distributions for fits with $q_0=0.5,  \; 1.0$ and $1.2\;\rm{GeV}$. }
\label{tab:fitParam}
\end{table}

 The description of the data for the 
 inclusive-DIS reduced cross section 
 is illustrated in Fig.~\ref{fig:HERAF2} 
 for the cases $q_0=0.5\;\rm{GeV}$ and $q_0=1.0\;\rm{GeV}$. 
As expected from  the results above,  the 
 differences between the sets do not influence very much  
 the description of the inclusive-DIS  data, and the quality of the 
different  fits  is similar. 
 The results for the fitted 
 parameters of the initial distributions 
 are shown in Tab.~\ref{tab:fitParam}.\footnote{One  
may note that for $q_0=0.5\;\rm{GeV}$ the parameter $D$ for $x\bar{U}$ is fixed to $0.0$. The reason for this is that the fit with the default parameterization of Eq.~(\ref{fitparametrization})  resulted in a negative value of this parameter, leading to a negative quark distribution. To avoid it, the fit was repeated with $D=0$ for $x\bar{U}$.}

\begin{figure}[!htb]
\begin{minipage}{0.31\linewidth}
\includegraphics[width=5.5cm]{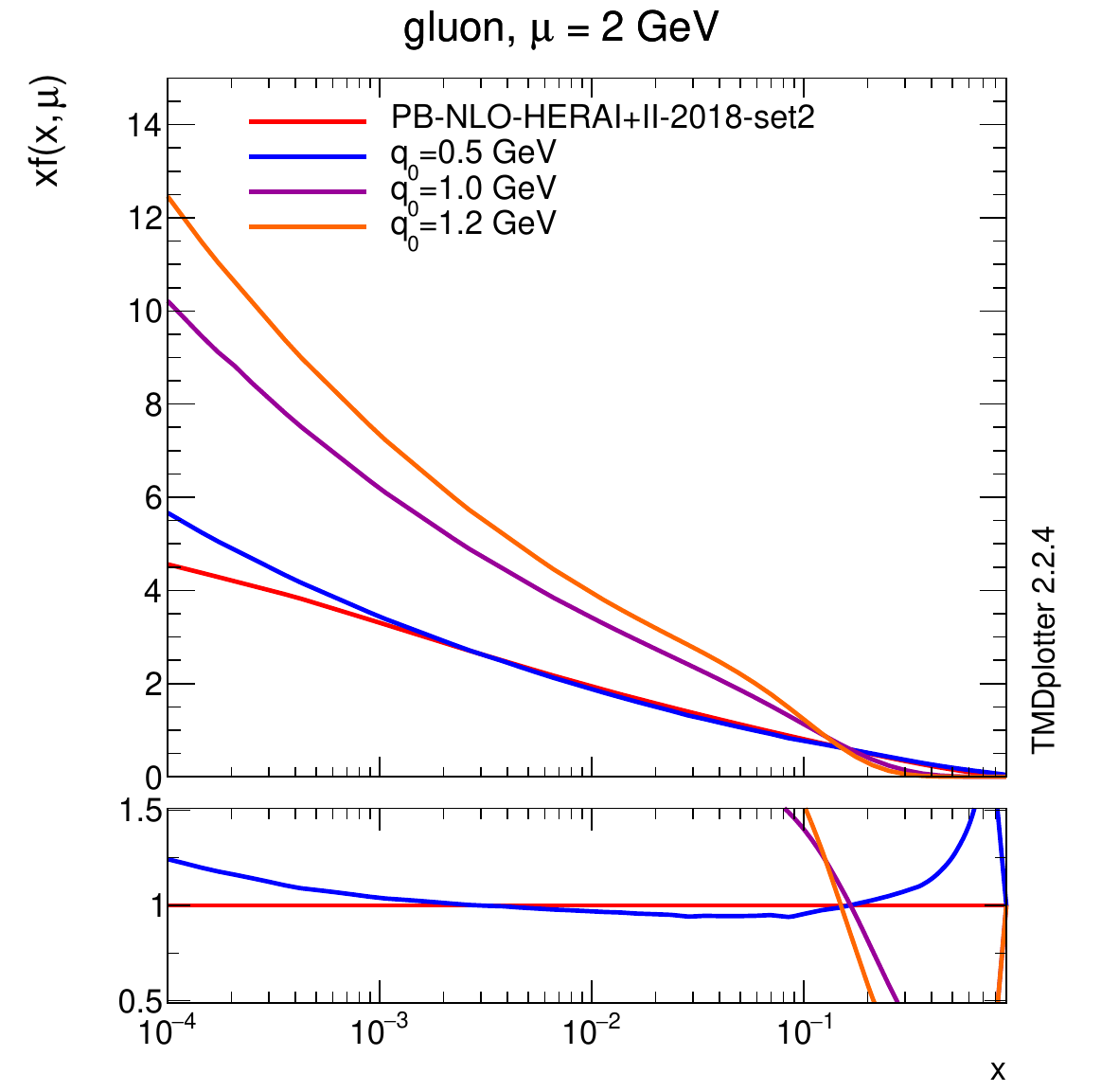}
\end{minipage}
\hfill
\begin{minipage}{0.31\linewidth}
\includegraphics[width=5.5cm]{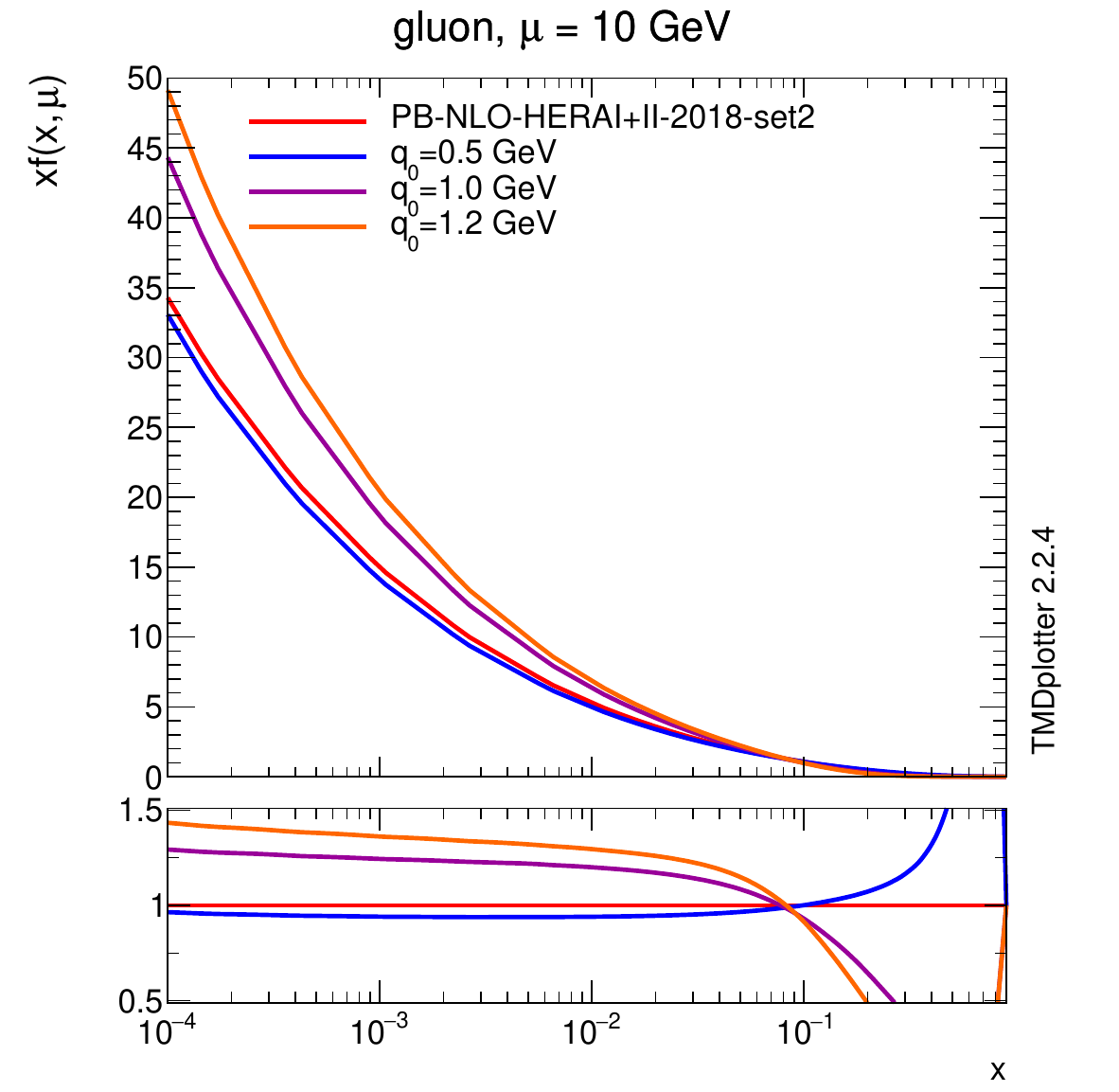}
\end{minipage}
\hfill
\begin{minipage}{0.31\linewidth}
\includegraphics[width=5.5cm]{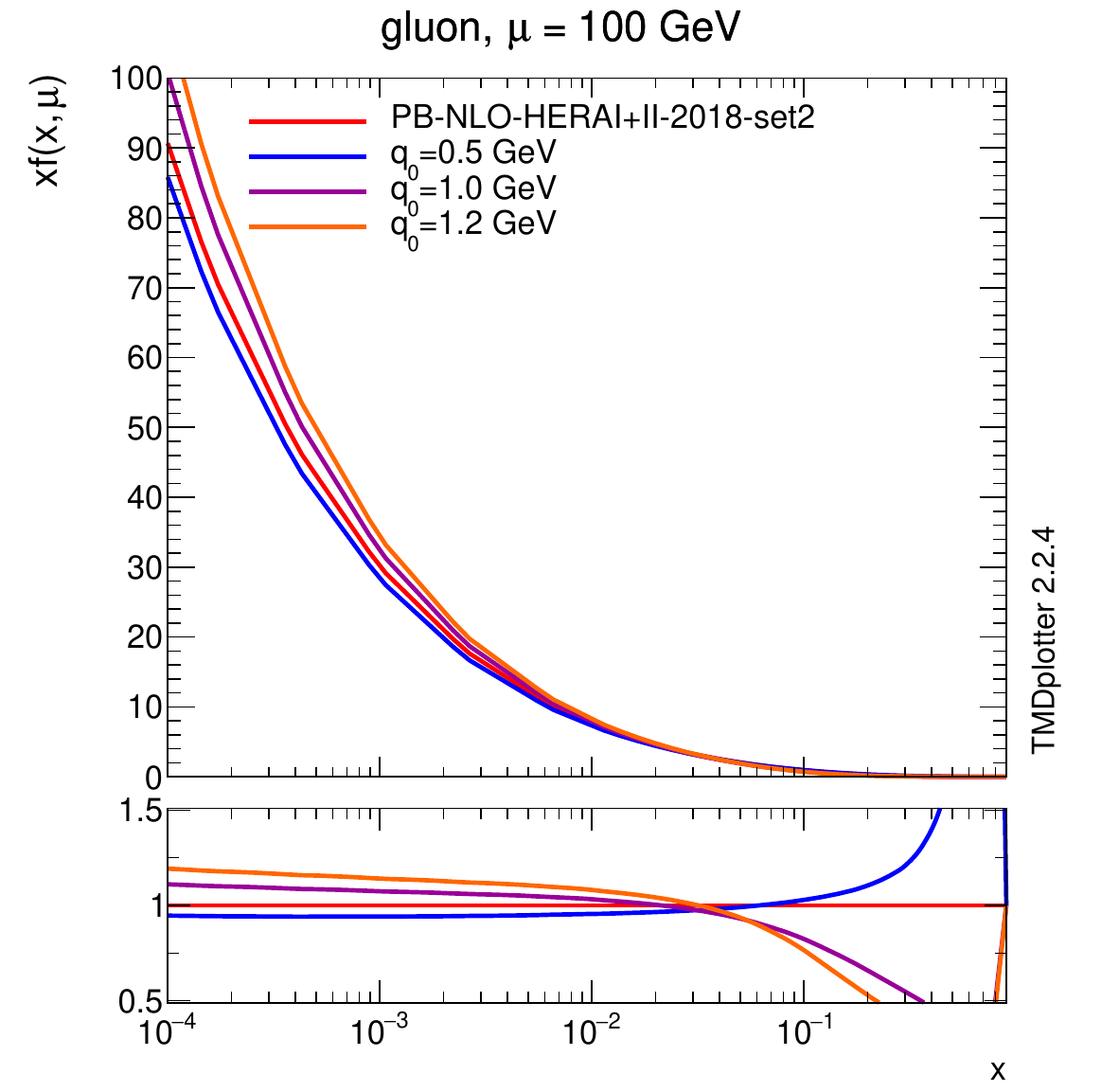}
\end{minipage}
\hfill
\begin{minipage}{0.31\linewidth}
\includegraphics[width=5.5cm]{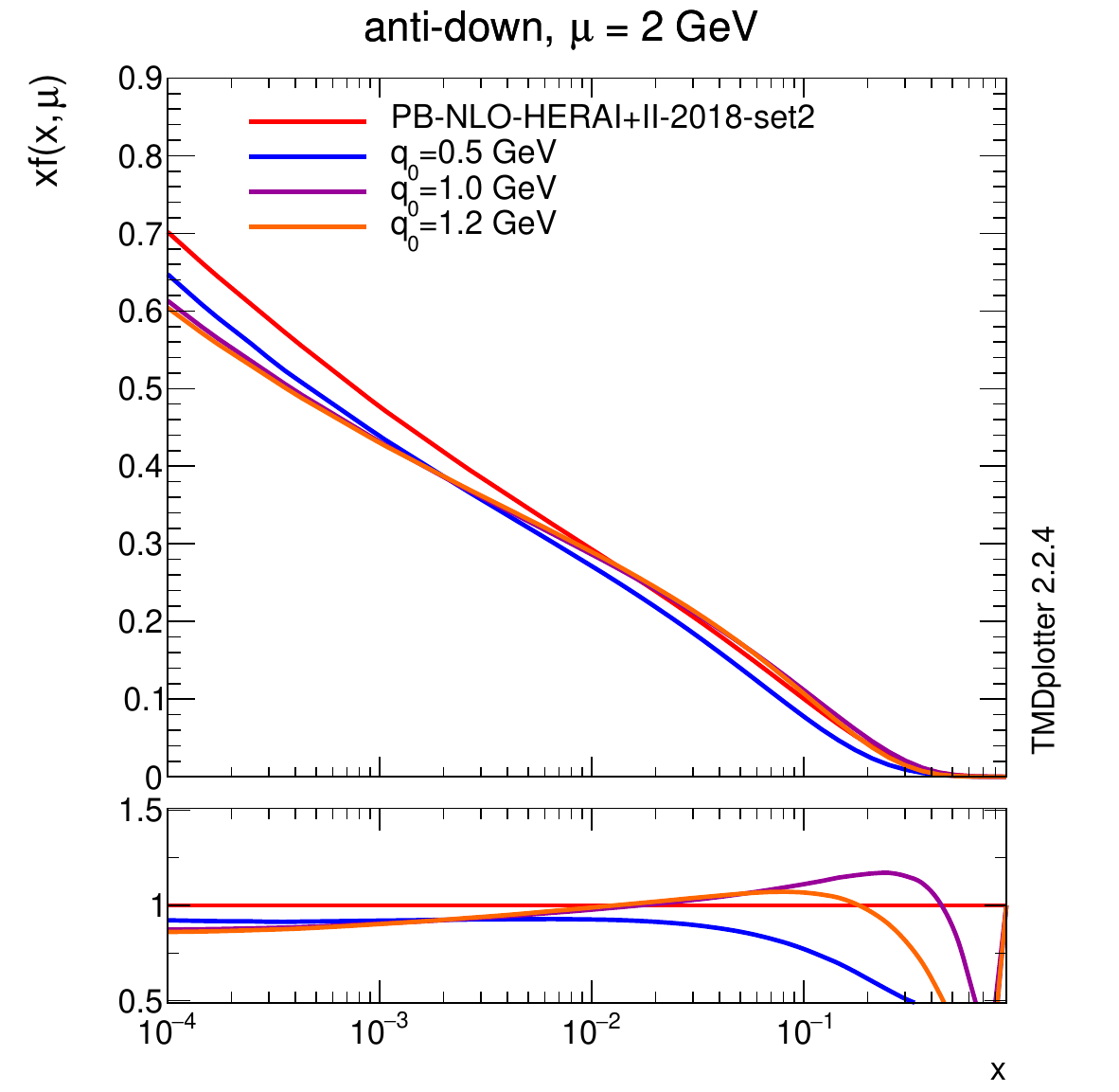}
\end{minipage}
\hfill
\begin{minipage}{0.31\linewidth}
\includegraphics[width=5.5cm]{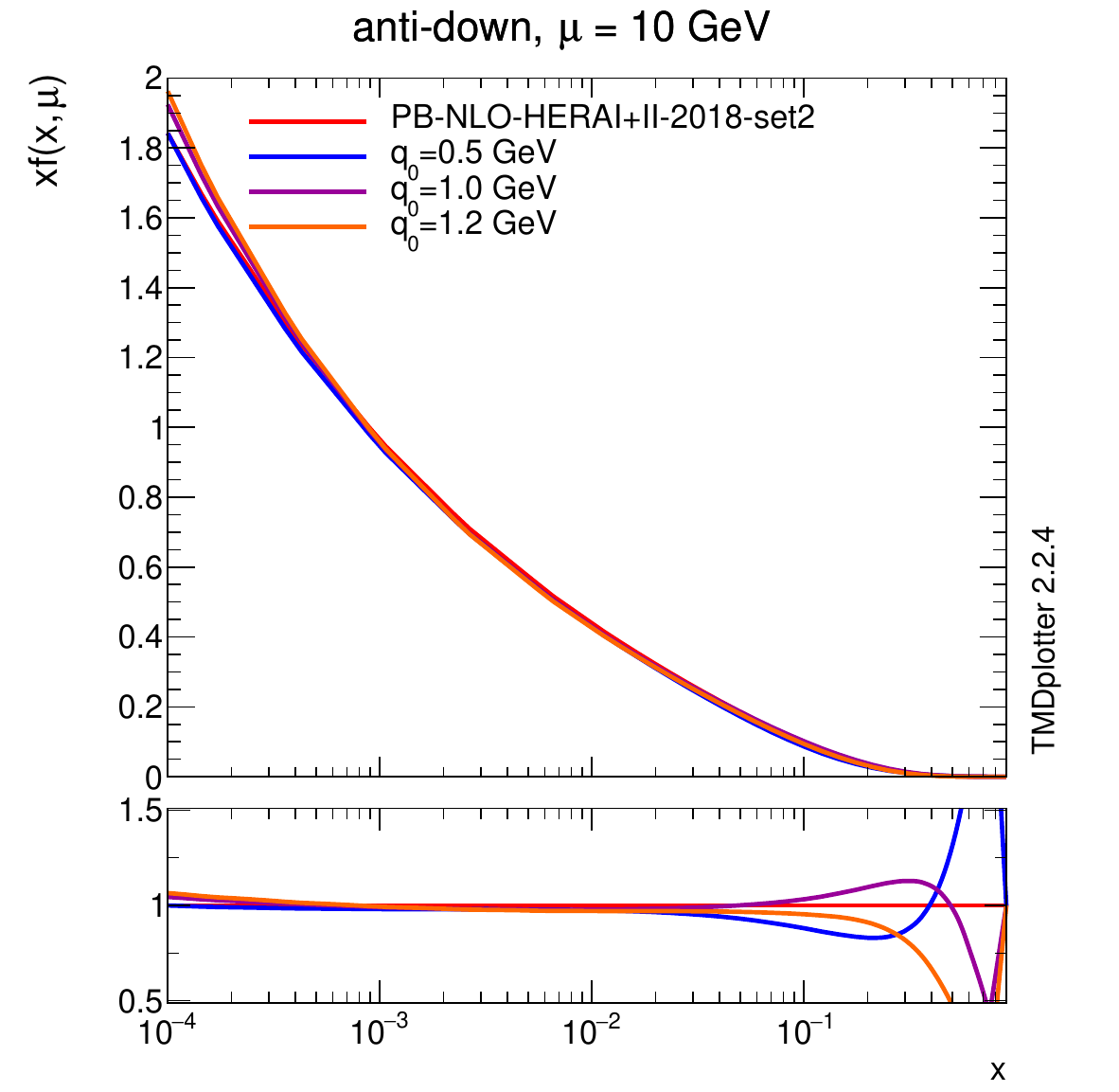}
\end{minipage}
\hfill
\begin{minipage}{0.31\linewidth}
\includegraphics[width=5.5cm]{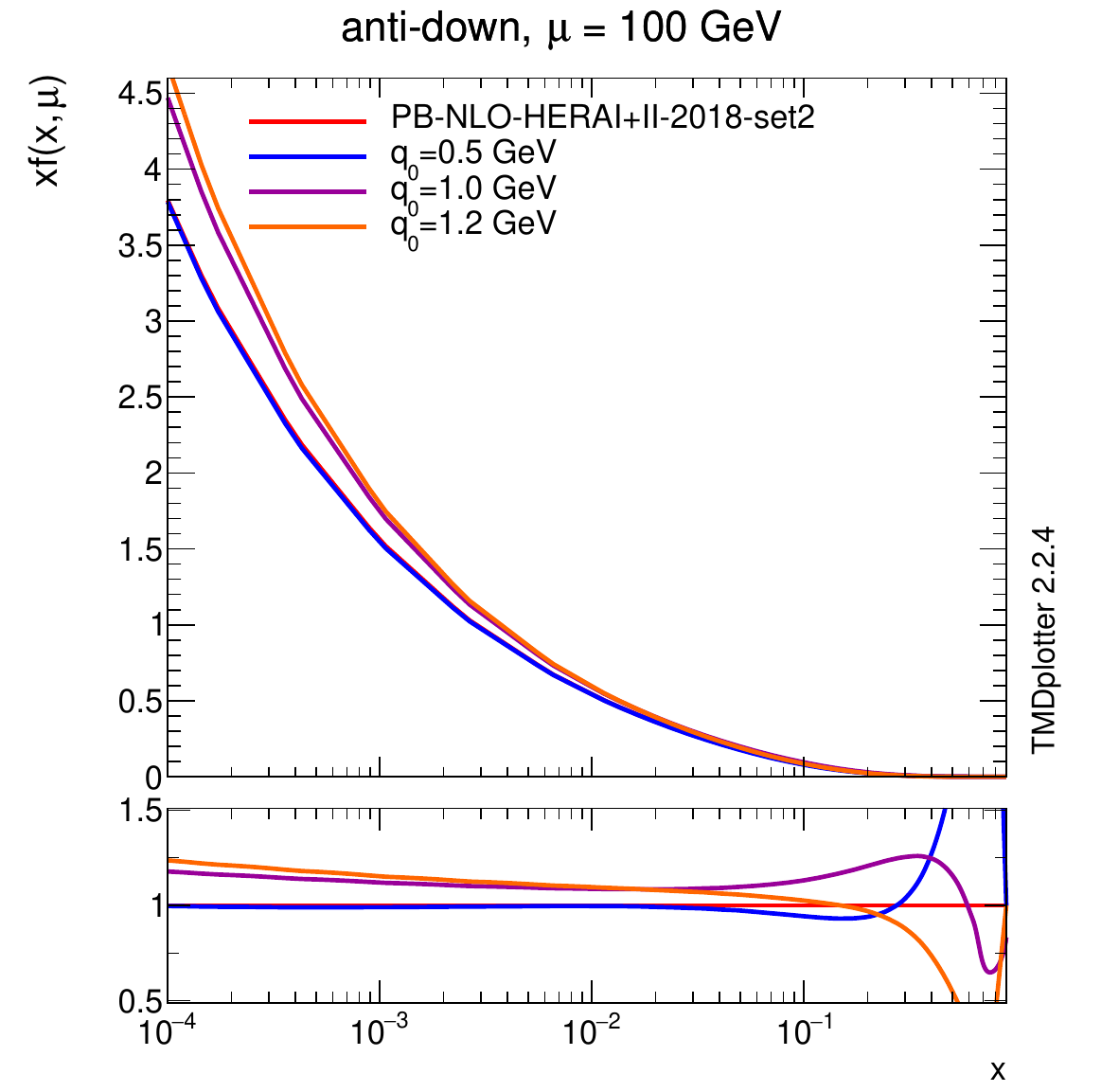}
\end{minipage}
\hfill
\begin{minipage}{0.31\linewidth}
\includegraphics[width=5.5cm]{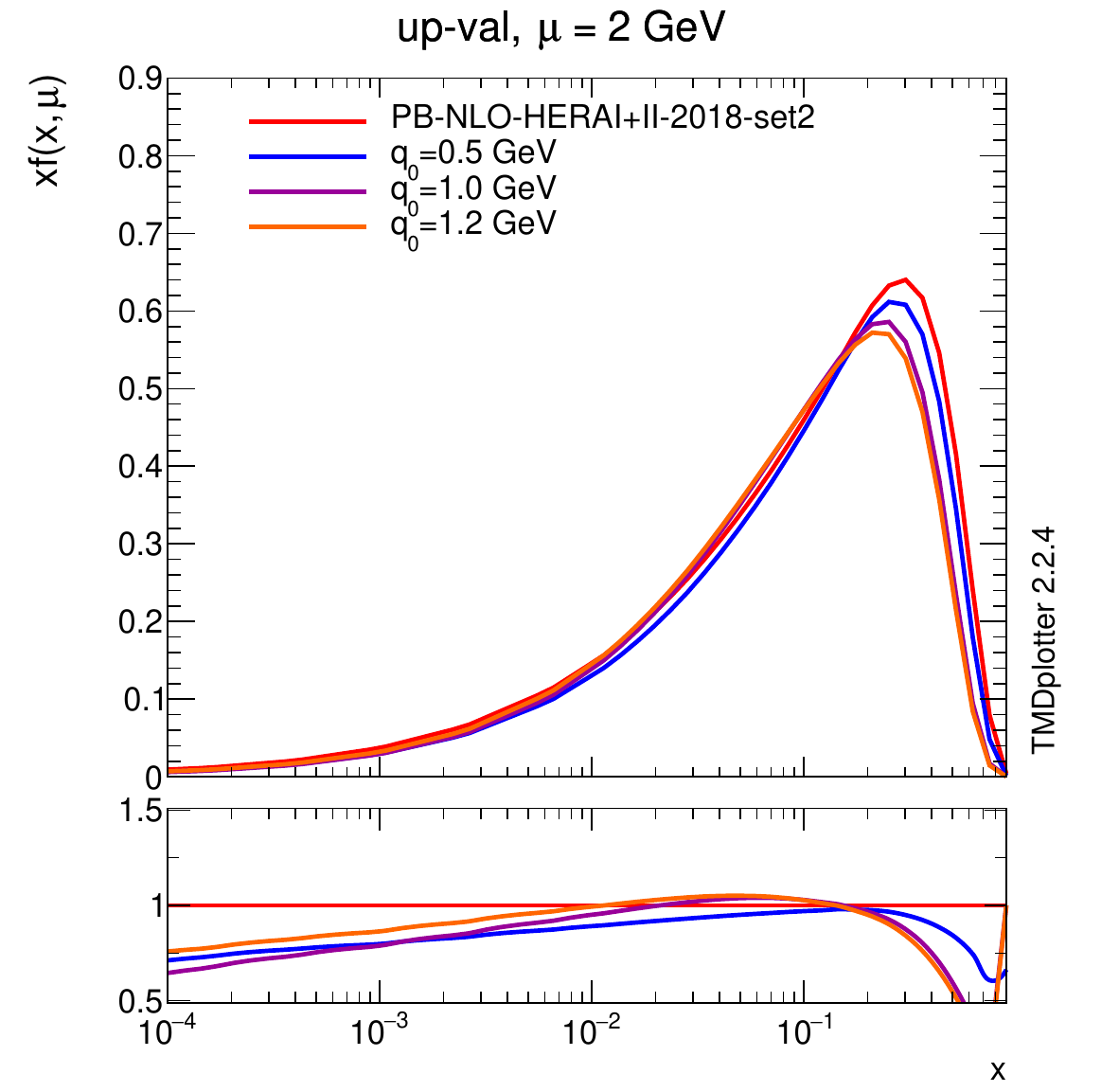}
\end{minipage}
\hfill
\begin{minipage}{0.31\linewidth}
\includegraphics[width=5.5cm]{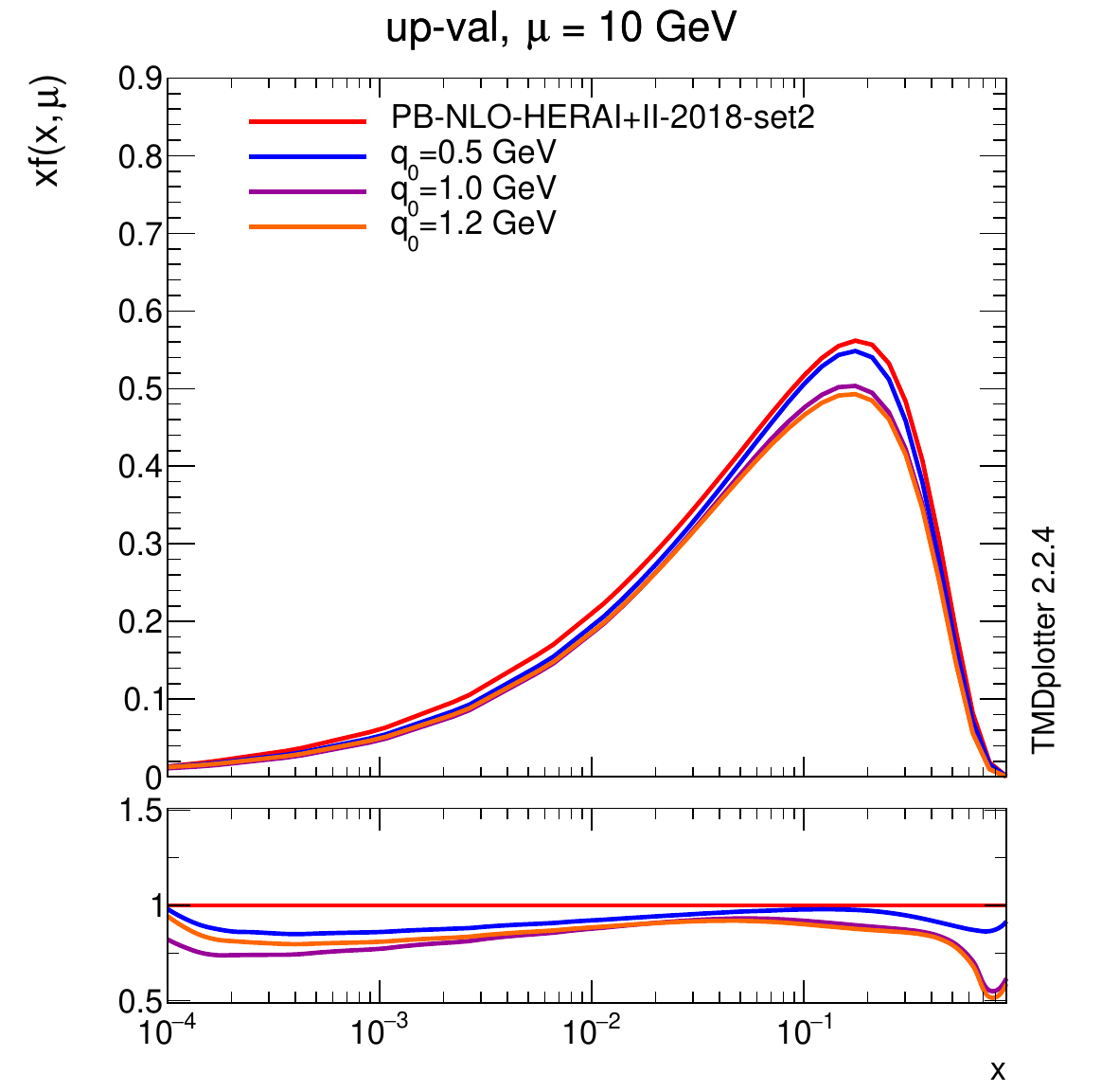}
\end{minipage}
\hfill
\begin{minipage}{0.31\linewidth}
\includegraphics[width=5.5cm]{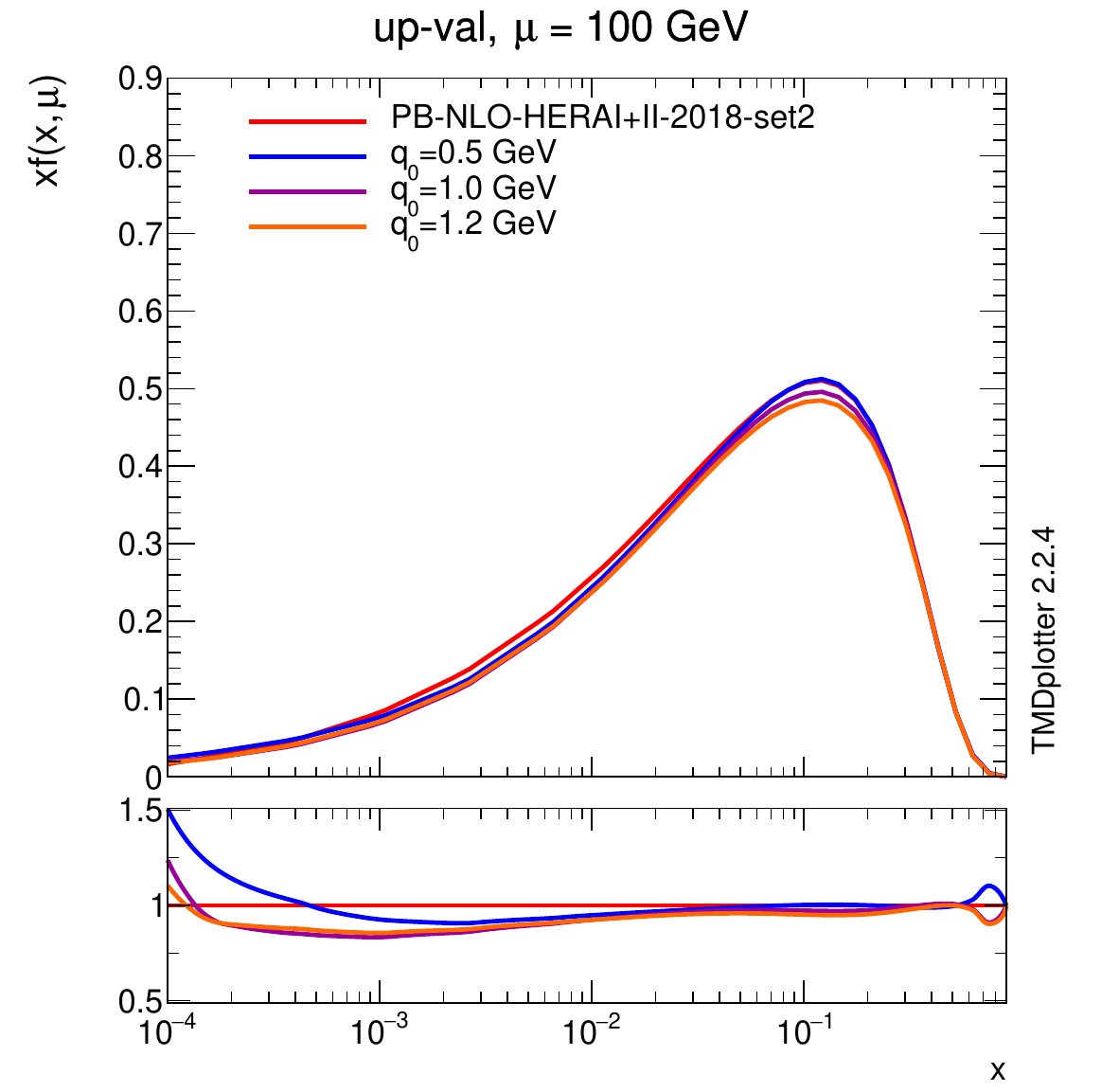}
\end{minipage}
\hfill
\caption[]{The iTMD distributions for different parton species as  functions of $x$ 
at $\mu=2,\;10,\;100  $ GeV obtained from the NLO dynamical-$z_M$ 
fits with  $q_0=0.5,  \;1.0\;\rm{and}\;1.2 $  GeV. 
The fixed-$z_M$ set PB-NLO-2018-Set2  is 
included for comparison.}
\label{fig:iTMDvsx-Diffq0}
\end{figure}

In Fig.~\ref{fig:iTMDvsx-Diffq0} the iTMD distributions 
corresponding to the three fits  with the different values of  $q_0$ 
are shown for several evolution scales and parton flavors.  The result for 
the fit PB-NLO-2018-Set2 of Ref.~\cite{BermudezMartinez:2018fsv}, 
which is based on fixed resolution scale $z_M$ and $q_c = 1 $ GeV, 
is shown as well and is used as a reference in the ratio plots. 
The distributions obtained with different $q_0$ values are different. Since the 
effect of $z_M$ accumulates in each branching, the distributions differ also at 
high scales.  For gluon, the differences are larger than for quarks. 
The gluon distribution in the fit with 
$q_0=0.5\;\rm{GeV}$ is however  similar to PB-NLO-2018-Set2.

\begin{figure}[!htb]
\begin{minipage}{0.31\linewidth}
\includegraphics[width=5.0cm]{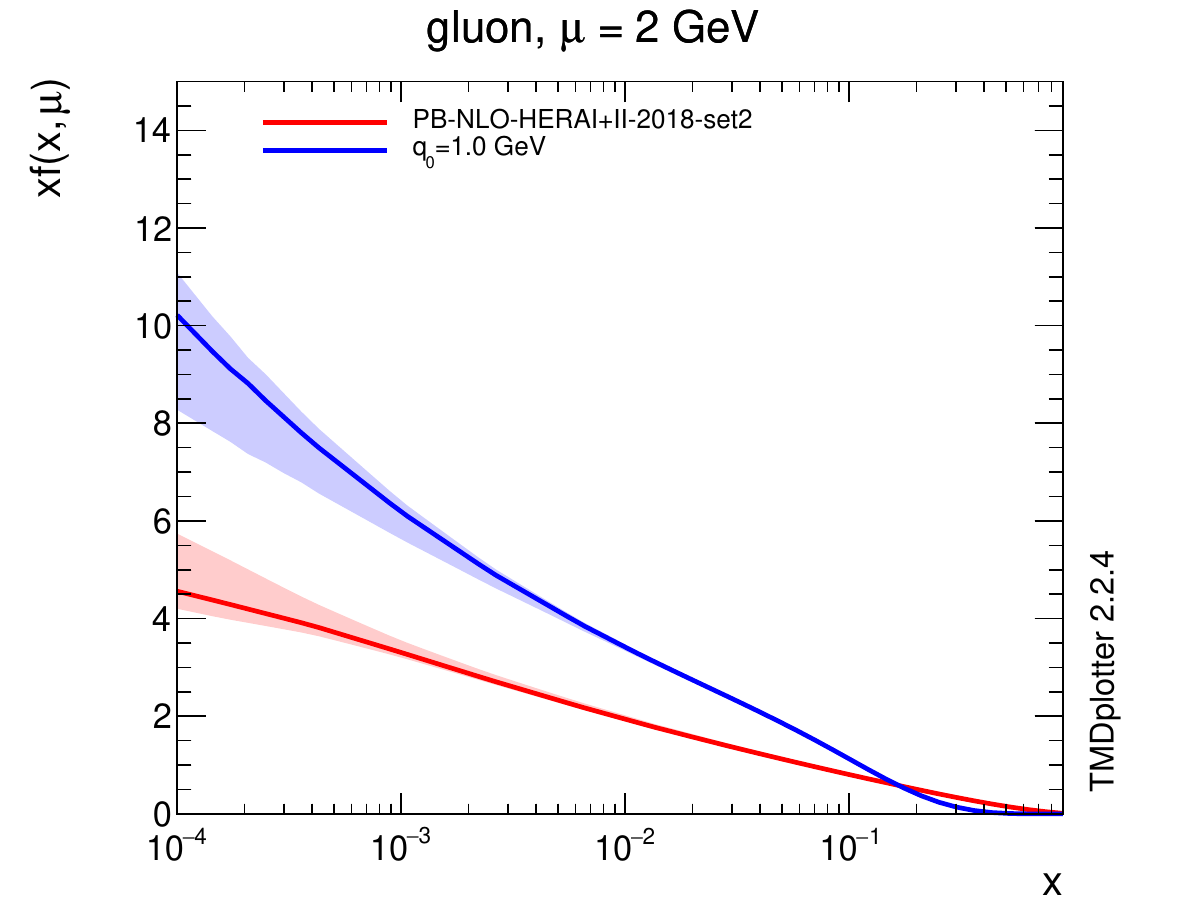}
\end{minipage}
\hfill
\begin{minipage}{0.31\linewidth}
\includegraphics[width=5.0cm]{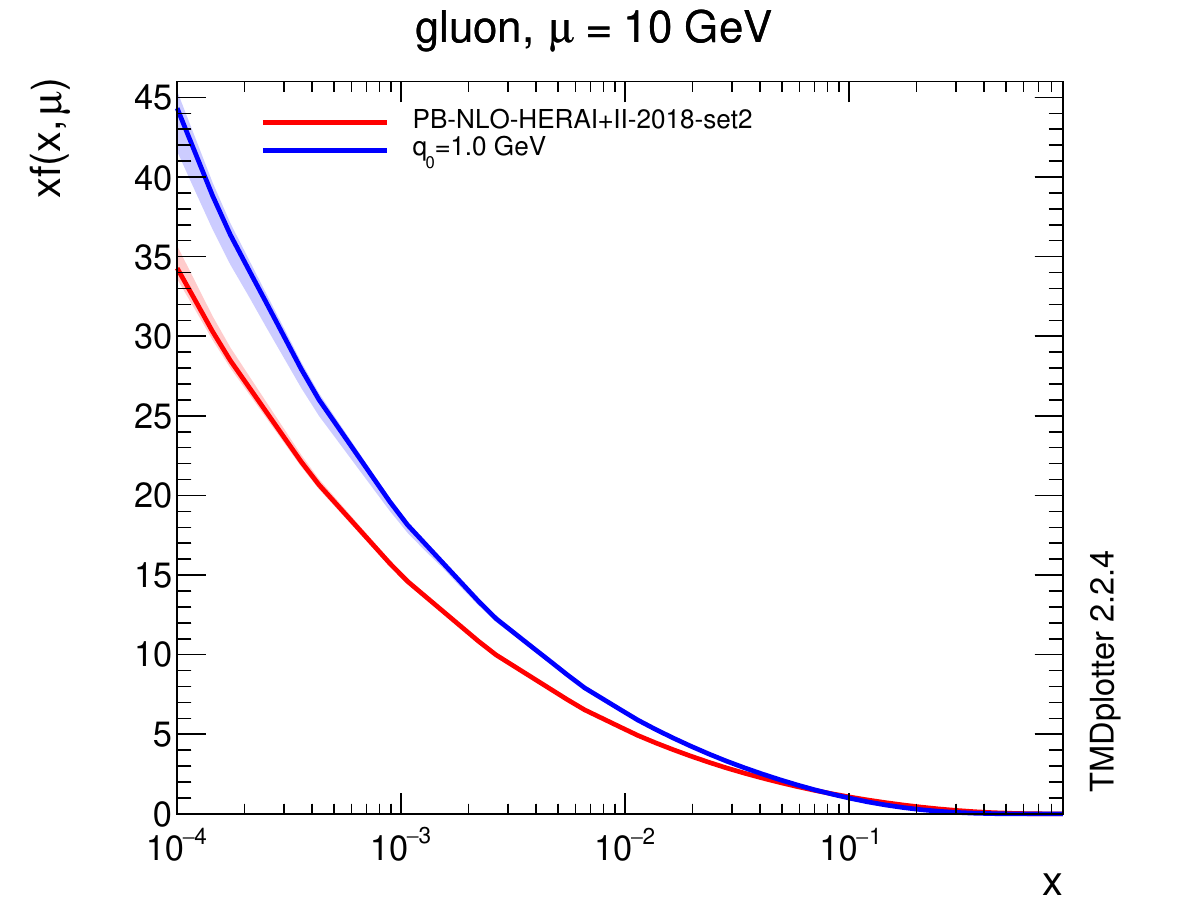}
\end{minipage}
\hfill
\begin{minipage}{0.31\linewidth}
\includegraphics[width=5.0cm]{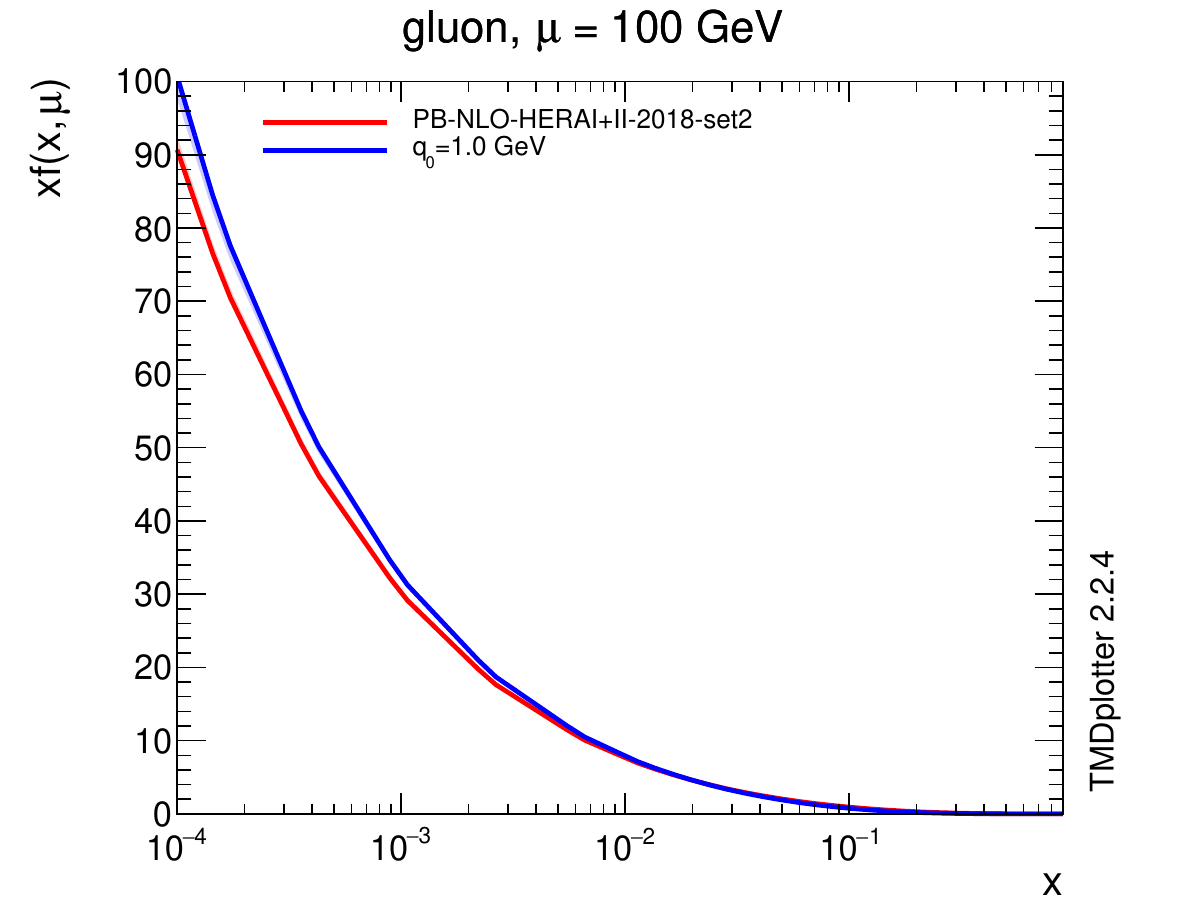}
\end{minipage}
\hfill
\begin{minipage}{0.31\linewidth}
\includegraphics[width=5.0cm]{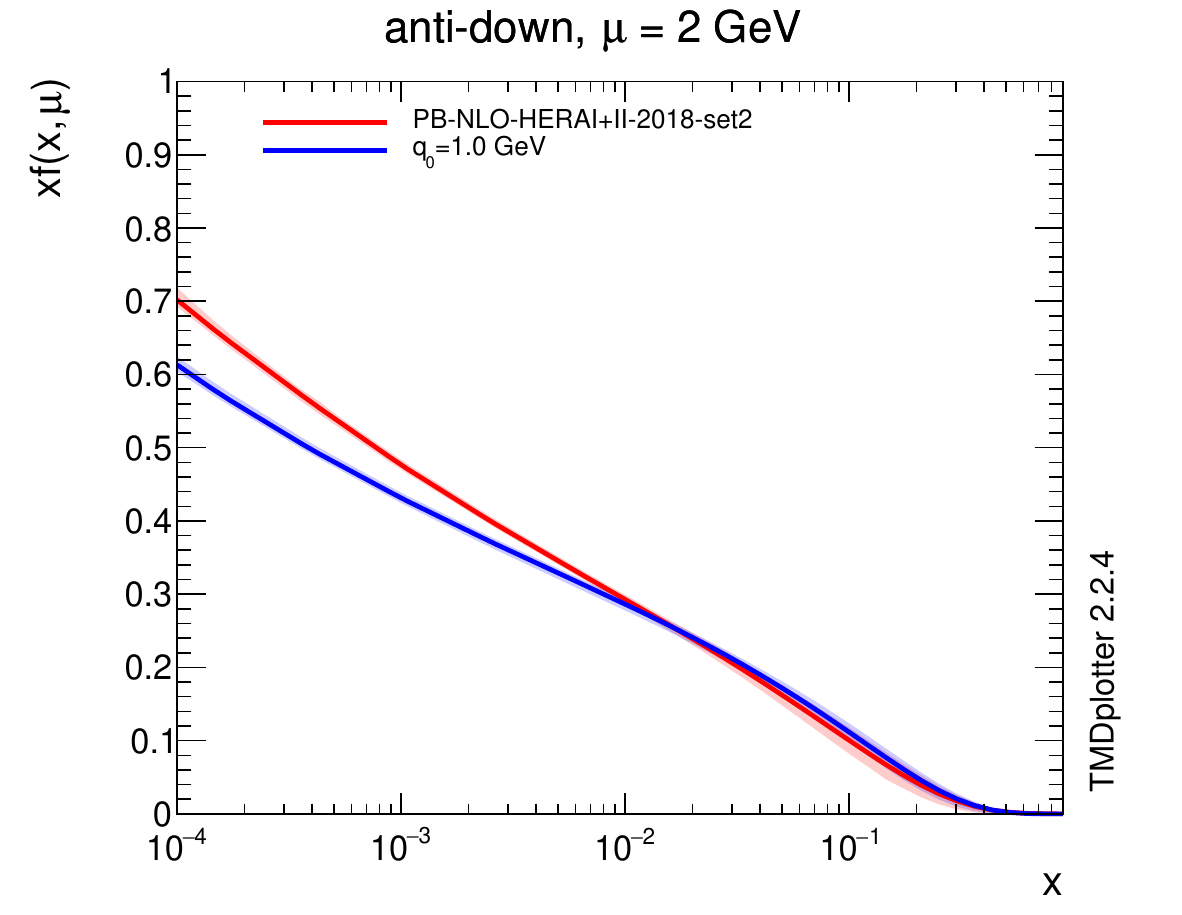}
\end{minipage}
\hfill
\begin{minipage}{0.31\linewidth}
\includegraphics[width=5.0cm]{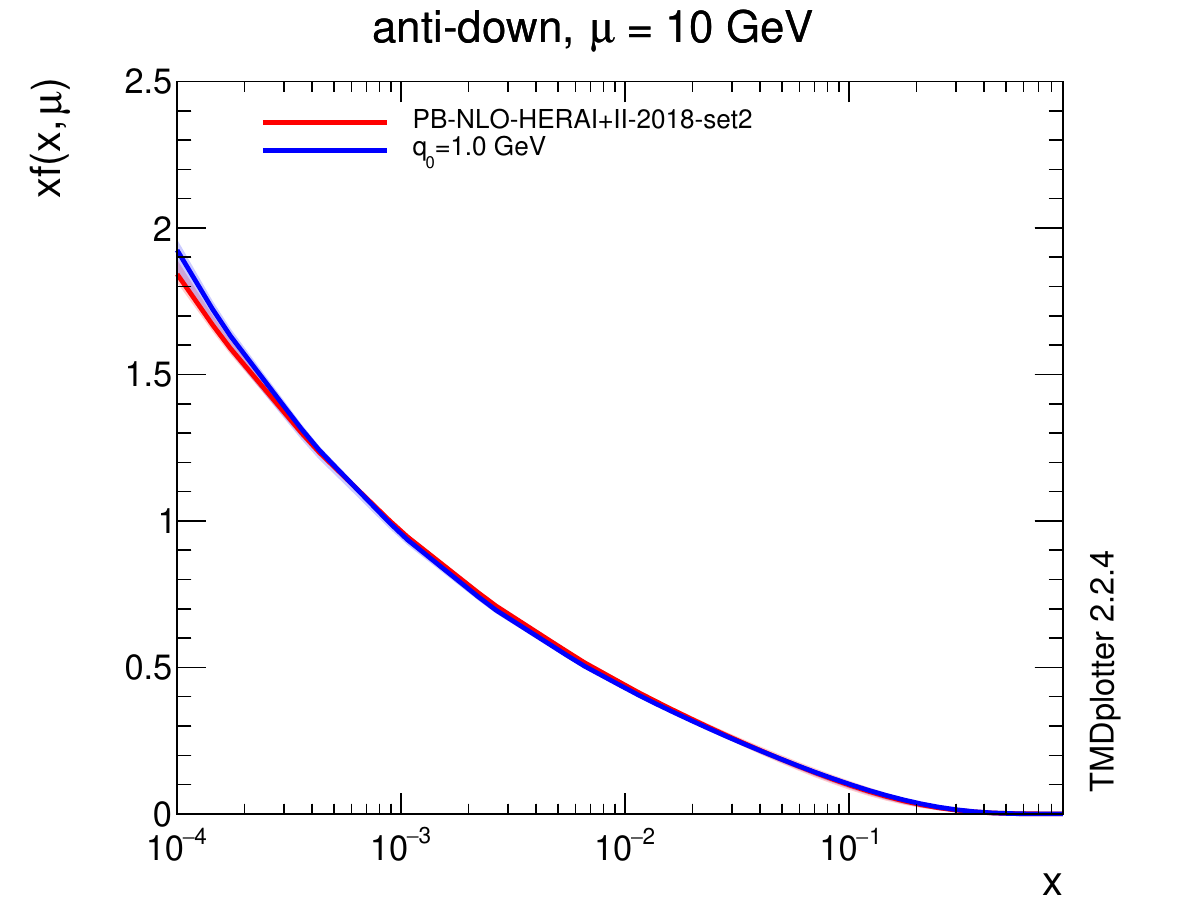}
\end{minipage}
\hfill
\begin{minipage}{0.31\linewidth}
\includegraphics[width=5.0cm]{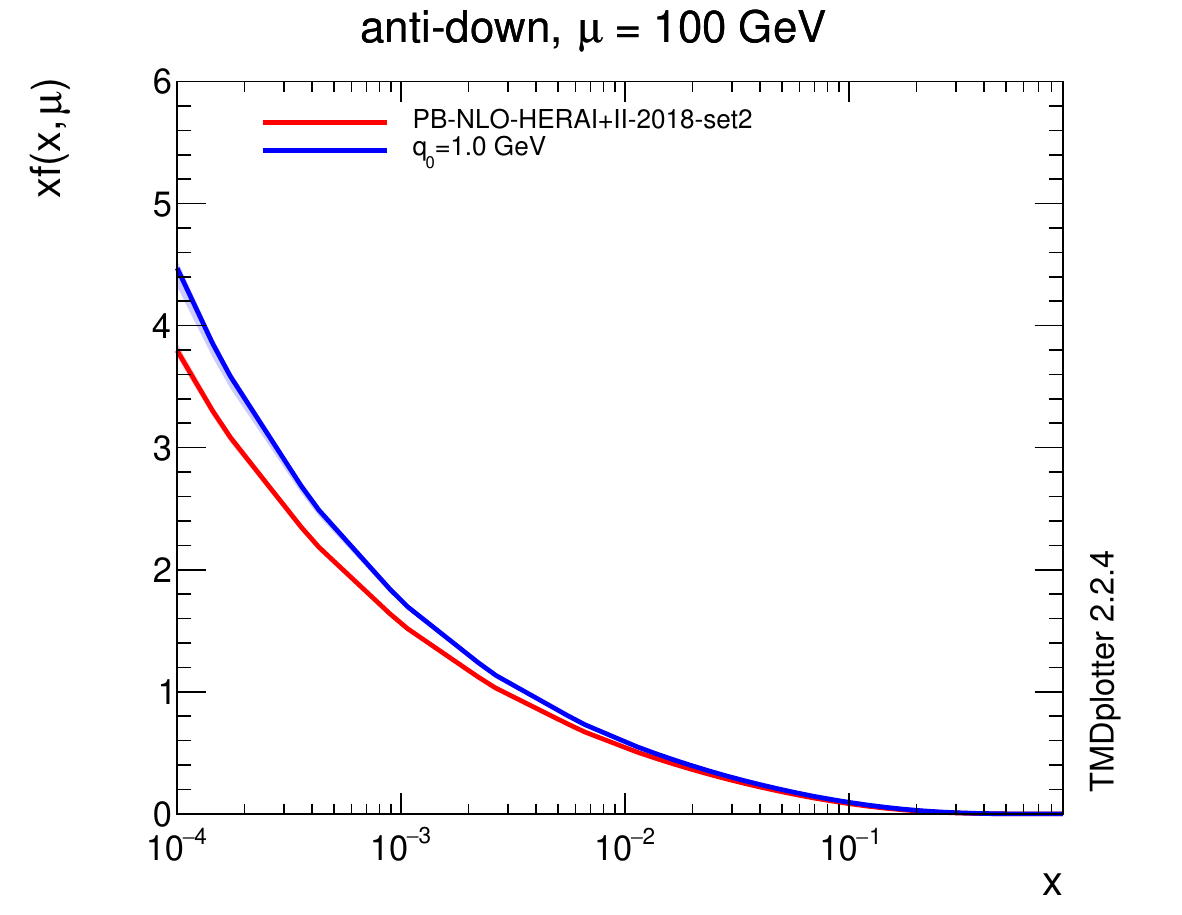}
\end{minipage}
\hfill
\begin{minipage}{0.31\linewidth}
\includegraphics[width=5.0cm]{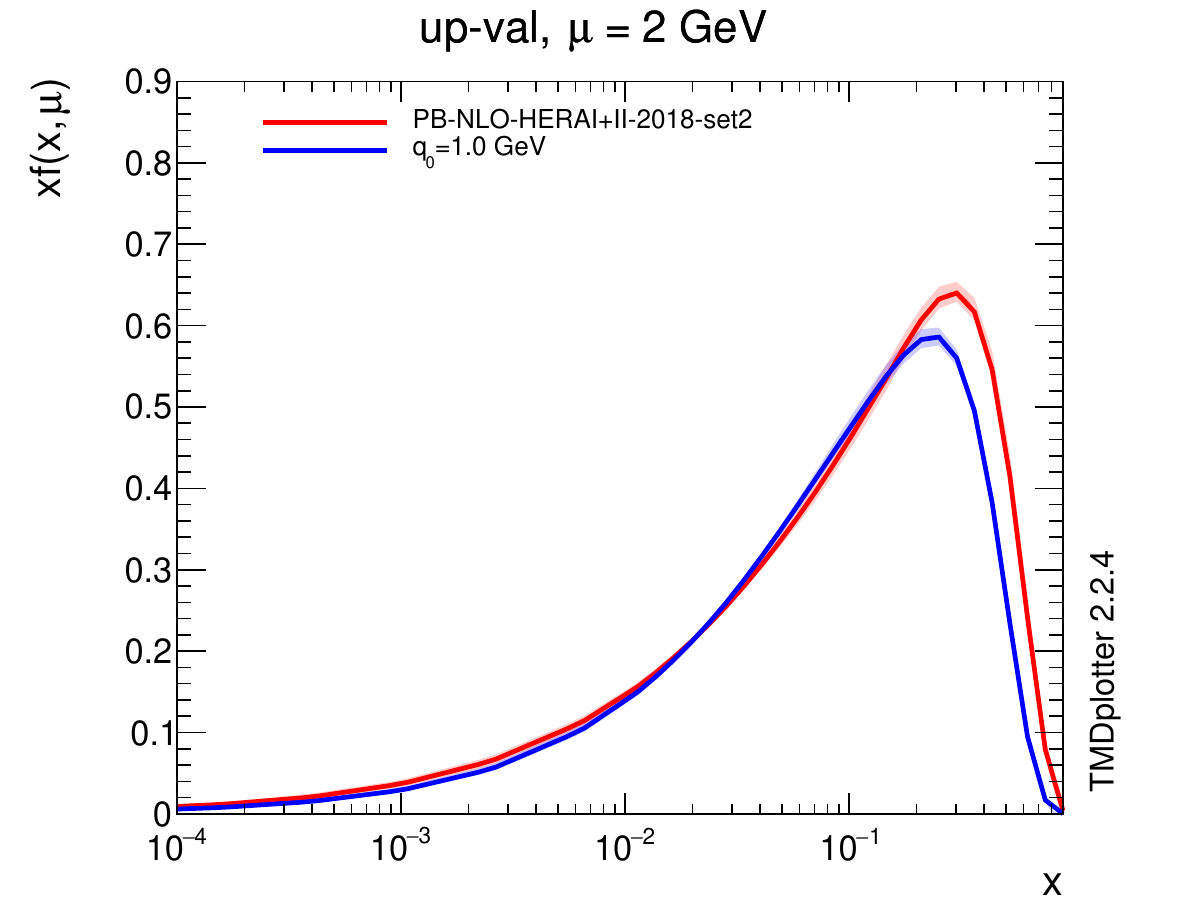}
\end{minipage}
\hfill
\begin{minipage}{0.31\linewidth}
\includegraphics[width=5.0cm]{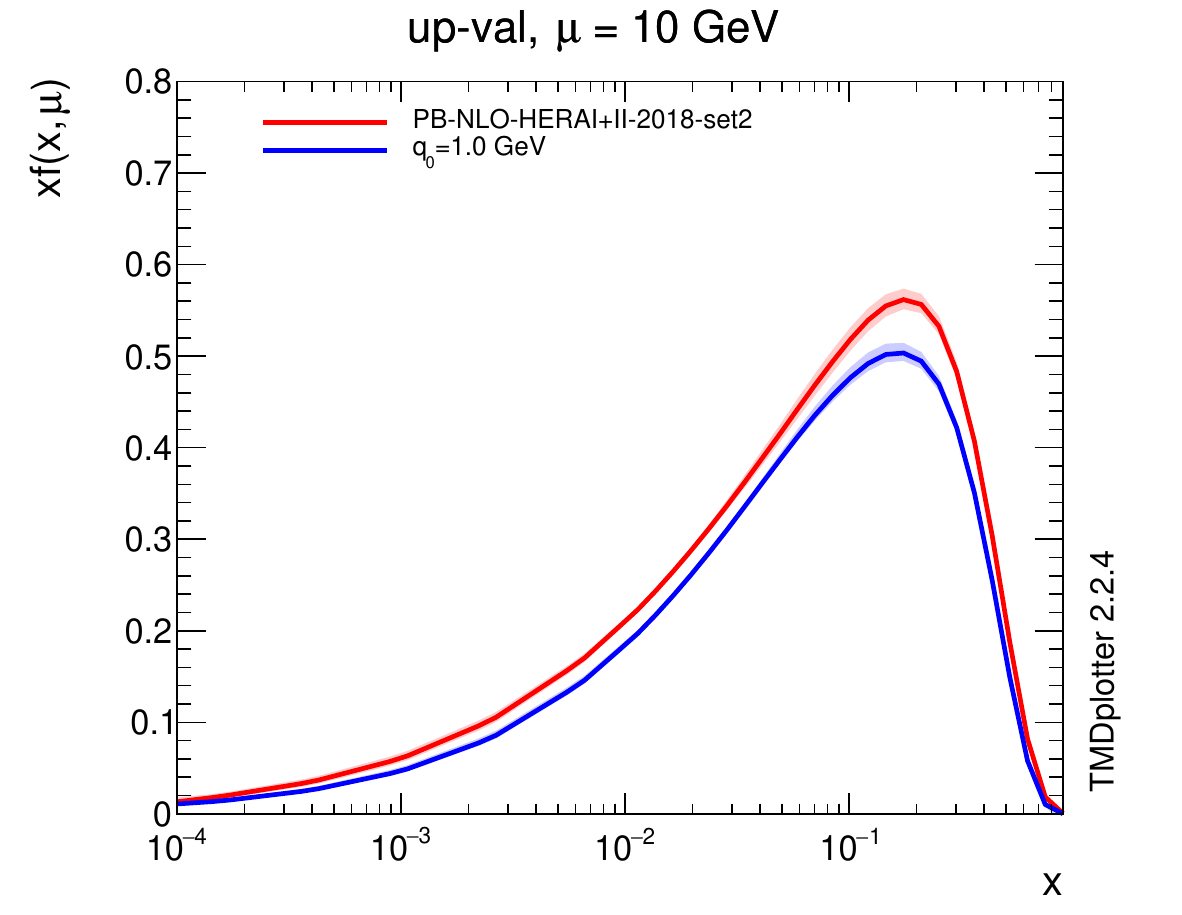}
\end{minipage}
\hfill
\begin{minipage}{0.31\linewidth}
\includegraphics[width=5.0cm]{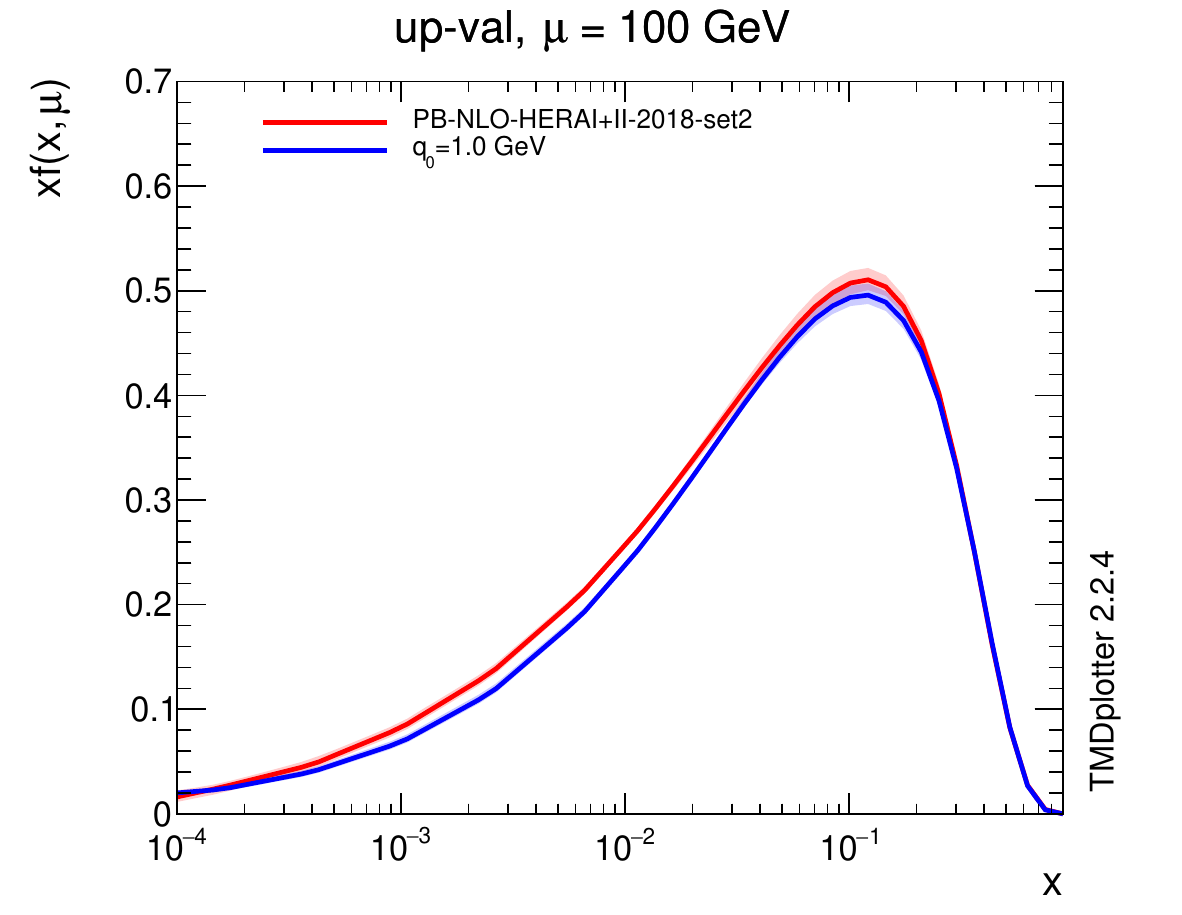}
\end{minipage}
\caption[]{Experimental and model uncertainties of the iTMD distributions 
for gluon, down and up quark as functions of $x$ at $\mu=2,\;10,\;100 \;\  \rm{GeV}$  
obtained from the dynamical-$z_M$ fit with  
$q_0=1.0\;\rm{GeV}$. 
The fixed-$z_M$ set PB-NLO-2018-Set2 is 
included for comparison.}
\label{fig:iTMDvsx-DynvsPBset2}
\end{figure}

In Fig.~\ref{fig:iTMDvsx-DynvsPBset2},   
experimental and model uncertainties of the fitted distributions are shown 
for the dynamical-$z_M$ set with  $q_0=1.0\;\rm{GeV}$, and compared to 
PB-NLO-2018-Set2. The differences 
between these two sets are generally larger 
than the uncertainty band from the fits. 
Also, the differences between the curves with 
different $q_0$ values from Fig.~\ref{fig:iTMDvsx-Diffq0} are 
larger than the uncertainty band for $q_0=1.0\;\rm{GeV}$.

\begin{figure}[!htb]
\begin{minipage}{0.24\linewidth}
\includegraphics[width=4.1cm]{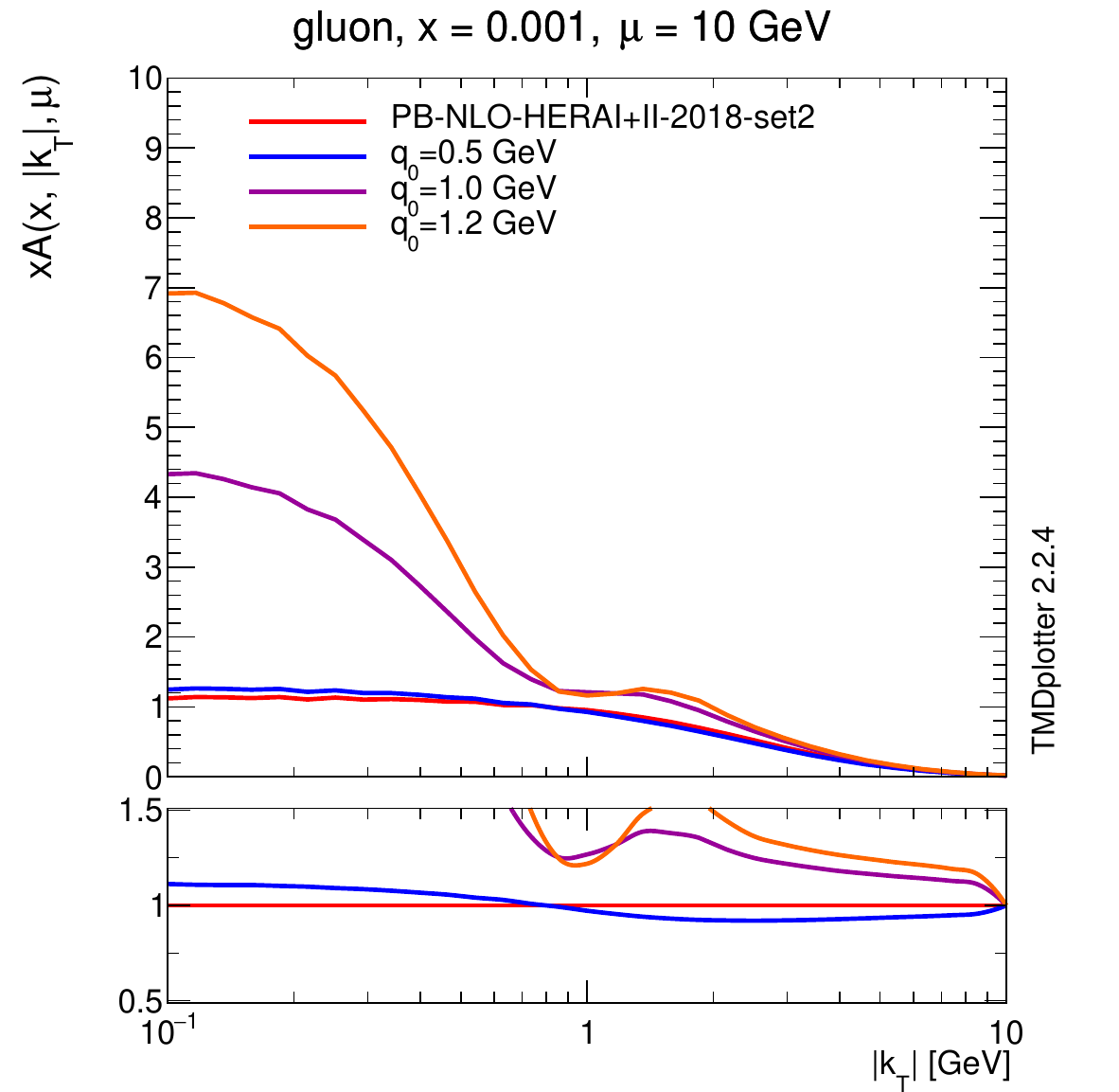}
\end{minipage}
\hfill
\begin{minipage}{0.24\linewidth}
\includegraphics[width=4.1cm]{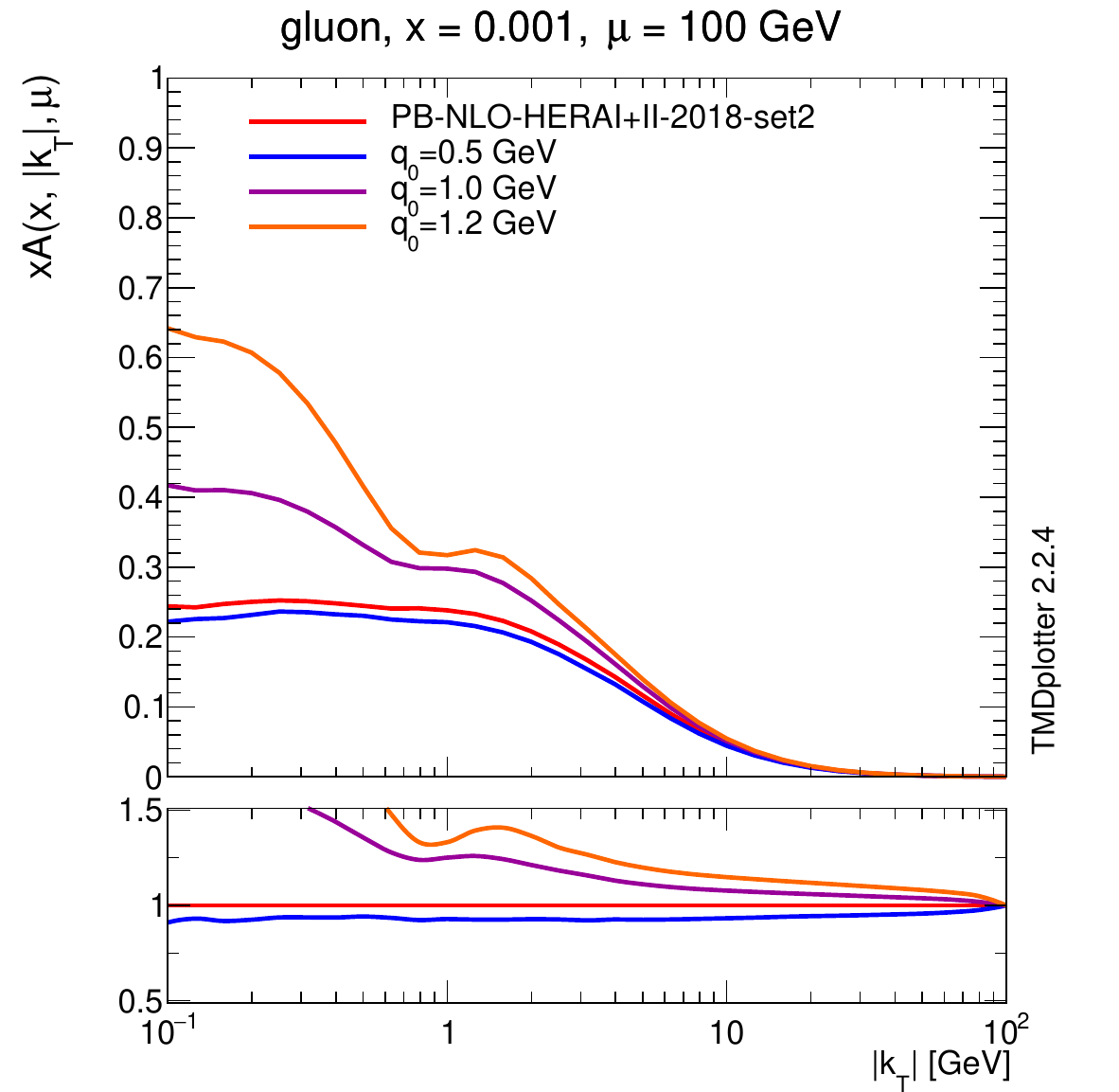}
\end{minipage}
\hfill
\begin{minipage}{0.24\linewidth}
\includegraphics[width=4.1cm]{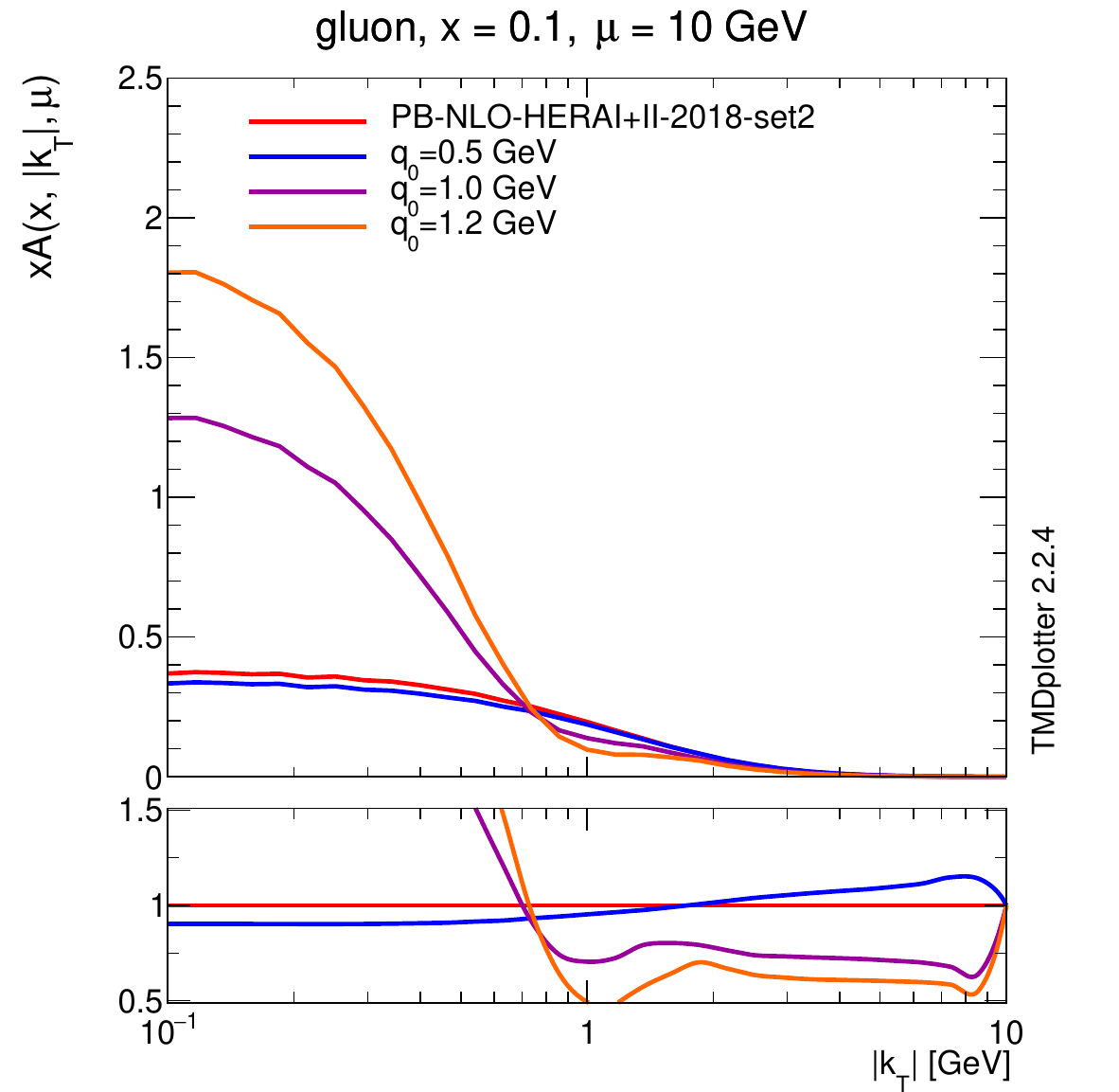}
\end{minipage}
\hfill
\begin{minipage}{0.24\linewidth}
\includegraphics[width=4.1cm]{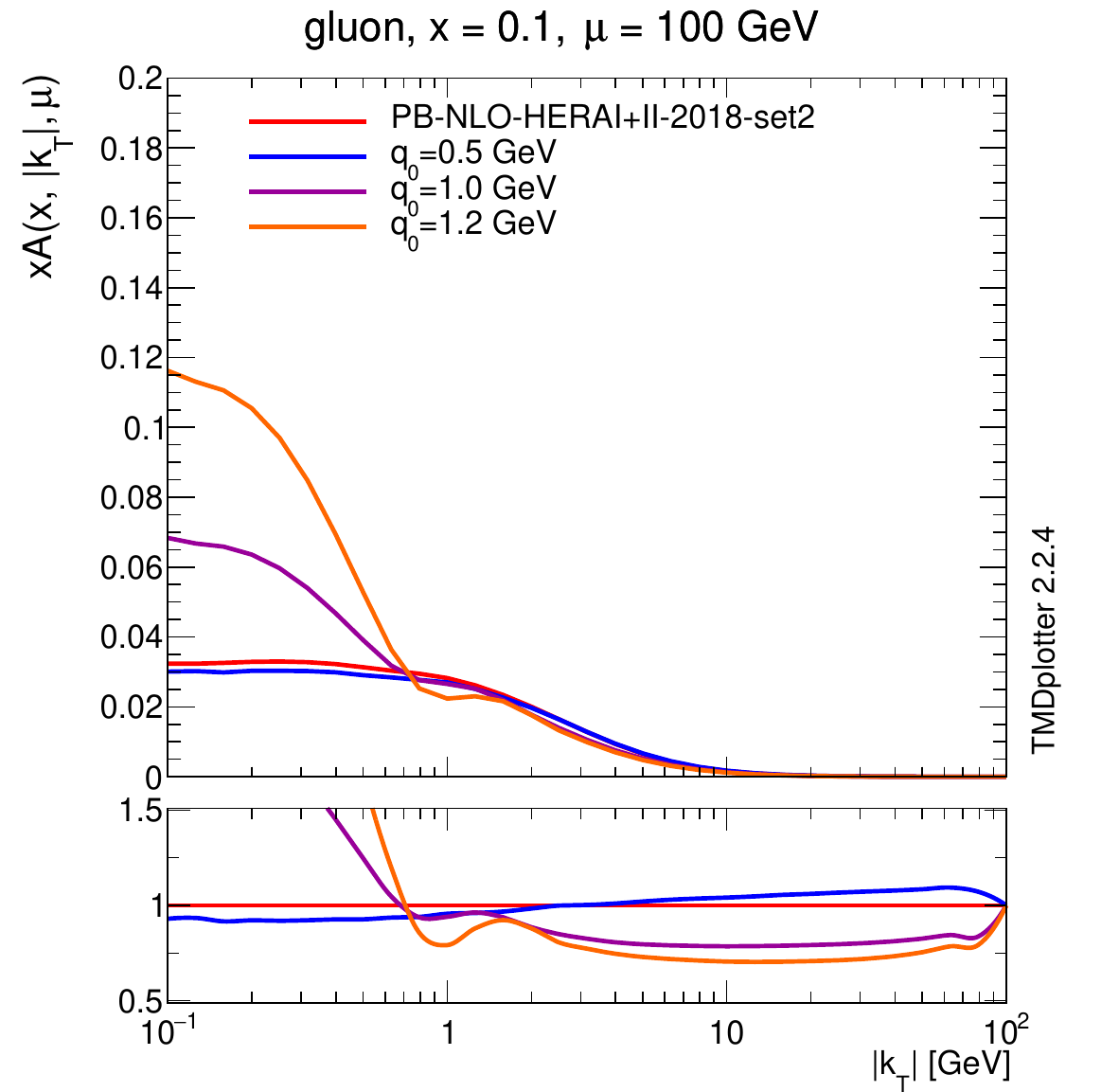}
\end{minipage}
\hfill
\begin{minipage}{0.24\linewidth}
\includegraphics[width=4.1cm]{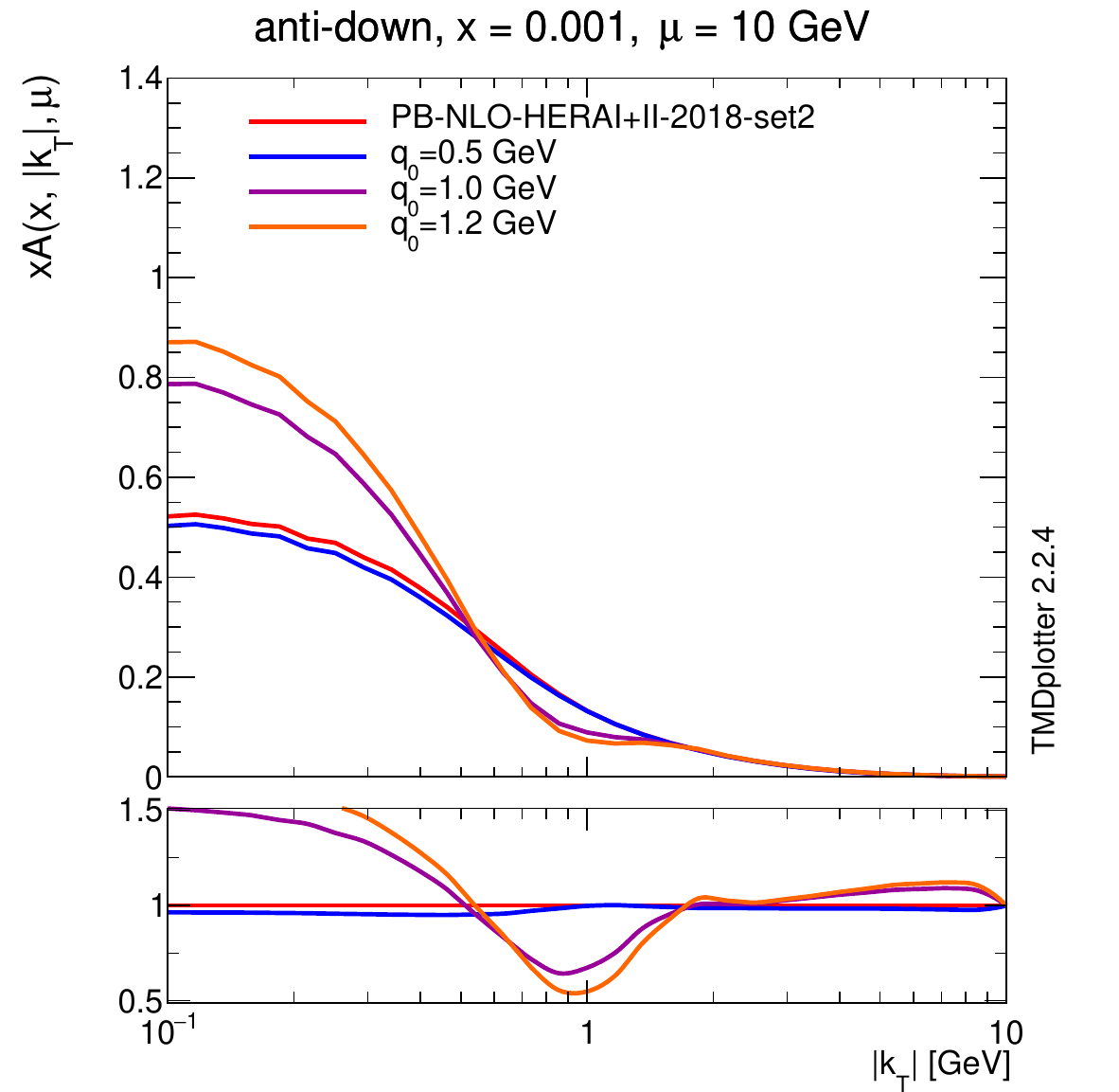}
\end{minipage}
\hfill
\begin{minipage}{0.24\linewidth}
\includegraphics[width=4.1cm]{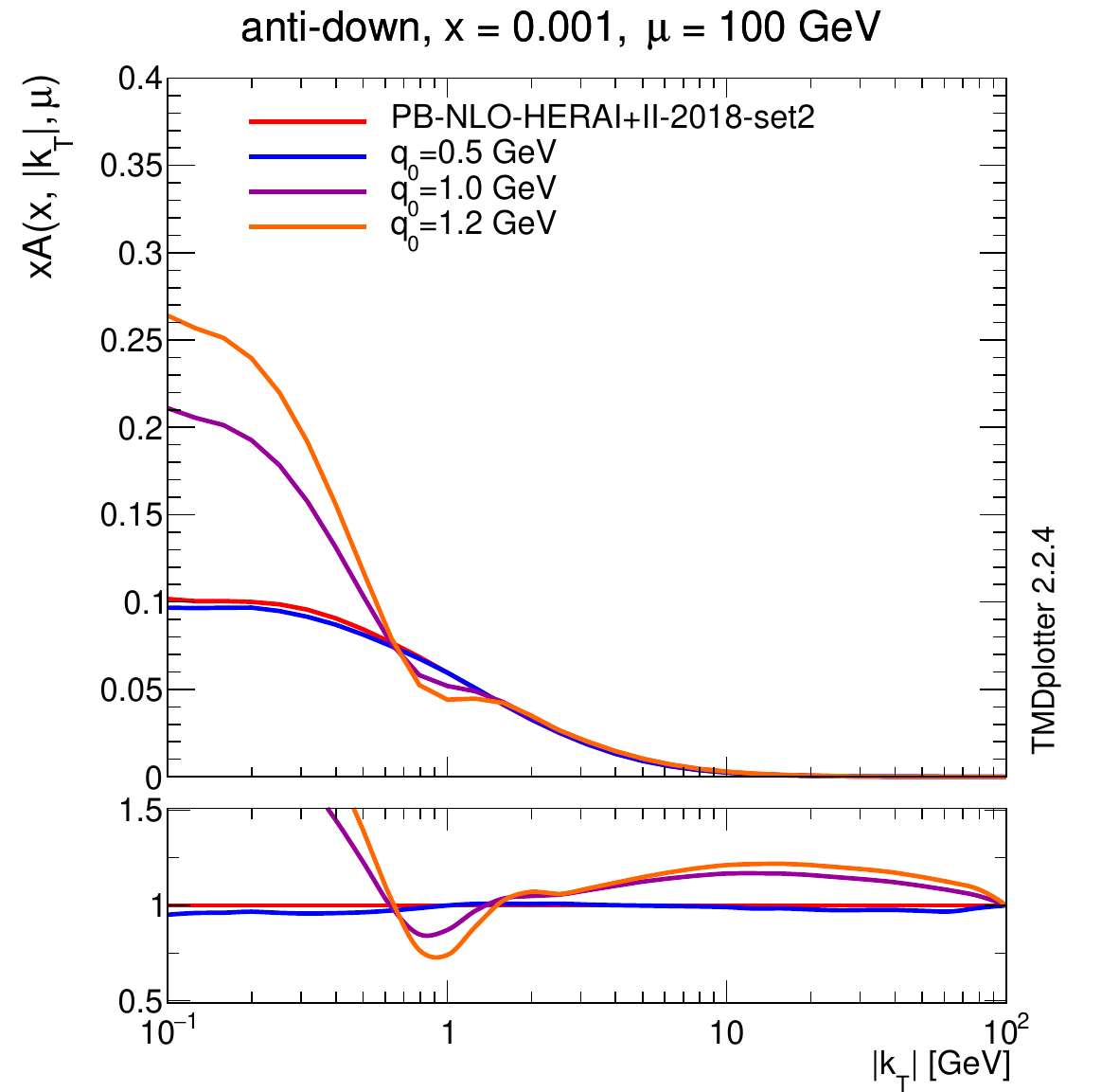}
\end{minipage}
\hfill
\begin{minipage}{0.24\linewidth}
\includegraphics[width=4.1cm]{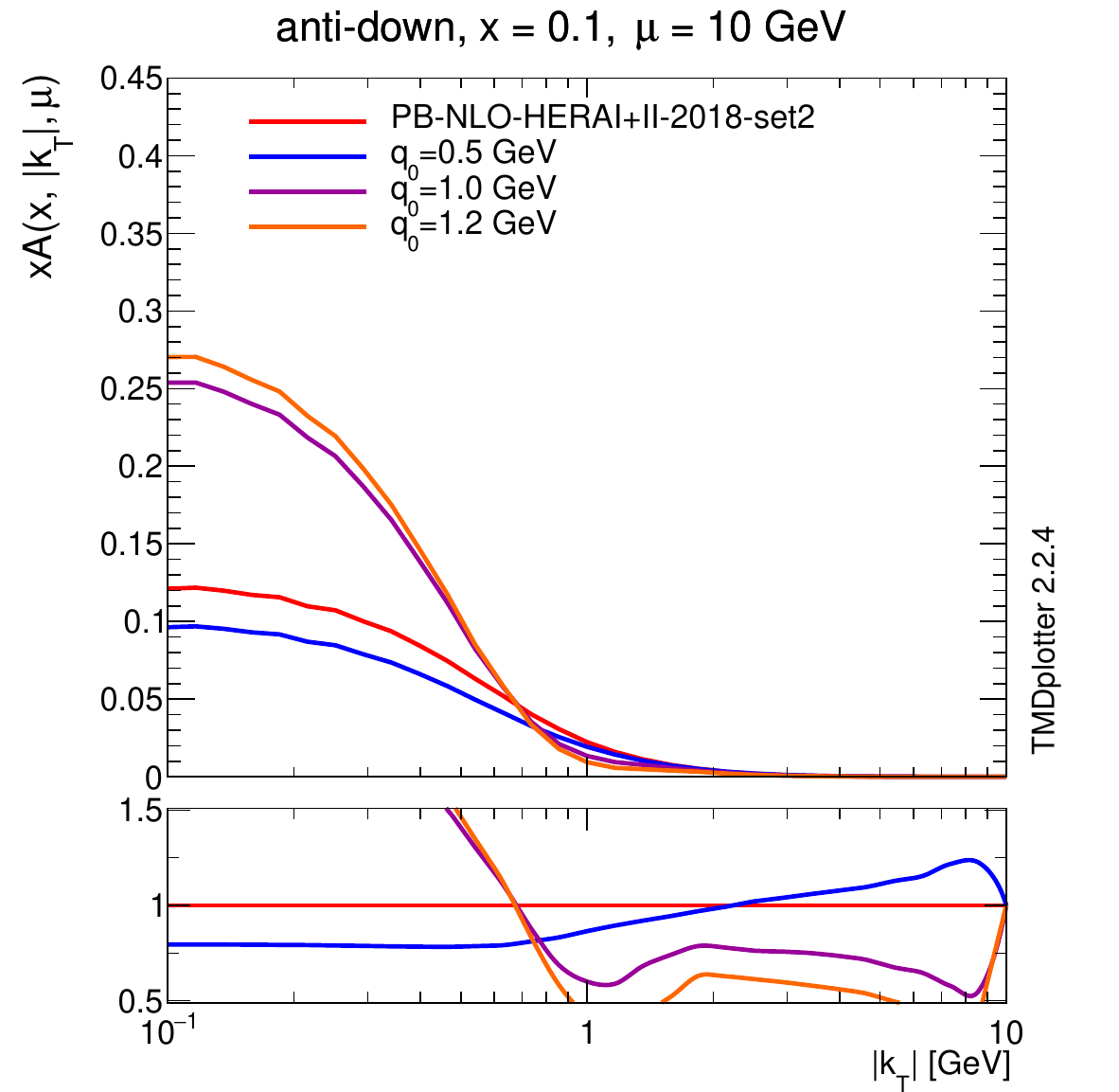}
\end{minipage}
\hfill
\begin{minipage}{0.24\linewidth}
\includegraphics[width=4.1cm]{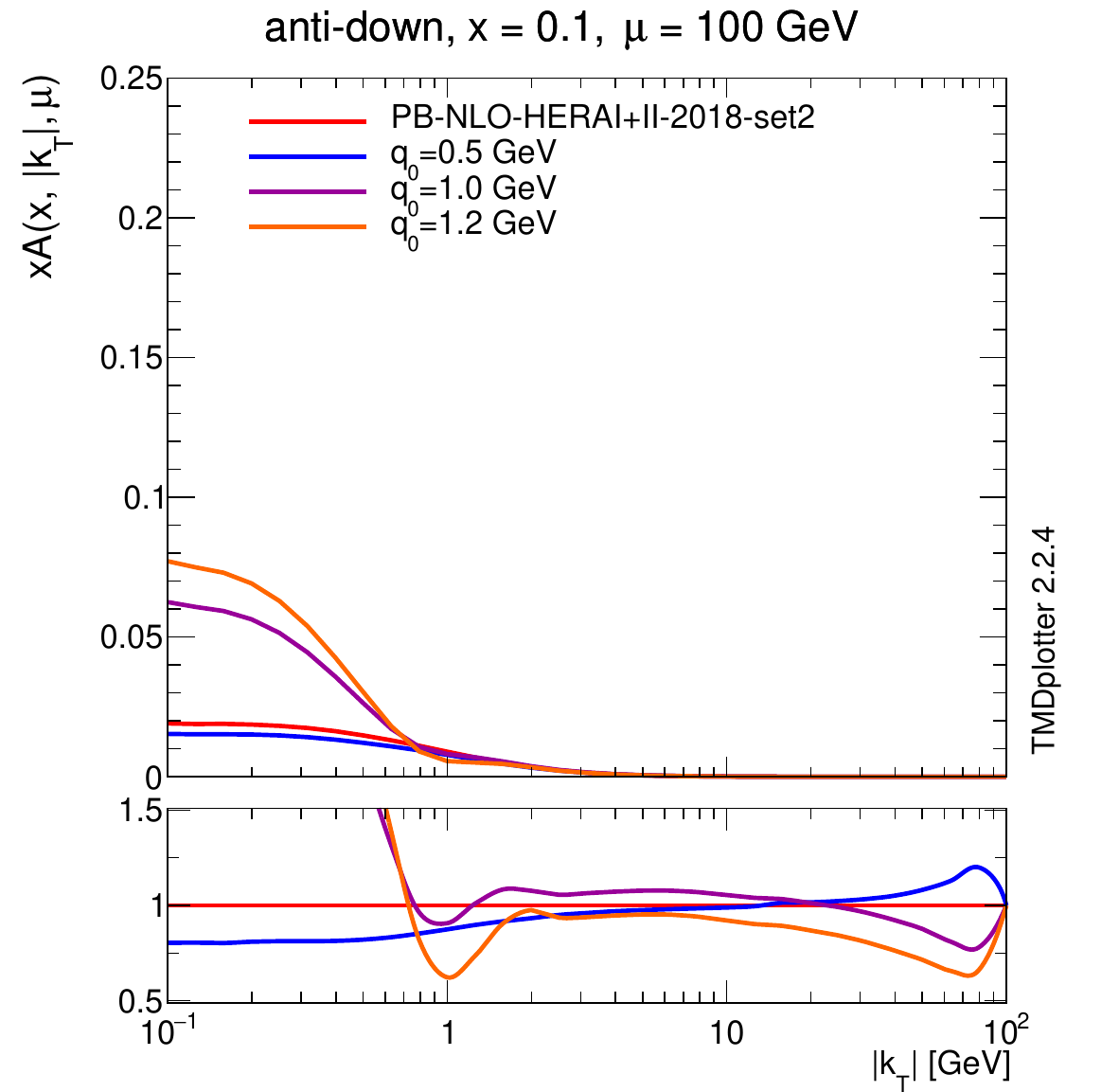}
\end{minipage}
\hfill
\caption[]{The $k_T$   dependence of 
TMD distributions for gluon and anti-down quark 
obtained from the NLO dynamical-$z_M$ fits with different $q_0$ values,
for $\mu=$  10 and 100 GeV and  $x=0.001\;\rm{and}\;0.1$.   
The fixed-$z_M$ set PB-NLO-2018-Set2 is 
included for comparison.}
\label{fig:TMDsDiffq0}
\end{figure}

As discussed above, following the procedure in 
Refs.~\cite{Hautmann:2013tba,Hautmann:2014uua,BermudezMartinez:2018fsv}
the TMD distributions are obtained from a 
convolution of the TMD kernel with the starting distribution. 
In Fig.~\ref{fig:TMDsDiffq0} the TMD distributions as  
functions of the transverse momentum $k_T$ are shown 
for gluon and anti-down quark, at values of longitudinal momentum 
fractions $x=0.1\;\;\rm{and}\;0.001$ and at evolution 
scales $\mu=10\;\rm{GeV}$ and $\mu=100\;\rm{GeV}$,  for different $q_0$ values. 
PB-NLO-2018-Set2  is shown as well and 
is used as a reference on the ratio plots. 
The distributions obtained with different $q_0$ values differ 
especially in the low-$k_T$ region. At larger $k_T$ the 
shape of the distributions is  similar. It is worth noting that for very 
low $|k_{\bot}|\approx 1\;\rm{GeV}$ 
a dip  may arise in 
the region of the minimal emitted transverse momentum due to 
the matching between the intrinsic transverse 
momentum and the transverse momentum generated by the evolution.  
The dip is less pronounced for 
lower $q_0$ values, when  more branchings are generated. 
Similarly to the iTMD case, the result with 
$q_0=0.5\;\rm{GeV}$ is quite similar to PB-NLO-2018-Set2.

\begin{figure}[!htb]
\begin{minipage}{0.24\linewidth}
\includegraphics[width=4.1cm]{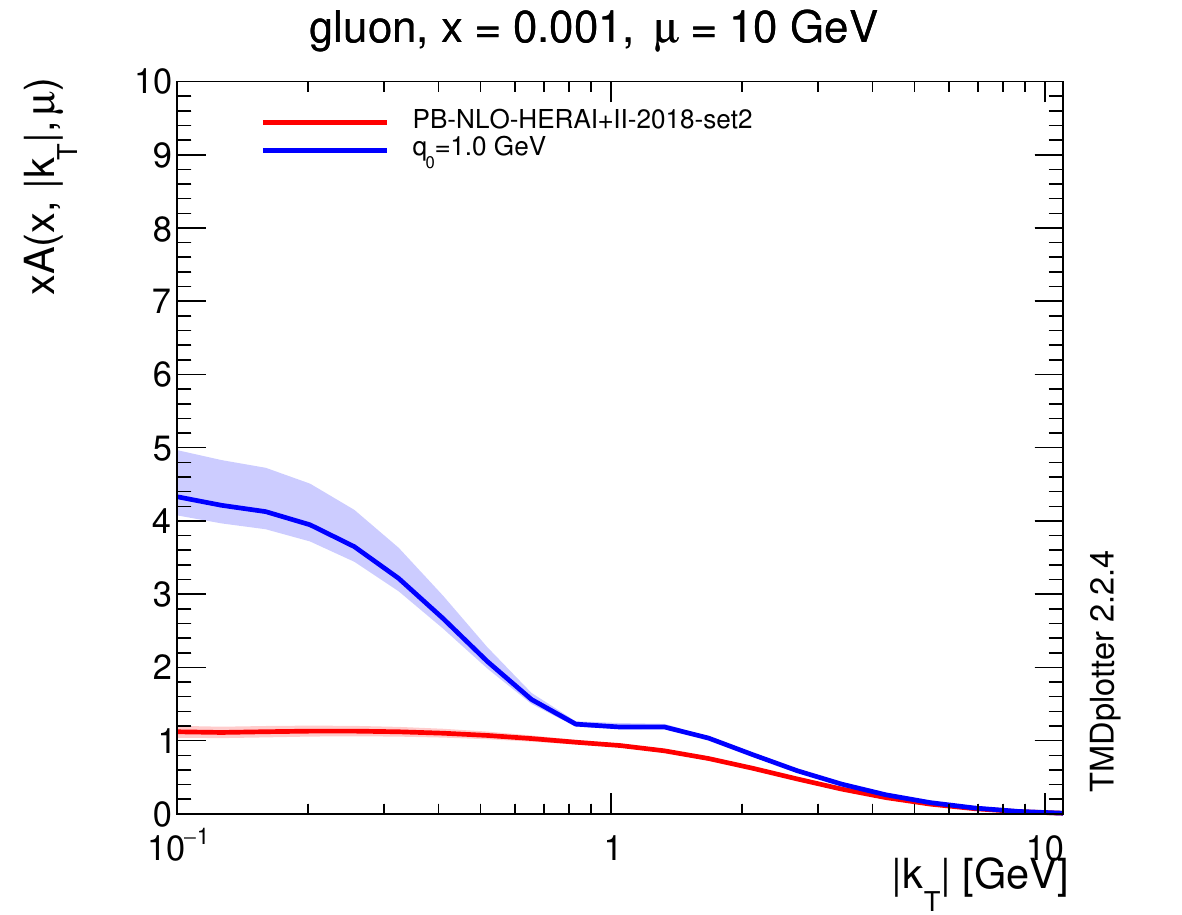}
\end{minipage}
\hfill
\begin{minipage}{0.24\linewidth}
\includegraphics[width=4.1cm]{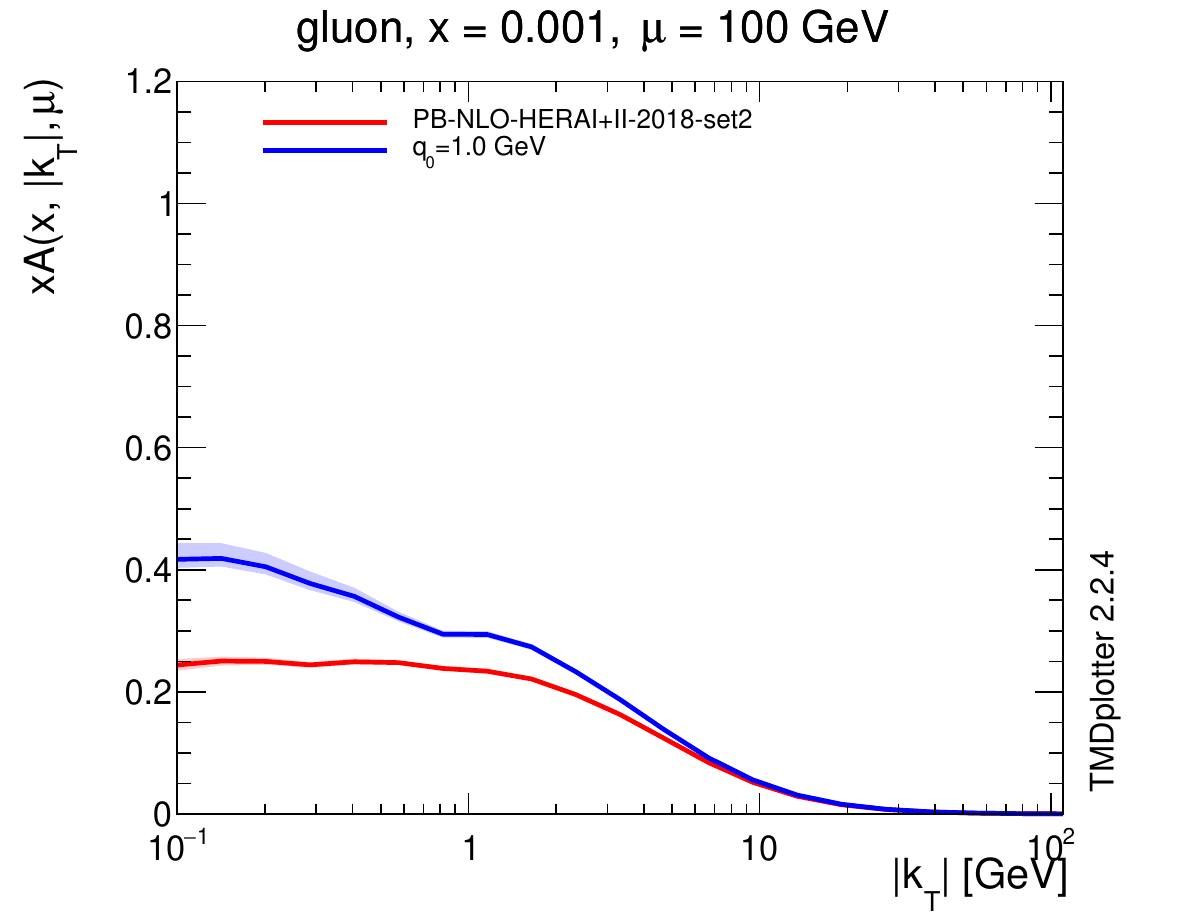}
\end{minipage}
\hfill
\begin{minipage}{0.24\linewidth}
\includegraphics[width=4.1cm]{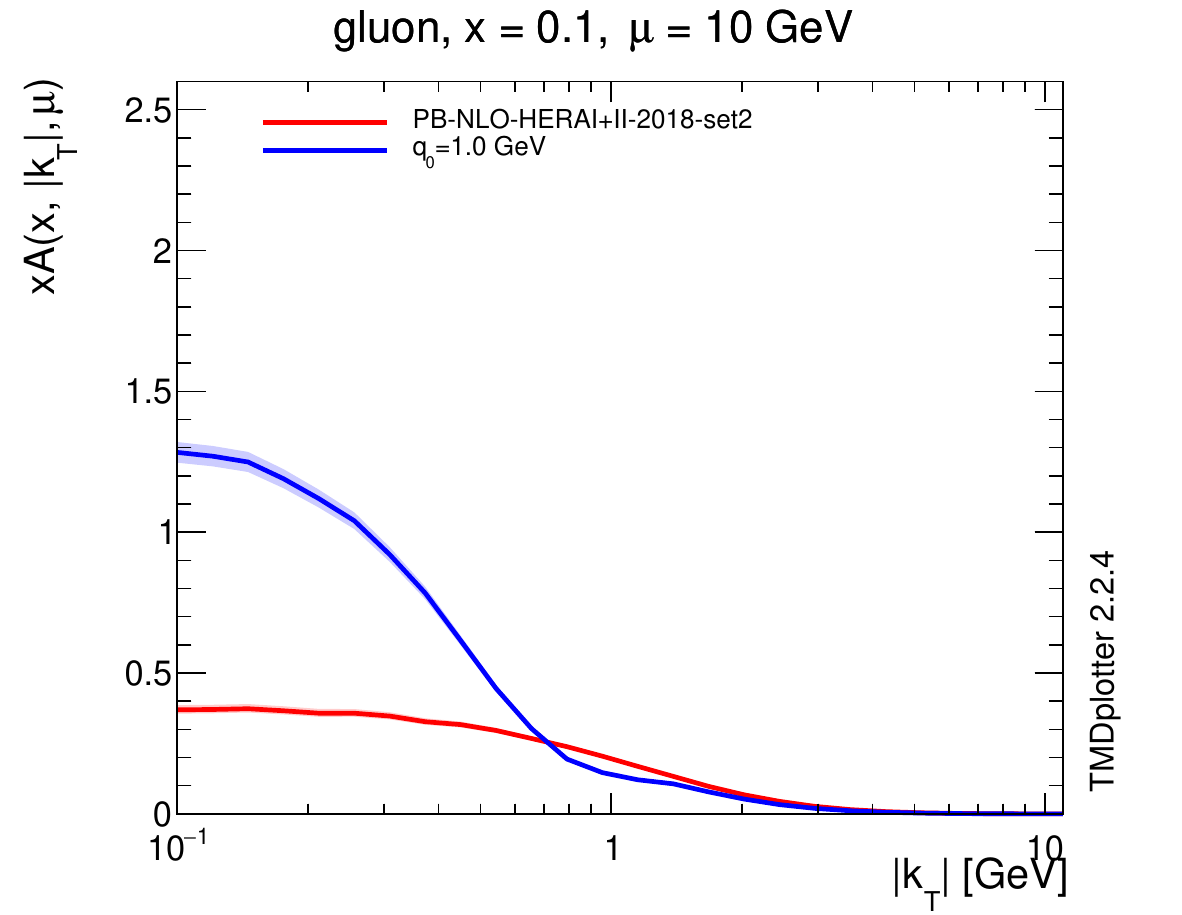}
\end{minipage}
\hfill
\begin{minipage}{0.24\linewidth}
\includegraphics[width=4.1cm]{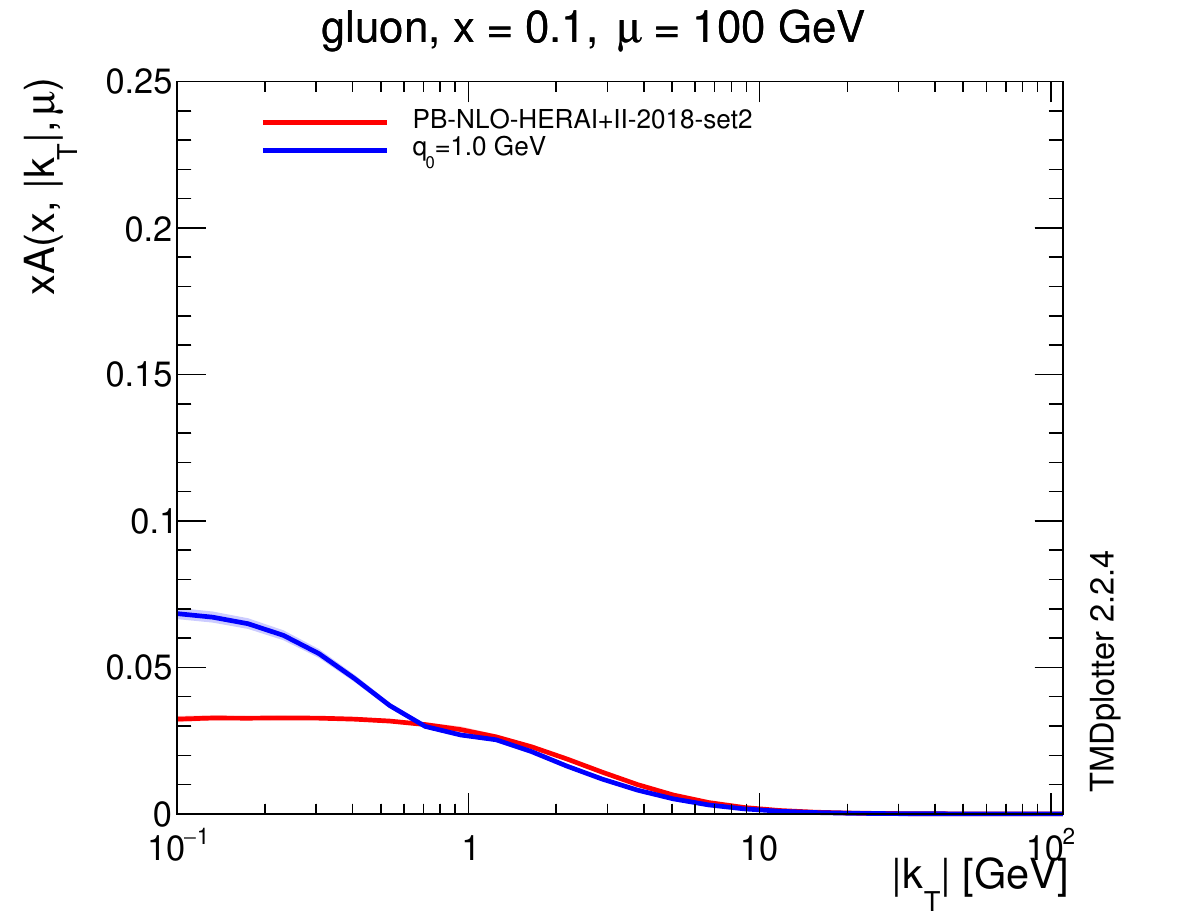}
\end{minipage}
\hfill
\begin{minipage}{0.24\linewidth}
\includegraphics[width=4.1cm]{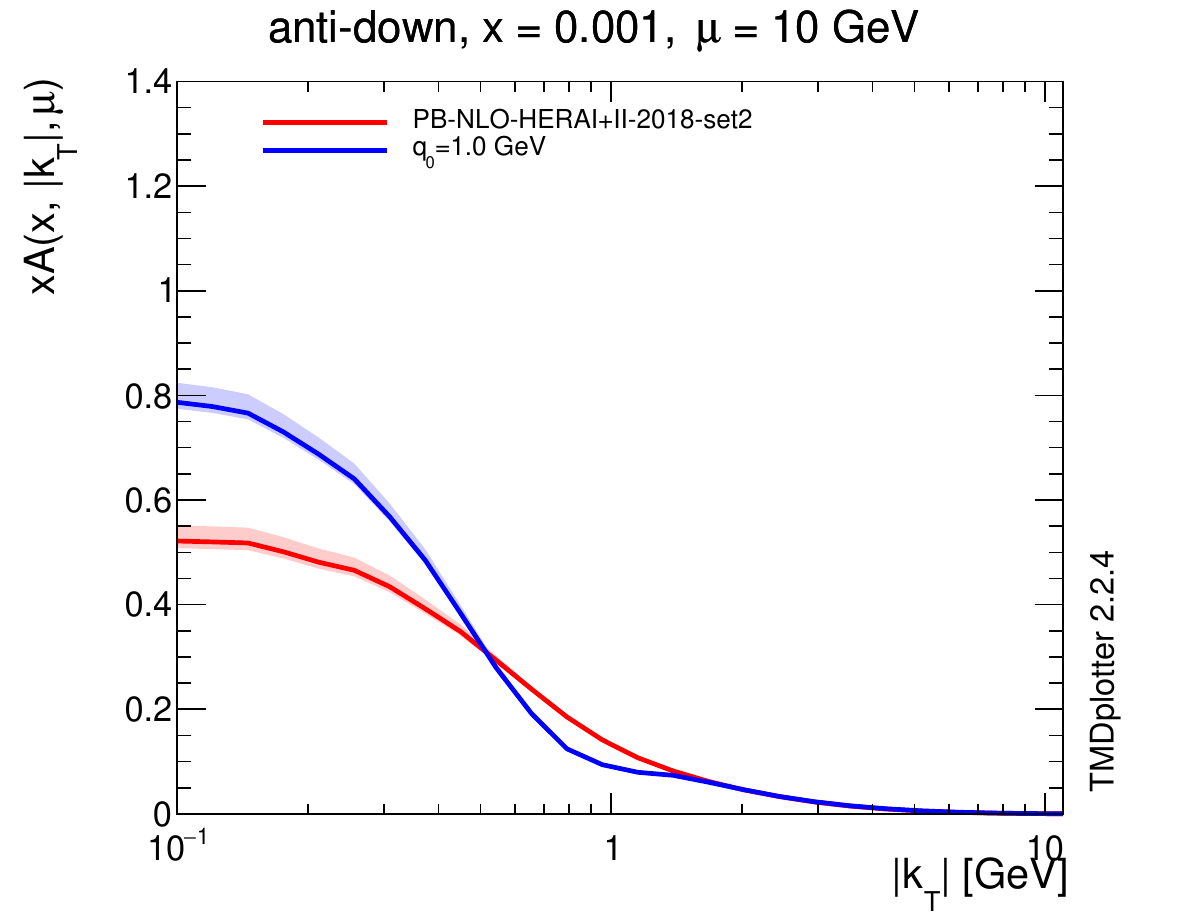}
\end{minipage}
\hfill
\begin{minipage}{0.24\linewidth}
\includegraphics[width=4.1cm]{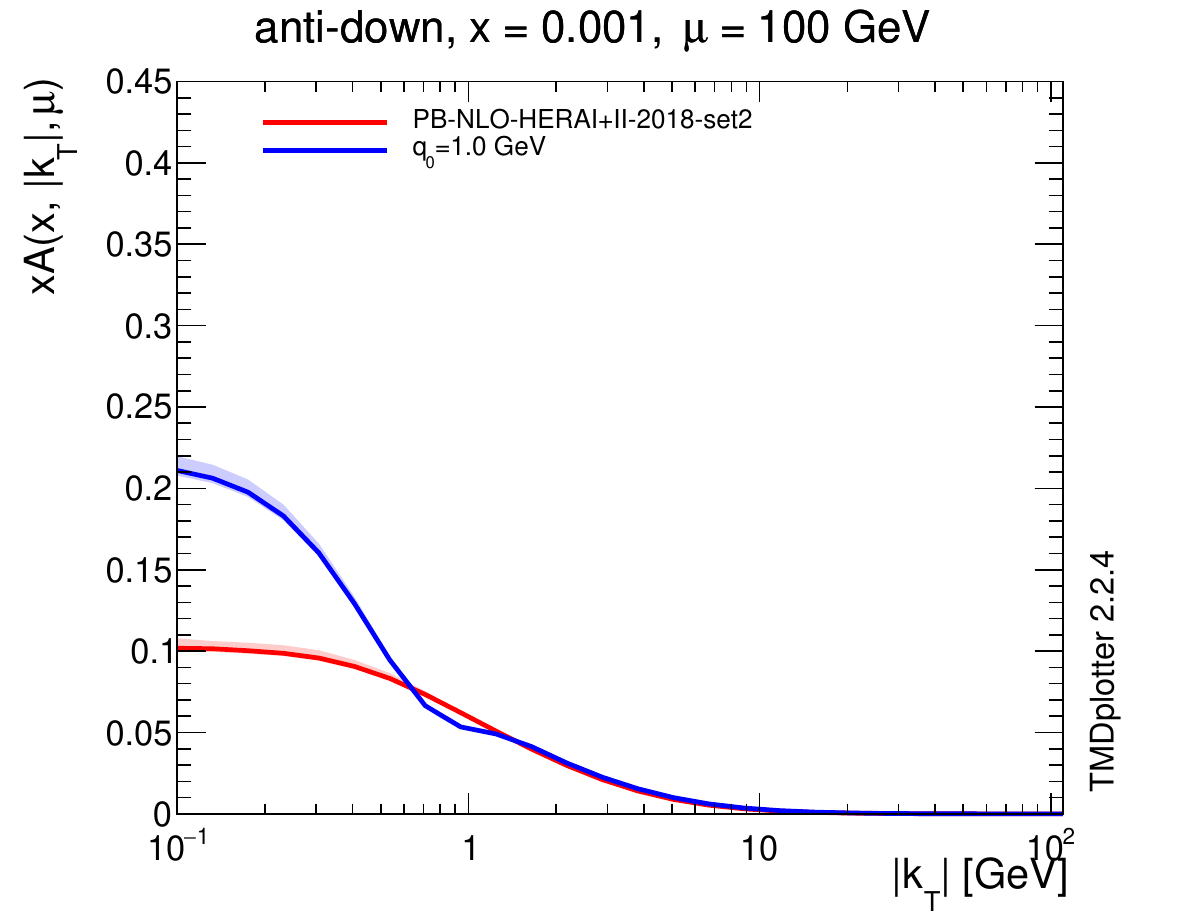}
\end{minipage}
\hfill
\begin{minipage}{0.24\linewidth}
\includegraphics[width=4.1cm]{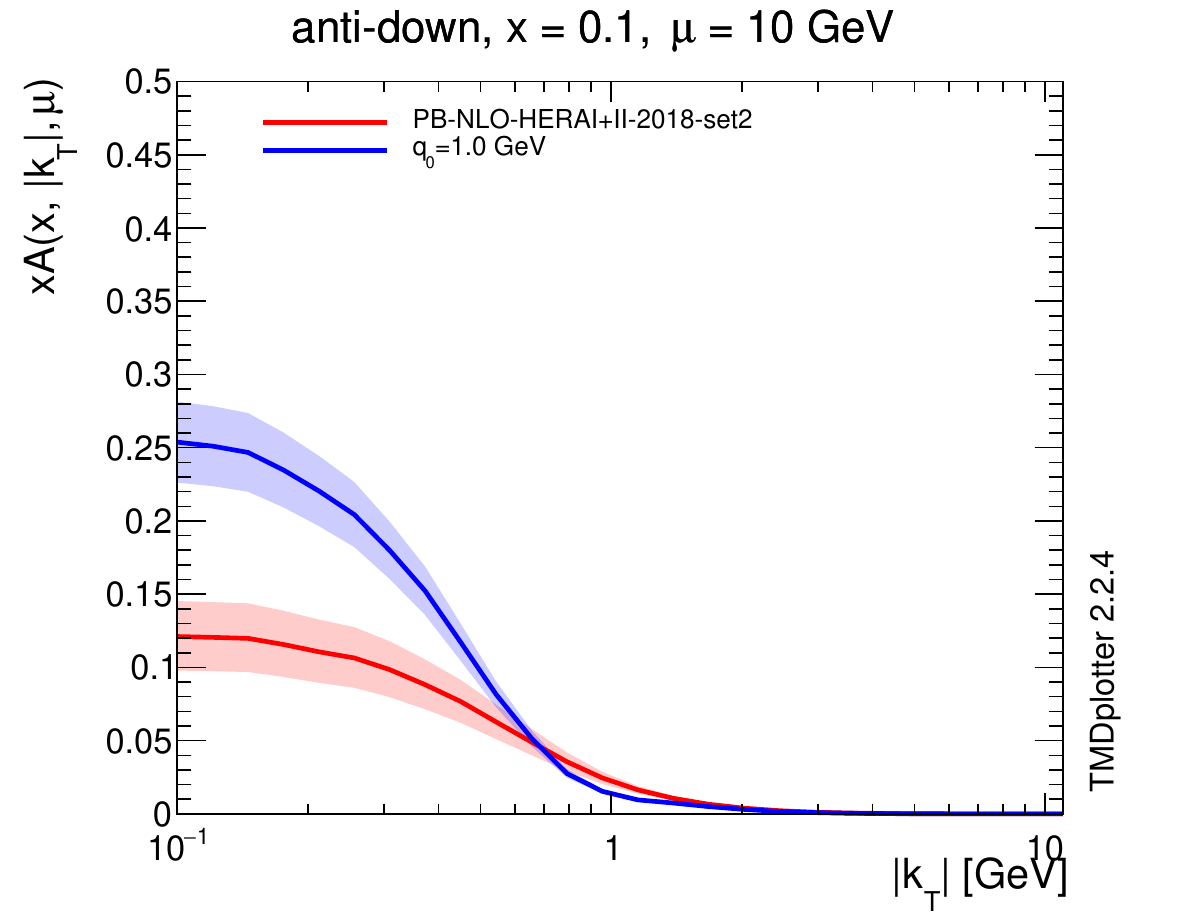}
\end{minipage}
\hfill
\begin{minipage}{0.24\linewidth}
\includegraphics[width=4.1cm]{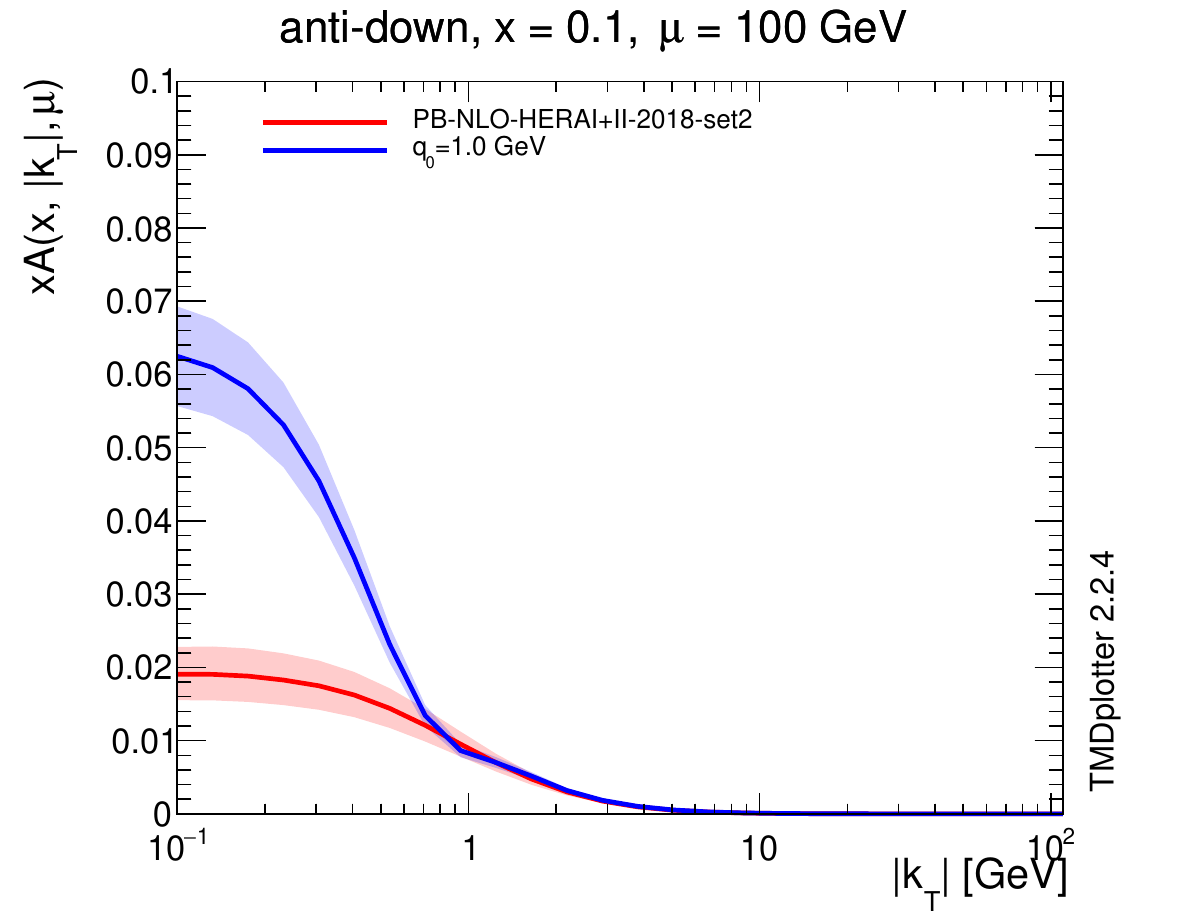}
\end{minipage}
\hfill
\caption[]{Experimental and model uncertainties for TMD distributions of 
gluon and anti-down quark as functions of $k_T$ at  
$\mu=$  10 and 100 GeV  and  $x=0.001\;\rm{and}\;0.1$, obtained from the dynamical-$z_M$ fit with $q_0 = 1 $ GeV. 
The fixed-$z_M$ set PB-NLO-2018-Set2  is 
included for comparison.}
\label{fig:TMDsDynZmvsPBset2}
\end{figure}

In Fig.~\ref{fig:TMDsDynZmvsPBset2}, the experimental and model 
uncertainty bands are shown as functions of $k_T$  for 
the TMD distributions obtained from the fit with $q_0=1.0\;\rm{GeV}$.  The 
PB-NLO-2018-Set2 result is also shown for comparison. Similarly to the 
iTMD case, the distributions are different, with the differences exceeding  
the uncertainty band.

\subsection{Leading-order fits}
\label{app:remarks}

We now present fits performed with the 
same approach as the ones in the previous subsection 
but at leading order (LO) in perturbation theory, that is, 
using LO splitting functions and hard-scattering matrix elements, 
and one-loop running coupling.\footnote{For the LO fits, 
the parameterizations of the initial distributions are kept 
as in Eq.~(\ref{fitparametrization})  except for the gluon, for 
which, following discussions in~\cite{Alekhin:2014irh,H1:2015ubc}, we 
use a parameterization with three parameters instead of five.}   
 LO fits may be useful, given the potential application of dynamical 
$z_M$ to parton showers using LO splitting functions. 
Besides, this study allows us to investigate the 
impact of perturbative order on DIS fits.

\begin{figure}[!htb]
\begin{center}
\includegraphics[width=8cm]{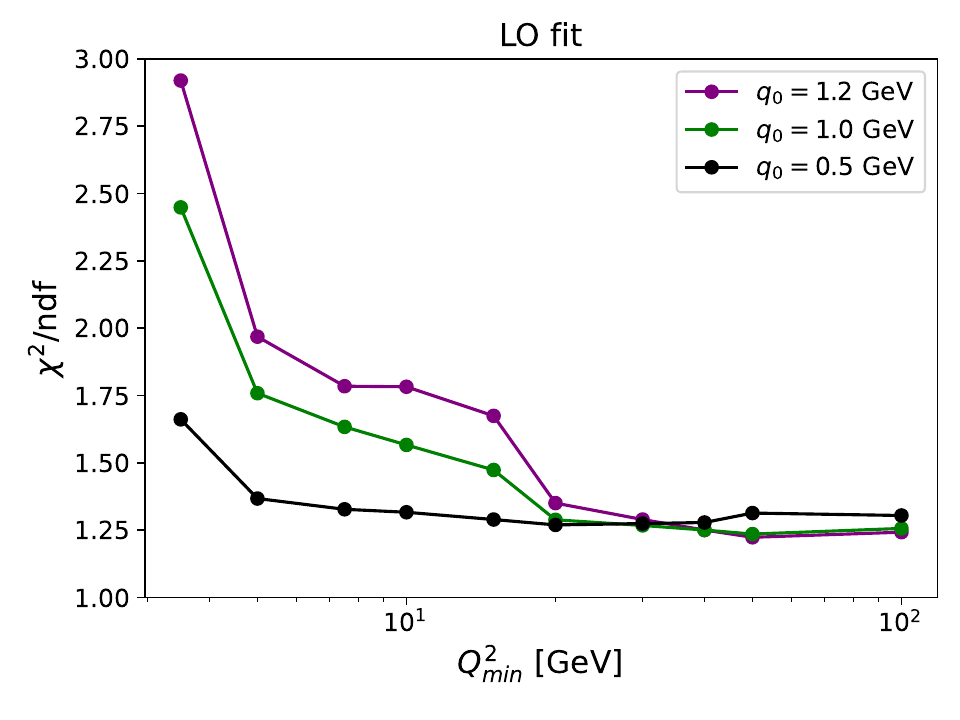}
\end{center}
\caption{The $\chi^2/\rm{n.d.f.}$ value of the LO dynamical-$z_M$ fit 
  to DIS precision data~\protect\cite{H1:2015ubc}  
as a function of the minimum $Q^2$ of 
the data included in the fit, for different values of $q_0$.}
\label{fig:chi2vsQ2minLO}
\end{figure}

The  results for the 
 $\chi^2/\rm{n.d.f.}$ values 
 of the LO fits, corresponding 
 to the   three  different values of $q_0$ 
 as in the previous subsection, 
 are reported in Fig.~\ref{fig:chi2vsQ2minLO}  
 as a function of $Q^2_{\rm{min}}$. 

By comparing  Fig.~\ref{fig:chi2vsQ2minLO}
with  Fig.~\ref{fig:chi2vsQ2minNLO},  we see that 
the $\chi^2/\rm{n.d.f.}$ is significantly better in the 
NLO case than the LO case for all $q_0$ values, and 
that the difference among fits with different $q_0$ values 
is stronger in the LO case. 

\begin{figure}[!htb]
\begin{center}
\includegraphics[width=4.9cm]{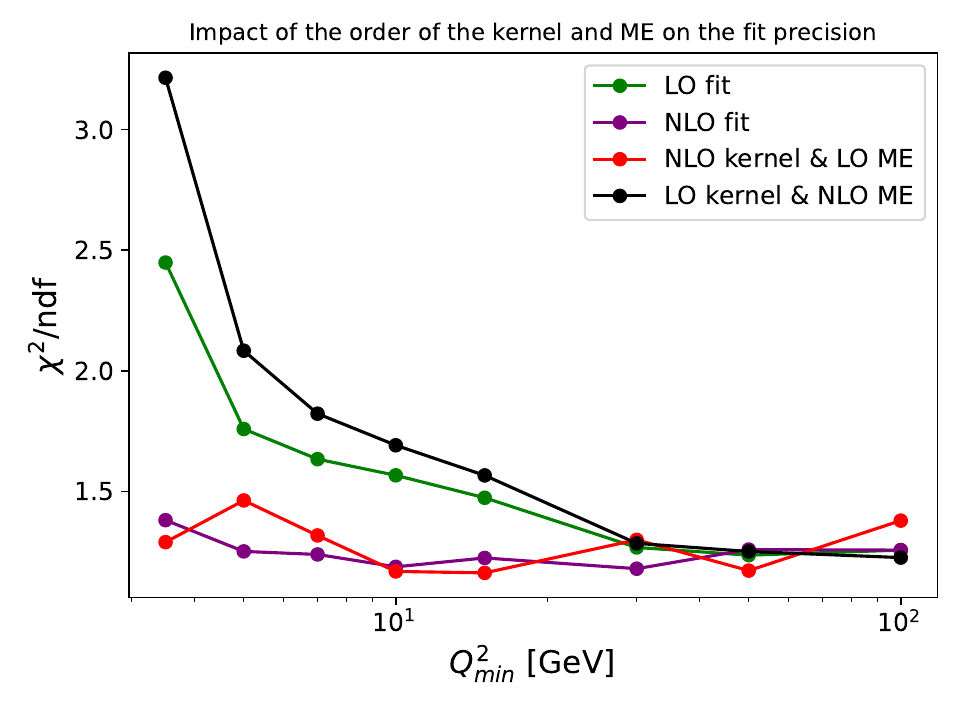}
\includegraphics[width=4.9cm]{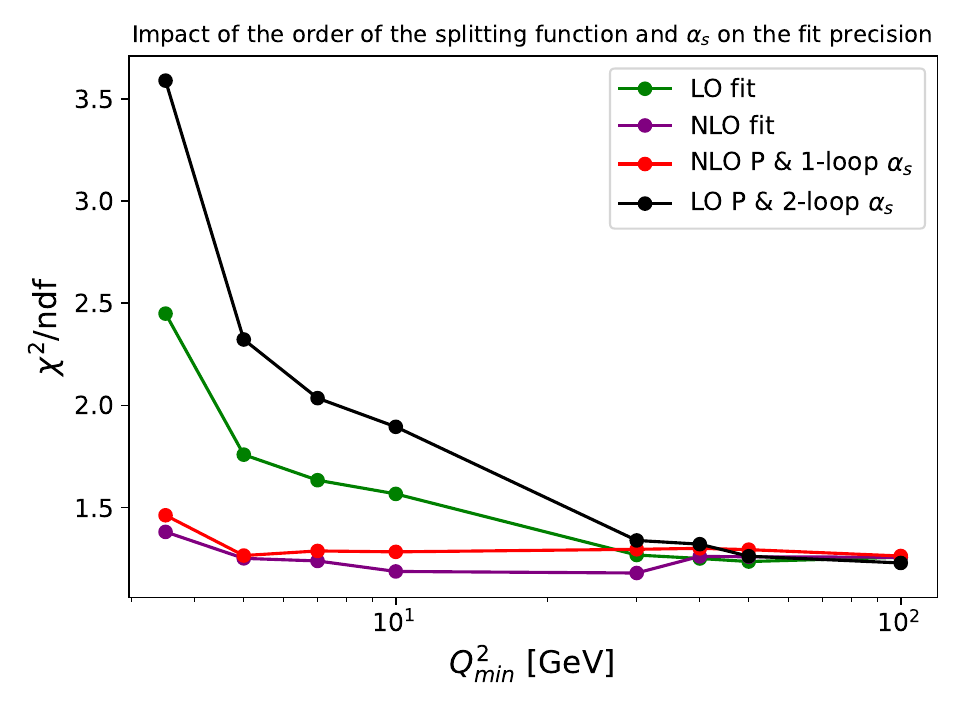}
\includegraphics[width=4.9cm]{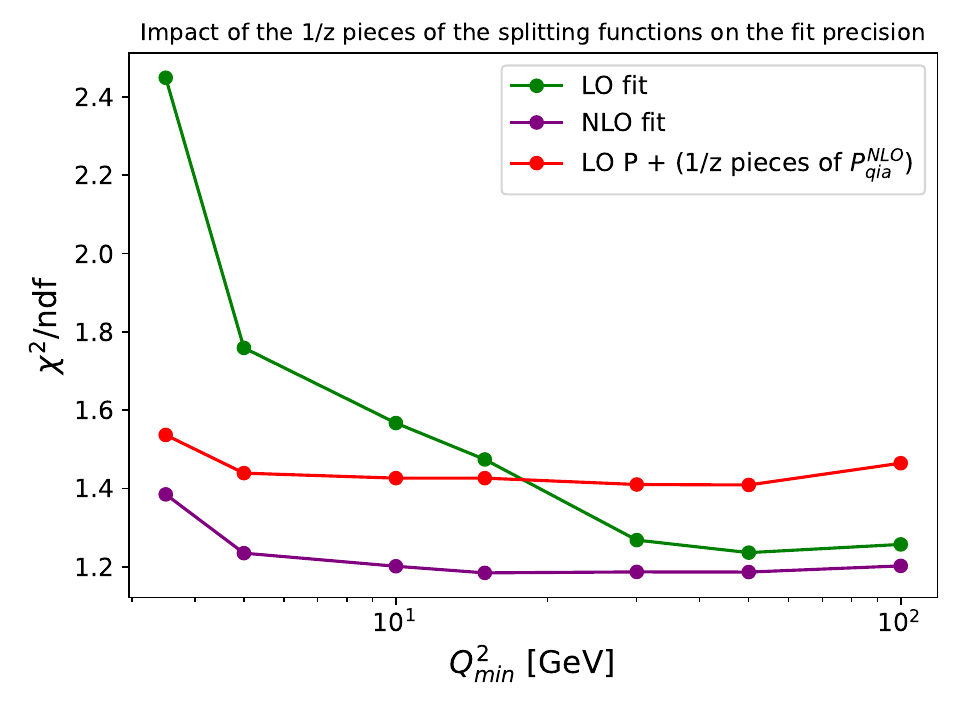}
\end{center}
\caption{The $\chi^2/\rm{n.d.f.}$ of the $q_0 = 1$ GeV fit 
versus $Q^2_{\rm{min}}$ for different scenarios: 
(left) effect of LO and NLO contributions to evolution kernels 
and matrix elements; (middle) effect of running coupling and 
splitting functions to evolution kernels; (right) effect of small-$x$ and 
finite-$x$ parts of the splitting functions.}  
\label{fig:kernelvsMEfitLONLO}
\end{figure}

To  investigate the origin of the different behaviors 
and compare in more detail LO and NLO,  
we perform studies to separate the effects on the 
fits from (i) NLO corrections to evolution kernels and 
matrix elements, (ii) two-loop corrections  to the 
running coupling, (iii) small-$x$ and finite-$x$ 
contributions.\footnote{The inclusion 
 of small-$x$ contributions beyond fixed order into the PB formalism has been 
 explored in~\cite{TaheriMonfared:2019bop}, using 
 modified CCFM kernels~\cite{Hautmann:2008vd,Catani:1989sg,Ciafaloni:1987ur}, 
   and in~\cite{Hautmann:2022xuc}, 
 using TMD splitting functions~\cite{Catani:1993rn,Catani:1994sq,Hautmann:2012sh,Gituliar:2015agu,Hentschinski:2016wya,Hentschinski:2017ayz}. We 
 do not include these developments in the present study.}   
The results are reported in the three panels of Fig.~\ref{fig:kernelvsMEfitLONLO}. 

All computations in  Fig.~\ref{fig:kernelvsMEfitLONLO} are done with 
$q_0 = 1$ GeV. The green and purple curves in each panel 
are, respectively, the LO and NLO results. In the left panel, these are 
compared with the black and red curves obtained, respectively, by 
using LO evolution kernel with NLO matrix element and 
NLO evolution kernel with LO matrix element. The 
closeness of the red and purple curves, on one hand, and 
of the green and black curves, on the other hand, 
suggests that it is the effect  
included in the evolution kernel that  
drives the difference between LO and NLO fits.  

The middle panel in Fig.~\ref{fig:kernelvsMEfitLONLO} examines  
the contributions to the evolution kernel from running coupling 
and splitting functions. Here, the black and red curves are obtained, 
respectively, by using LO splitting functions with two-loop running coupling, 
and NLO splitting functions with one-loop running coupling. 
The closeness of the red and purple curves, on one hand, and 
of the green and black curves, on the other hand, 
suggests that it is the effect included in the splitting functions, rather than 
$\alpha_s$,  that drives the difference between LO and NLO fits. 

The right panel of Fig.~\ref{fig:kernelvsMEfitLONLO} examines the 
contributions to the NLO splitting functions from small $x$ and from 
finite $x$. 
Here the red curve is obtained by adding to the LO splitting functions 
only the $x \to 0$ singular parts of NLO corrections. The 
 behavior of the red curve for small 
$Q^2_{\rm{min}}$ is qualitatively similar to the NLO curve, rather 
 than to the LO curve, suggesting that it is the effect included 
 in the small-$x$ parts of the  NLO splitting functions that drives 
 the difference between LO and NLO fits, particularly in the 
 region of low $Q^2$. This result, besides 
 helping to interpret the behavior observed in 
  Figs.~\ref{fig:chi2vsQ2minNLO}, \ref{fig:chi2vsQ2minLO}, 
  may be useful for 
 studies of small-$x$ dynamics in inclusive-DIS data, {\em e.g.}  including 
saturation~\cite{Armesto:2022mxy,Beuf:2020dxl} and 
 resummation~\cite{xFitterDevelopersTeam:2018hym,Ball:2017otu,Altarelli:2008aj,Ellis:1995gv,Ciafaloni:2007gf,Lipatov:2024xni,White:2006yh}. As discussed earlier, 
 fits with dynamical $z_M$ at varying $q_0$ scale, as performed in this paper,  
 probe the impact of the soft Sudakov radiation region; they can provide a 
 useful tool for  investigations  
 of Sudakov and small-$x$ effects   
 particularly in low-$Q^2$ DIS measurements. 
 
\subsection{A fit with $q_0 \neq q_c$}
\label{subsec:q0notqc} 

As mentioned earlier, while in the fits presented in 
Subsecs.~\ref{subsec:NLO-DynZm-fits} and \ref{app:remarks} 
the showering scale $q_0$ in Eq.~(\ref{running-zM}) and 
the scale $q_c$ in the running coupling in Eq.~(\ref{freeze})   
 are set to be equal, the two scales  are conceptually 
 distinct, and may be set to different values. In this subsection we 
 present a fit in which $q_0=$ 0.5 GeV, $q_c = $ 1 GeV. 
 This parameter choice may be useful to 
 disentangle the effects from the soft-gluon 
 resolution scale defined by the parameter $q_0$ and 
 the $\alpha_s$ freezing scale $q_c$ in the 
 pre-confinement picture. 
In keeping with the notation at the beginning of 
 Subsec.~\ref{subsec:NLO-DynZm-fits}, we will 
 name this fit  PB-NLO-2025-DynZm-q0-0$\_5$B. 
 
 With $q_0=$ 0.5 GeV and $q_c = $ 1 GeV, 
 we find that the \texttt{xFitter}  fit converges 
 $\chi^2/n.d.f. = 1.26$.

\begin{figure}[!htb]
\begin{minipage}{0.31\linewidth}
\includegraphics[width=5.0cm]{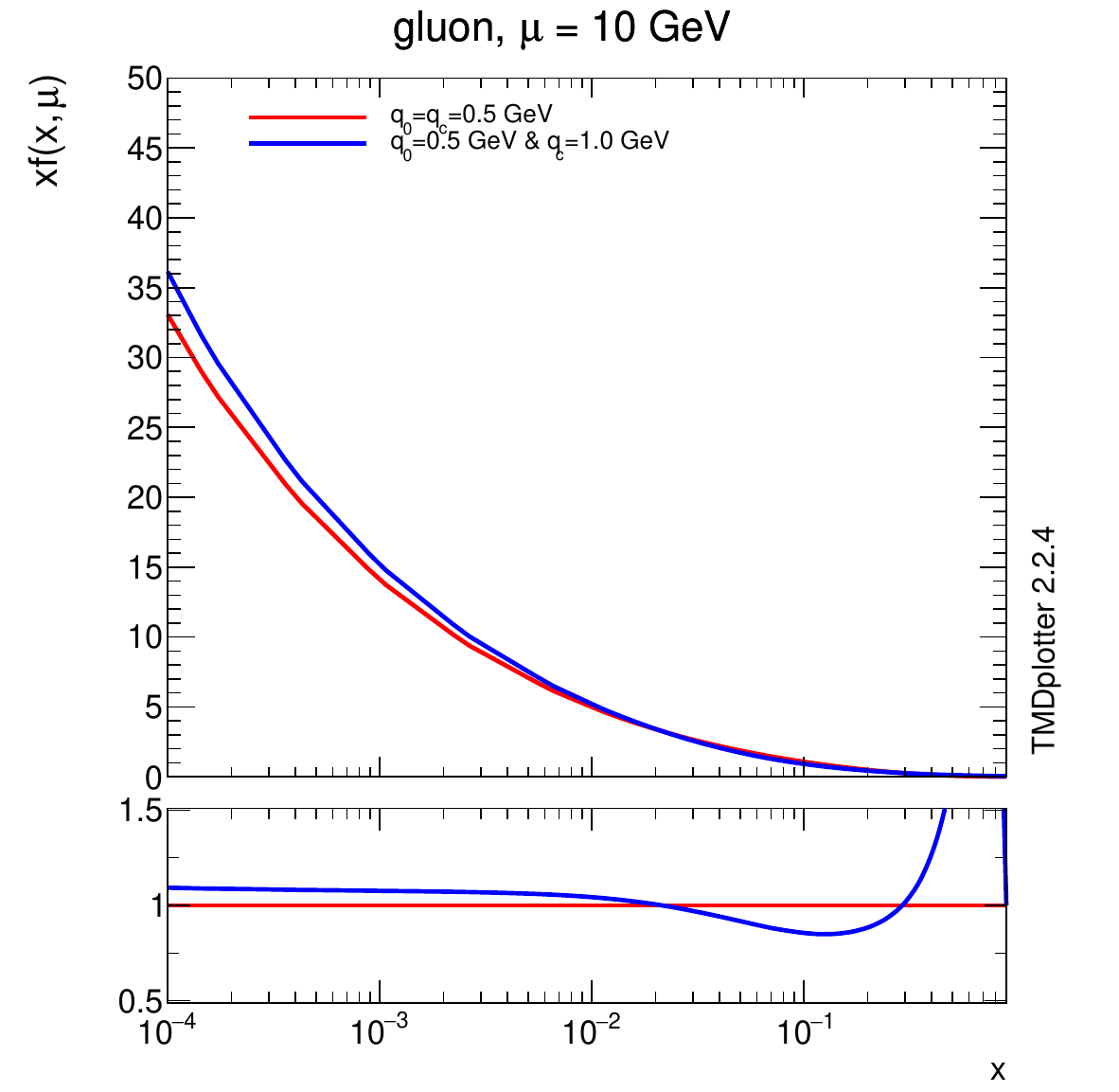}
\end{minipage}
\hfill
\begin{minipage}{0.31\linewidth}
\includegraphics[width=5.0cm]{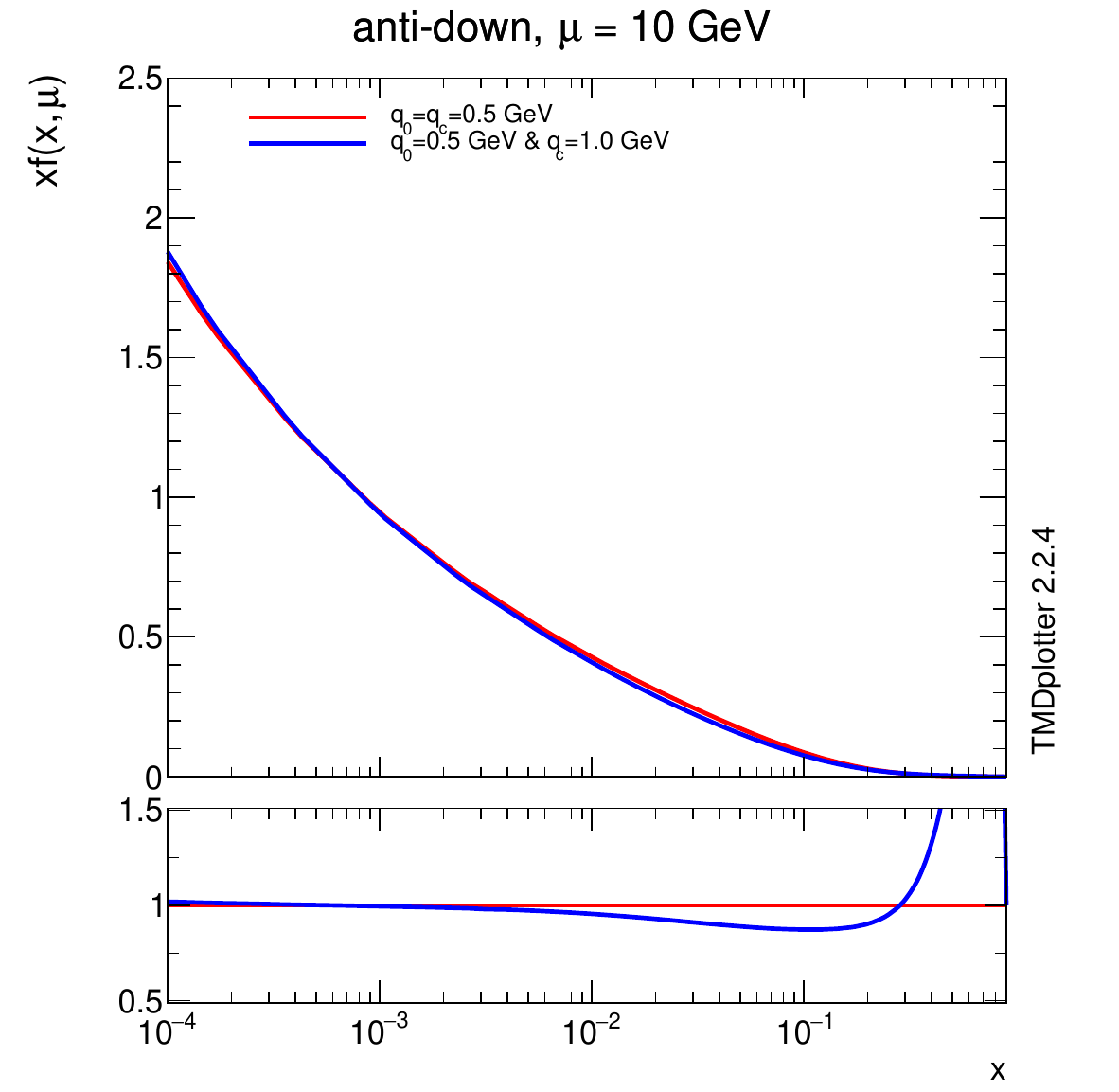}
\end{minipage}
\hfill
\begin{minipage}{0.31\linewidth}
\includegraphics[width=5.0cm]{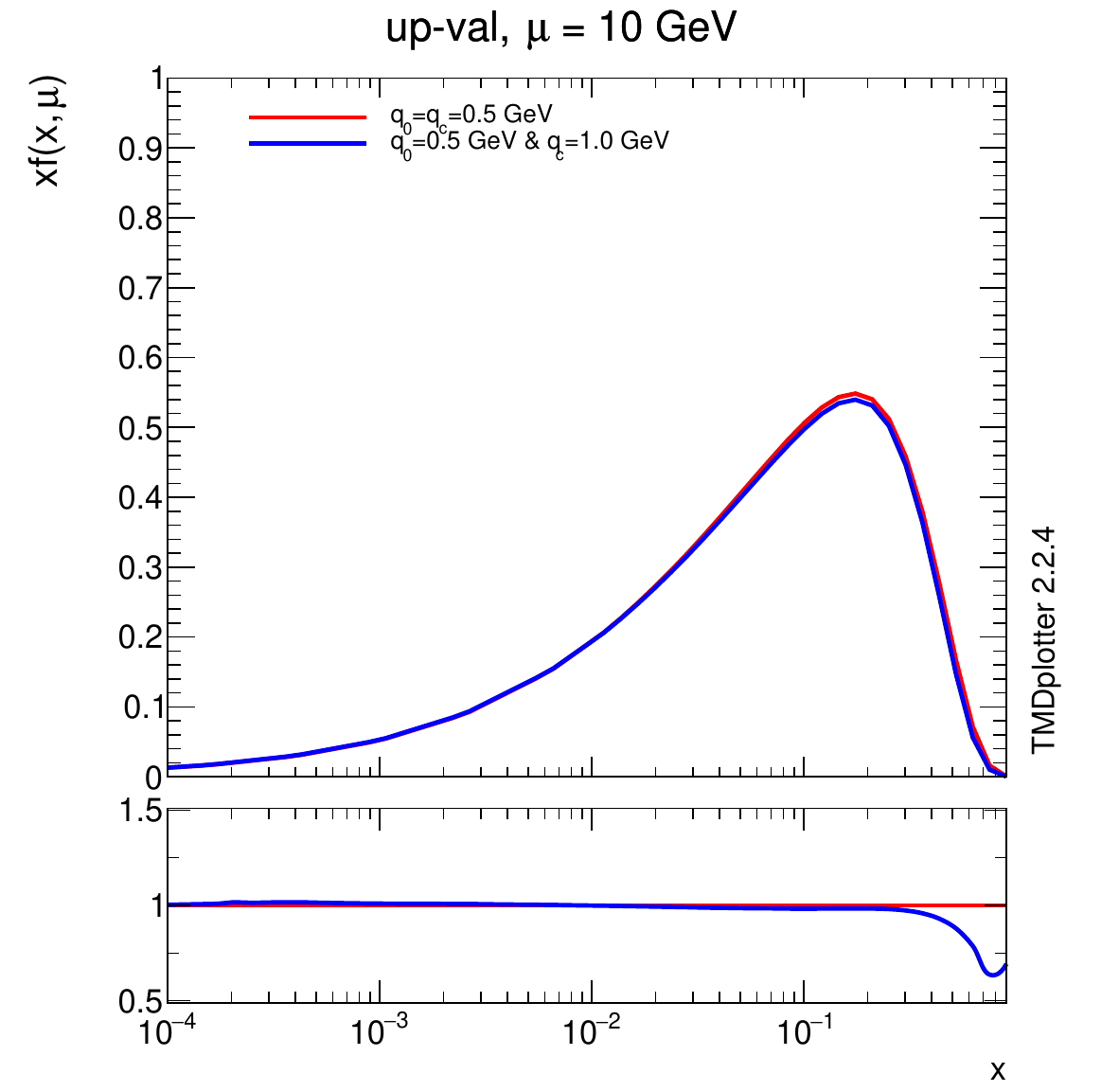}
\end{minipage}

\hfill
\caption[]{The  $x$ dependence of iTMD 
distributions for different parton species 
at $\mu=$ 10 GeV obtained from 
the NLO $q_0=0.5 \; \  \rm{GeV}$ fits 
with   $q_c=1   \; \  \rm{GeV}$ (blue curves) and 
$q_c=0.5 \; \ \rm{GeV}$ (red curves).}
\label{fig:qcnotqc}
\end{figure}

\begin{figure}[!htb]
\begin{minipage}{0.24\linewidth}
\includegraphics[width=4.1cm]{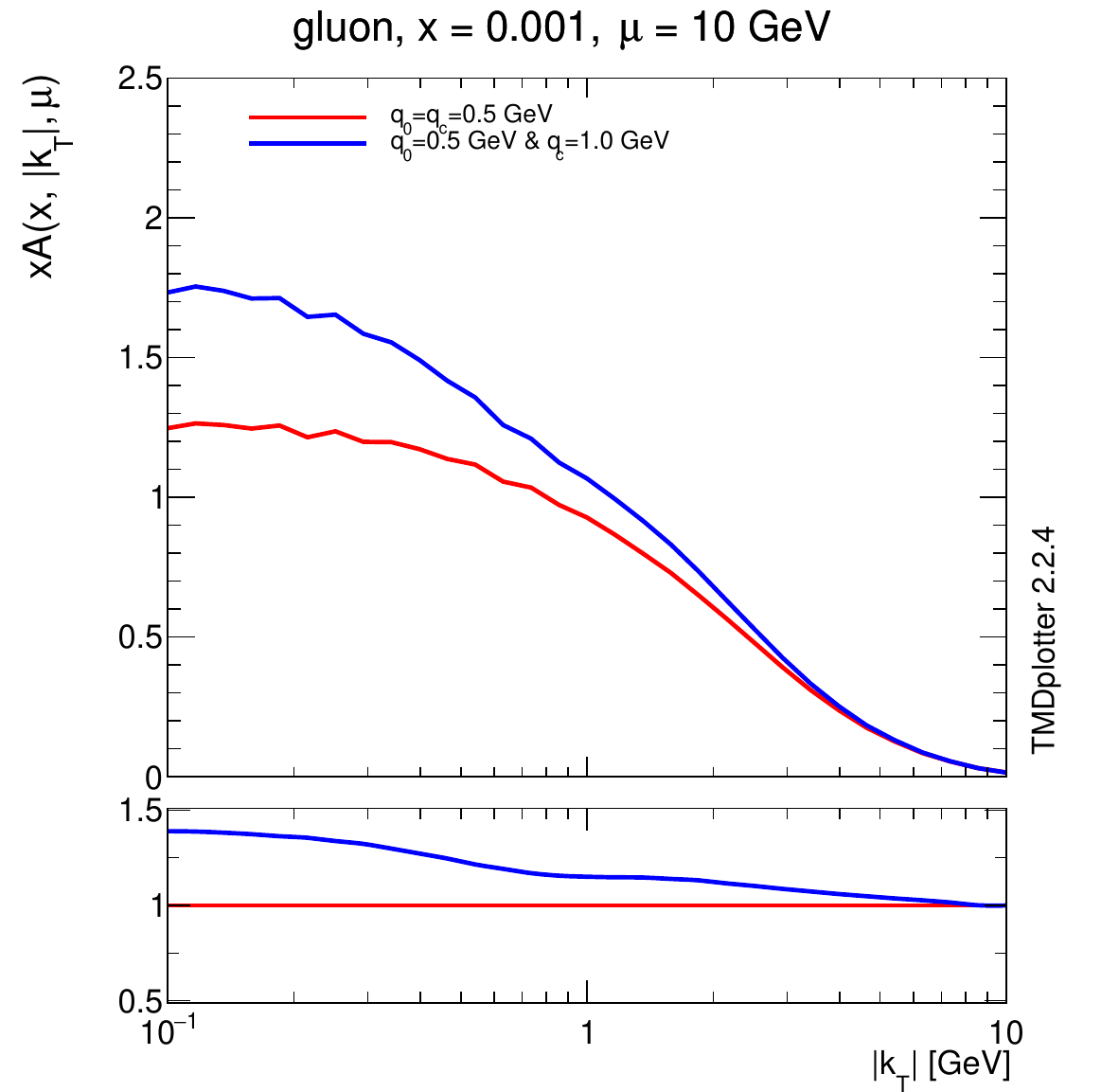}
\end{minipage}
\hfill
\begin{minipage}{0.24\linewidth}
\includegraphics[width=4.1cm]{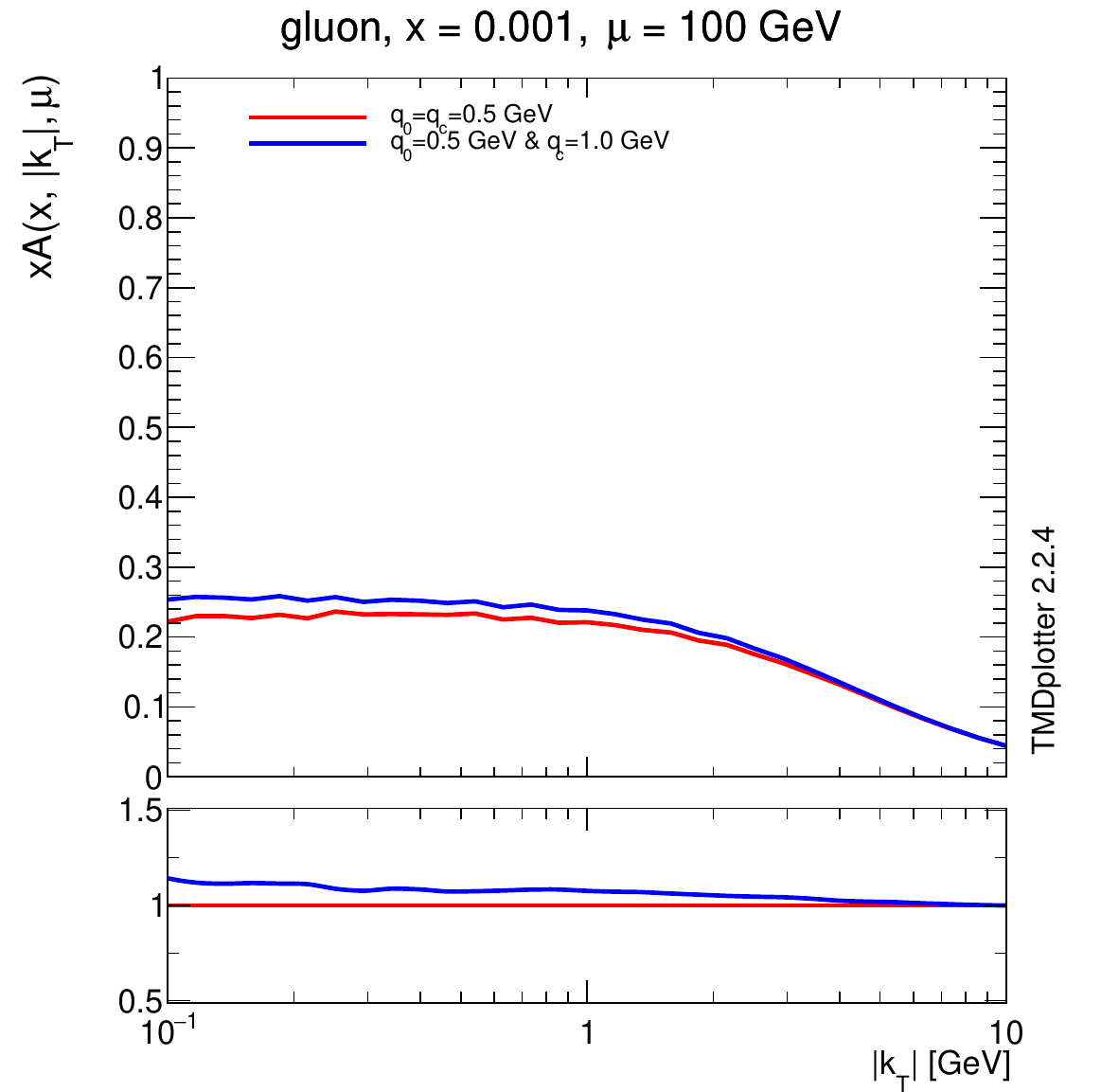}
\end{minipage}
\hfill
\begin{minipage}{0.24\linewidth}
\includegraphics[width=4.1cm]{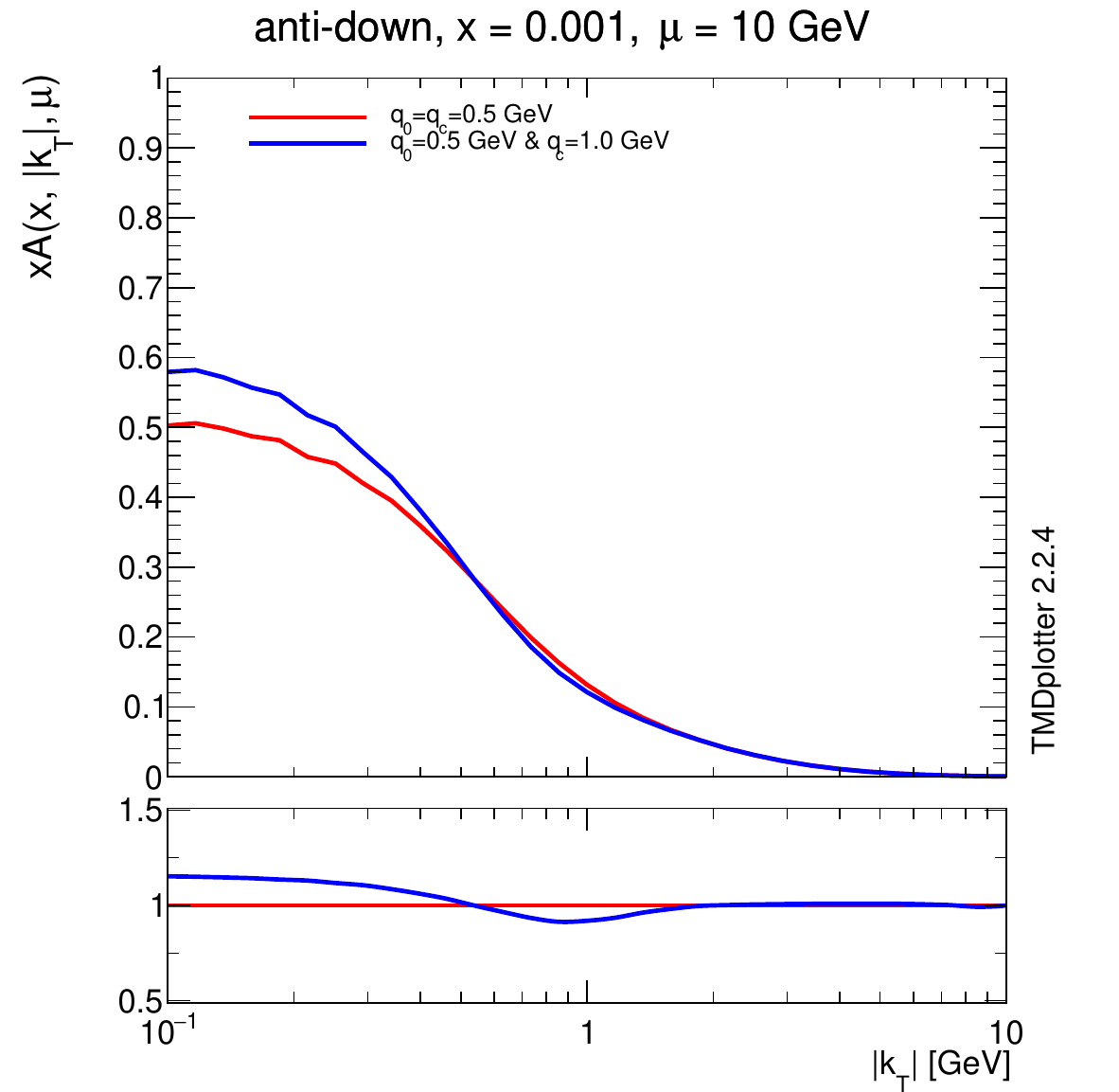}
\end{minipage}
\hfill
\begin{minipage}{0.24\linewidth}
\includegraphics[width=4.1cm]{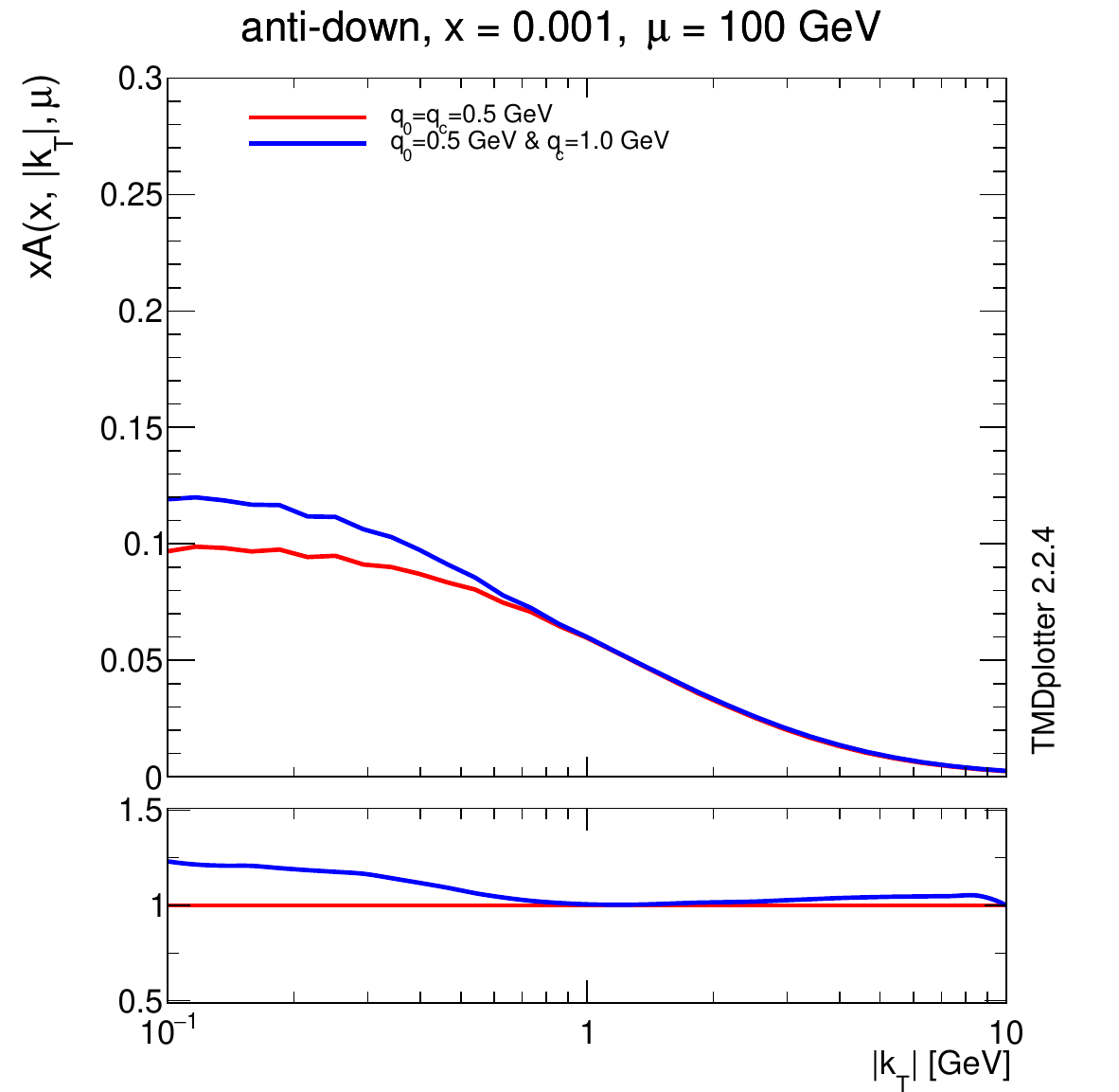}
\end{minipage}
\hfill
\caption[]{The $k_T$   dependence of 
TMD distributions for gluon and anti-down quark, 
 at  $x=0.001$ and   $\mu=$  10 and 100 GeV, 
obtained from the NLO $q_0=0.5  \;\  \rm{GeV}$ fits 
with    $q_c=1   \; \  \rm{GeV}$ (blue curves) and 
$q_c=0.5 \; \ \rm{GeV}$ (red curves).}
\label{fig:qcnotqcTMDs}
\end{figure}

Fig.~\ref{fig:qcnotqc} and  Fig.~\ref{fig:qcnotqcTMDs}
show, respectively, a comparison of the iTMD and TMD 
distributions for the sets PB-NLO-2025-DynZm-q0-0\_5 and 
PB-NLO-2025-DynZm-q0-0$\_5$B.  The results are 
similar for most of the kinematic regions and parton 
channels; however, non-negligible differences arise 
especially in the gluon distributions and for small transverse 
momenta.

%% file: sec-zmaxfit2_rvsd.tex
\section{Application to DY transverse momentum}
\label{sec-zmaxfit2}

As an application of the DIS fits  
with dynamical resolution scale 
$z_M$ described in Sec.~\ref{sec-zmaxfit1}, 
in this section we compute theoretical predictions for the 
DY transverse momentum distribution, following the 
method developed in 
Refs.~\cite{BermudezMartinez:2019anj,BermudezMartinez:2020tys}, 
and  
compare them 
 with  
experimental data from the LHC and lower-energy experiments. 
In Subsec.~\ref{subsec4-1} we present the 
main results, and in  Subsec.~\ref{subsec4-2} we 
give comments and point to further applications.

\subsection{PB results with dynamical $z_M$}
\label{subsec4-1}

For the purpose of this study, we concentrate on 
the set with $q_0 = $ 1 GeV given in Sec.~\ref{sec-zmaxfit1}. 
 Following the procedure  in 
 Refs.~\cite{BermudezMartinez:2019anj,BermudezMartinez:2020tys}, we first 
use the integrated TMDs 
determined in Sec.~\ref{sec-zmaxfit1} 
to generate the DY hard-scattering matrix element  at NLO 
from  MCatNLO in the LHE format~\cite{Alwall:2006yp};  we then 
 supplement it  with the transverse momentum generated according 
 to the TMD distributions of Sec.~\ref{sec-zmaxfit1} 
by employing the Monte Carlo event generator 
{\sc Cascade}3~\cite{CASCADE:2021bxe}.   
The  matching of NLO matrix element and PB TMD distributions 
is performed by using 
the {\sc Herwig}6 \cite{Corcella:2002jc,Corcella:2000bw} subtraction 
term in MCatNLO (a 
detailed study of the procedure using this subtraction term is performed in the 
appendix of Ref.~\cite{Yang:2022qgk}).   
 Theoretical uncertainties from perturbation theory 
 are estimated  within MCatNLO by variation of the 
renormalization and factorization scale. Besides,  uncertainties 
from the PB TMD distributions as determined in Sec.~\ref{sec-zmaxfit1} 
are included. The events thus generated are analyzed 
by using the {\sc Rivet} package~\cite{Buckley:2010ar}.

\begin{figure}
\begin{centering}
\includegraphics[width=8cm]{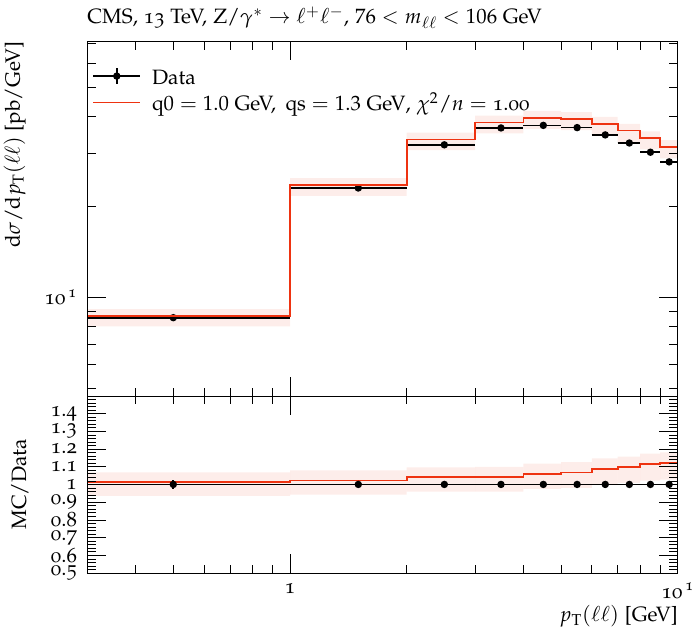}
\caption[]{Predictions for the lepton-pair transverse momentum  
obtained with MCatNLO matched with dynamical-$z_M$ PB TMD 
distributions with  $q_0=1\;\rm{GeV}$,  compared to $13\;\rm{TeV}$ CMS data.
Uncertainties from perturbative scale variations in MCatNLO 
are shown. }
\label{fig:DYCMS}
\end{centering}
\end{figure}

In Fig.~\ref{fig:DYCMS} the theoretical predictions thus obtained 
are compared with $13\;\rm{TeV}$ CMS data \cite{CMS:2022ubq}.
Similarly to the finding in Ref.~\cite{BermudezMartinez:2019anj} for 
the case of  fixed $z_M$,  
the predictions obtained from PB TMD evolution with NLO matching 
are capable of describing the data well  for low and moderate $p_T$, 
while  for  higher $p_T$ contributions from multiple hard-parton emissions 
beyond NLO  become important and need to be taken into account, 
as illustrated, {\em e.g.}, through the TMD multi-jet merging technique in   
Refs.~\cite{BermudezMartinez:2021lxz,BermudezMartinez:2021zlg,BermudezMartinez:2022bpj}. 
In the present study we  focus on the low-$p_T$ region, in which the 
NLO-matched predictions are sufficient. 
 
Next, we       investigate 
further  the predictions obtained with 
 dynamical resolution scale $z_M$. In particular,  
we wish to examine the sensitivity of  DY measurements  
to the intrinsic-$k_{T}$ parameter $q_s$ of the TMD distribution 
in the dynamical-$z_M$ scenario, by performing a fit of 
the theoretical predictions 
to the 13 TeV CMS data~\cite{CMS:2022ubq} 
for DY transverse momentum. 
We focus on the DY mass region around the $Z$-boson peak. 
We leave a complete intrinsic-$k_{T}$ analysis, 
extending to all DY mass regions   
measured in~\cite{CMS:2022ubq} 
(from 50 GeV to 1 TeV)   and including a full treatment of 
correlations (as done in  Ref.~\cite{Bubanja:2023nrd} for 
fixed $z_M$), to future work.   

\begin{figure}[h]
  \begin{centering}
     \includegraphics[width=0.69\textwidth]{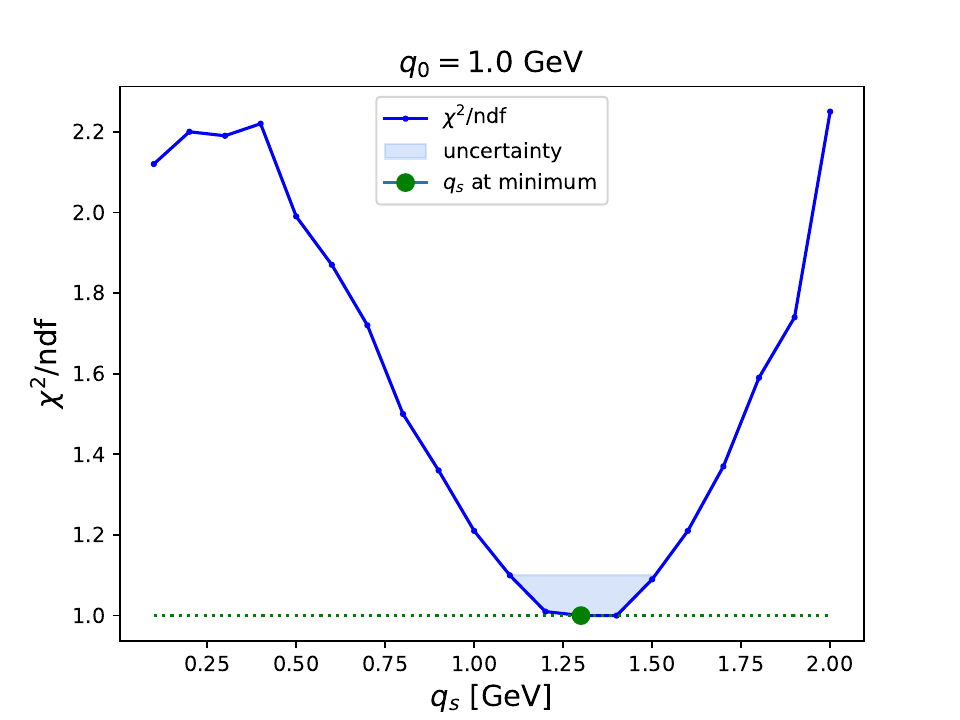}   
    \caption{The $\chi^2/n$ values versus the $q_s$ parameter    
    from a comparison of the NLO-matched dynamical-$z_M$ predictions 
     at $q_0 = 1.0$ GeV with the CMS measurements~\protect\cite{CMS:2022ubq}. 
     }
    \label{fig:amine-chi}
  \end{centering}
\end{figure}

In order to determine the intrinsic $k_{T}$ that describes the data best,  
we vary the value of the $q_s$ parameter 
and for each value we calculate the $\chi^2$ 
to quantify the model agreement with the measurement.   
We consider the region of DY transverse momenta  
  $ p_T \leq p_{T , {\rm{max}}} $, for different  values of $p_{T , {\rm{max}}} $. 
The results for the $\chi^2$ per degree of freedom versus $q_s$  are shown 
in  Fig.~\ref{fig:amine-chi} for the case $p_{T , {\rm{max}}}  = $ 10 GeV. 
We have verified the robustness of the result for $q_s$ by varying 
$p_{T , {\rm{max}}} $ in the range between 5 ad 15 GeV.

The results of the 
calculation reported in Fig.~\ref{fig:amine-chi}  
indicate that the dynamical-$z_M$ distributions, fitted to precision DIS data 
in Sec.~\ref{sec-zmaxfit1},  successfully  
describe the LHC DY data~\cite{CMS:2022ubq} at low $p_T$  and that sensitivity to the  nonperturbative 
intrinsic-$k_{T}$   parameter $q_s$ can be achieved from these low-$p_T$ data.  

\begin{figure}[h]
  \begin{centering}
     \includegraphics[width=0.79\textwidth]{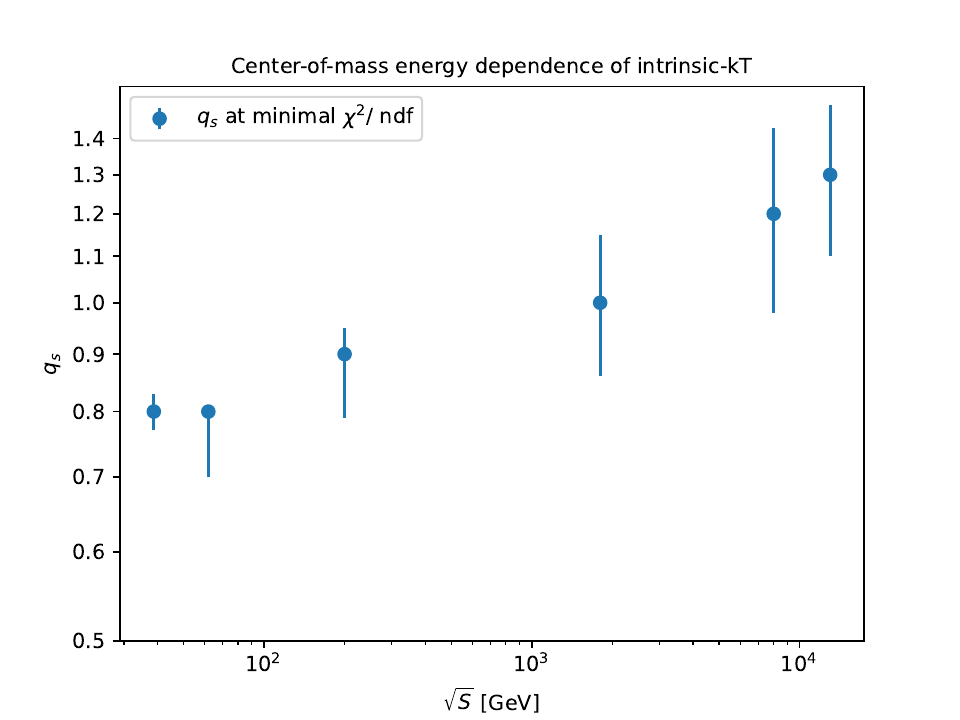}   
    \caption{The values of the  $q_s$ parameter  obtained   
    from a comparison of the NLO-matched dynamical-$z_M$ predictions 
     at $q_0 = 1.0$ GeV with measurements at different 
     energies~\protect\cite{CMS:2022ubq,ATLAS:2015iiu,D0:1999jba,PHENIX:2018dwt,Antreasyan:1981eg,NuSea:2003qoe,Webb:2003bj}. 
     }
    \label{fig-qsVsSdyn8may}
  \end{centering}
\end{figure}

To embed these results in a broader context 
and further explore the description of DY transverse momentum 
by predictions with  dynamical resolution scale $z_M$, 
we turn to DY measurements from lower-energy experiments, 
 and repeat the analysis  above in the cases of 
 DY data sets at different  energies. 
Similarly to what is done in 
 Refs.~\cite{BermudezMartinez:2020tys,Bubanja:2023nrd} 
  for the fixed-$z_M$ scenario, 
   we consider data from the measurements by  
 ATLAS at 8 TeV~\cite{ATLAS:2015iiu},  D0 at 1.8 TeV~\cite{D0:1999jba}, 
 PHENIX at 200 GeV~\cite{PHENIX:2018dwt}, 
 R209 at 62 GeV~\cite{Antreasyan:1981eg}, 
 NuSea at 38 GeV~\cite{NuSea:2003qoe,Webb:2003bj}. 
 In each of these cases, we apply the same method as above, and 
 determine the intrinsic $k_{T}$ that describes the data best by 
 varying the value of the $q_s$ parameter 
and computing  $\chi^2 $  for every value. We 
find 
good $\chi^2 $ for each of the data sets. 
The $q_s$ values corresponding to the 
best $\chi^2$ for each energy are shown 
in Fig.~\ref{fig-qsVsSdyn8may}. 

We observe from Fig.~\ref{fig-qsVsSdyn8may} 
an increase in the value of the 
intrinsic-$k_{T}$   parameter $q_s$ 
with increasing center-of-mass energy. 
This observation supports the physical 
picture suggested in Ref.~\cite{Bubanja:2023nrd} 
and discussed at the end of Sec.~\ref{sec-pb} in this paper,  
in which the behavior of the intrinsic $k_{T}$  is 
correlated with non-perturbative Sudakov effects 
near the soft-gluon resolution boundary. In particular, the 
$q_s$  increase  with energy can be associated with 
dynamical $z_M$. The results of Fig.~\ref{fig-qsVsSdyn8may} 
are obtained with showering scale $q_0 = 1$ GeV, and suggest 
that it will be 
relevant to systematically investigate correlations of 
$q_0$ and $q_s$.

\subsection{Comments and future applications}
\label{subsec4-2}

The results of the previous subsection 
enable further investigations to be carried out 
in various directions. 
Since 
a detailed uncertainty breakdown is available for the 
measurements~\cite{CMS:2022ubq} and   
correlations between bins of the measurement are included 
 for each uncertainty source separately,  
  the results of this paper  can be extended 
 by carrying  
 out   a complete analysis  including the full covariance matrix.  
 This can then be compared  
with the analogous analysis  
performed, for the fixed-$z_M$ case,  
in   Ref.~\cite{Bubanja:2023nrd} with  DY masses 
from 50 GeV to 1 TeV.

Also, the 
extraction of $q_s$ from  dynamical-$z_M$ fits  
to DY measurements at  
different  center-of-mass energies, presented in 
Fig.~\ref{fig-qsVsSdyn8may} for $q_0 = 1 $ GeV,  
can be explored further by exploiting 
 the possibility to vary  
 the emissions' minimum transverse momentum $q_0$  in the 
 soft-gluon resolution $z_M$. This can help shed light on 
 the transition between the distinct  energy behaviors of $q_s$  
 illustrated  by the left hand side panel and right hand side panel in 
  Fig.~\ref{fig:extract-intrins}. 

As discussed in Sec.~\ref{sec-pb}, the use of the 
dynamical resolution scale allows 
 nonperturbative Sudakov contributions to be explored. 
Ref.~\cite{Martinez:2024mou} presents 
the PB TMD evaluation of the Collins-Soper kernel as a function of 
transverse distance $b$. This is to be compared   
with recent evaluations of the kernel 
 from lattice calculations~\cite{Ebert:2018gzl,Ebert:2019tvc,Shanahan:2020zxr,Shanahan:2021tst,Avkhadiev:2024mgd,Avkhadiev:2023poz,Bollweg:2024zet,Bollweg:2025iol,LatticeParton:2020uhz,LatticePartonLPC:2022eev,LatticePartonLPC:2023pdv,Li:2021wvl,Schlemmer:2021aij,Shu:2023cot} and from fits to experimental 
data for transverse momentum spectra~\cite{Bacchetta:2022awv,Bacchetta:2024qre,Bury:2022czx,Moos:2023yfa}.   
In Ref.~\cite{Martinez:2024mou}, dynamical $z_M$ is found  to 
give rise to a flatter shape in the 
kernel at large distances $b$ compared to fixed $z_M$. 
This behavior is close to the spirit of parton saturation in the $s$-channel 
picture~\cite{Hautmann:2007cx,Hautmann:2000pw} for 
partonic distribution functions, and is phenomenologically relevant 
since  recent fits find that a flat large-$b$ behavior is preferred by 
data  for DY transverse momentum~\cite{Hautmann:2020cyp} and 
$e^+ e^-$ thrust distribution~\cite{Boglione:2023duo}. 
An improved understanding of these nonperturbative effects will be 
important for precision analyses of DY observables involving 
the  lowest  measured transverse momenta~\cite{Camarda:2023dqn,Camarda:2022qdg,Camarda:2021ict,Bertone:2024snr,Isaacson:2023iui,Neumann:2022lft,Chen:2022cgv,Ju:2021lah,Ebert:2020dfc,Billis:2024dqq,Becher:2020ugp}.

As a general comment, we note that the 
analysis of dynamical resolution scales  presented in this 
paper is applicable both to collinear-sensitive 
observables and to TMD-sensitive observables (in contrast with 
 other approaches which are  
specifically designed for one or the other class of observables): 
see e.g. the cases of 
 inclusive DIS structure functions 
 in Sec.~\ref{sec-zmaxfit1}   and 
 DY transverse momentum distributions  in this section.   
We  regard the  distributions with dynamical $z_M$ 
determined in this paper as well-suited to address open 
issues on    the use of parton distributions, both 
collinear~\cite{Nagy:2020gjv,prestel:2020,Mendizabal:2023mel,Frixione:2023ssx} 
and TMD~\cite{Jung:2024eam,Raicevic:2024obe},   in parton shower generators.

%% file: sec-concDyn_rvsd.tex
\section{Conclusions}
\label{sec:concl}

 In this work we investigate the role of   
 dynamical soft-gluon resolution scales $z_M$ in PB algorithms, 
 examining parton distributions at 
 both collinear and TMD level. 
 
 At collinear level, the motivation for this study 
 comes from the potential systematic mismatch arising in 
 parton shower Monte Carlo event generators due to the fact 
 that the resolution scale is used in the shower evolution 
 but not in the PDF evolution. At TMD level, the motivation 
 comes from the  interplay of the 
  resolution scale with the transverse momentum recoils  
  in the shower, and possibly with the non-perturbative 
  intrinsic transverse momentum.  
 
The main result of this work is that we 
 obtain a set of collinear and TMD parton distributions from 
  fits to precision DIS data performed by using PDF and TMD 
  evolution with dynamical $z_M$, as in parton showers. 
  This is done working  at NLO in perturbation 
  theory. The dynamical-$z_M$ set includes  
  experimental and model uncertainties.  
   We show that this  set not only gives a good fit to precision 
  DIS data but also is capable of describing  DY transverse 
  momentum data. 
  
  To illustrate this quantitatively, we  
  perform a fit to DY transverse momentum 
measurements  at the LHC and carry out 
the extraction of the non-perturbative TMD parameter $q_s$, representing the 
intrinsic transverse momentum,  for the first time 
in the presence of dynamical resolution scales. We further perform fits to 
DY transverse momentum measurements from lower-energy experiments at Tevatron, RHIC 
and fixed target, extract $q_s$ and present its energy dependence. 
   
 We expect the results of this paper to be useful for several applications. 
 The collinear   distributions with dynamical $z_M$ can be applied to explore 
 open   issues on the use of PDFs in parton shower generators. The TMD 
 distributions with dynamical $z_M$  can be applied to investigate   
 non-perturbative Sudakov contributions, and  address open questions on the 
 energy dependence as well as mass dependence 
 of intrinsic transverse momentum.

%% file: acknDyn.tex
\section*{Acknowledgments}

We thank  H.~Jung  for useful discussions.  
A.~Lelek acknowledges funding by 
Research Foundation-Flanders (FWO) (applications' numbers: 1272421N, 1278325N). 
S.~Sadeghi Barzani acknowledges funding by the 
University of Antwerp Research Fund (BOF). 
S.~Taheri Monfared acknowledges the support of the 
German Research Foundation (DFG) under grant number 467467041.